\documentclass[10pt, doublecolumn, nofonttune]{IEEEtran}
\ifCLASSINFOpdf
\else
\fi
\usepackage{algorithm}
\usepackage[noend]{algorithmic}
\usepackage{bbold}
\usepackage{bbm}
\usepackage{cite}
\DeclareMathAlphabet{\pazocal}{OMS}{zplm}{m}{n}
\usepackage{graphicx}
\usepackage{textcomp}
\usepackage{soul}
\usepackage{xcolor}
\usepackage{amsthm, amsmath,amssymb,amsfonts}
\usepackage{xspace}
\newtheorem{theorem}{Theorem}
\usepackage{pgfplots}
\pgfplotsset{compat=1.10} 
\usepackage{float}
\usepackage{booktabs}
\RequirePackage{xstring}
\RequirePackage{xparse}
\RequirePackage{acro}
\usepackage{multirow} 
\usepackage{url}
\usepackage{subcaption}
\usepackage[capitalise]{cleveref}
\usepackage{xcolor}
\crefname{section}{Sect.}{Sect.}
\crefname{figure}{Figure}{Fig.}
\crefname{table}{Tab.}{Tab.}
\crefname{equation}{Eq.}{Eq.}
\usepackage[letterpaper]{geometry}
\usepackage{tikz}
\usepackage{pgfplots}
\pgfplotsset{compat=1.18}
\usepackage{todonotes}
\usepackage{tikz}

\newlength\fheight
\newlength\fwidth
\usepackage[subtle]{savetrees}
\usepackage[acronyms,nonumberlist,nopostdot,nomain,nogroupskip,acronymlists={hidden}]{glossaries}
\newglossary[algh]{hidden}{acrh}{acnh}{Hidden Acronyms}

\newacronym{3gpp}{3GPP}{3rd Generation Partnership Project}
\newacronym{4g}{4G}{4th generation}
\newacronym{5g}{5G}{5th generation}
\newacronym{6g}{6G}{6th generation}
\newacronym{5gc}{5GC}{5G Core}
\newacronym{adc}{ADC}{Analog to Digital Converter}
\newacronym{aerpaw}{AERPAW}{Aerial Experimentation and Research Platform for Advanced Wireless}
\newacronym{ai}{AI}{Artificial Intelligence}
\newacronym{aimd}{AIMD}{Additive Increase Multiplicative Decrease}
\newacronym{am}{AM}{Acknowledged Mode}
\newacronym{amc}{AMC}{Adaptive Modulation and Coding}
\newacronym{amf}{AMF}{Access and Mobility Management Function}
\newacronym{aops}{AOPS}{Adaptive Order Prediction Scheduling}
\newacronym{api}{API}{Application Programming Interface}
\newacronym{apn}{APN}{Access Point Name}
\newacronym{ap}{AP}{Application Protocol}
\newacronym{aqm}{AQM}{Active Queue Management}
\newacronym{ausf}{AUSF}{Authentication Server Function}
\newacronym{avc}{AVC}{Advanced Video Coding}
\newacronym{awgn}{AGWN}{Additive White Gaussian Noise}
\newacronym{balia}{BALIA}{Balanced Link Adaptation Algorithm}
\newacronym{bbu}{BBU}{Base Band Unit}
\newacronym{bdp}{BDP}{Bandwidth-Delay Product}
\newacronym{ber}{BER}{Bit Error Rate}
\newacronym{bf}{BF}{Beamforming}
\newacronym{bler}{BLER}{Block Error Rate}
\newacronym{brr}{BRR}{Bayesian Ridge Regressor}
\newacronym{bs}{BS}{Base Station}
\newacronym{bsr}{BSR}{Buffer Status Report}
\newacronym{bss}{BSS}{Business Support System}
\newacronym{ca}{CA}{Carrier Aggregation}
\newacronym{caas}{CaaS}{Connectivity-as-a-Service}
\newacronym{cb}{CB}{Code Block}
\newacronym{cc}{CC}{Congestion Control}
\newacronym{ccid}{CCID}{Congestion Control ID}
\newacronym{cco}{CC}{Carrier Component}
\newacronym{cd}{CD}{Continuous Delivery}
\newacronym{cdd}{CDD}{Cyclic Delay Diversity}
\newacronym{cdf}{CDF}{Cumulative Distribution Function}
\newacronym{cdn}{CDN}{Content Distribution Network}
\newacronym{cli}{CLI}{Command-line Interface}
\newacronym{cn}{CN}{Core Network}
\newacronym{codel}{CoDel}{Controlled Delay Management}
\newacronym{comac}{COMAC}{Converged Multi-Access and Core}
\newacronym{cord}{CORD}{Central Office Re-architected as a Datacenter}
\newacronym{cornet}{CORNET}{COgnitive Radio NETwork}
\newacronym{cosmos}{COSMOS}{Cloud Enhanced Open Software Defined Mobile Wireless Testbed for City-Scale Deployment}
\newacronym{cots}{COTS}{Commercial Off-the-Shelf}
\newacronym{cp}{CP}{Control Plane}
\newacronym{cyp}{CP}{Cyclic Prefix}
\newacronym{up}{UP}{User Plane}
\newacronym{cpu}{CPU}{Central Processing Unit}
\newacronym{cqi}{CQI}{Channel Quality Information}
\newacronym{cr}{CR}{Cognitive Radio}
\newacronym{cran}{CRAN}{Cloud \gls{ran}}
\newacronym{crs}{CRS}{Cell Reference Signal}
\newacronym{csi}{CSI}{Channel State Information}
\newacronym{csirs}{CSI-RS}{Channel State Information - Reference Signal}
\newacronym{cu}{CU}{Central Unit}
\newacronym{d2tcp}{D$^2$TCP}{Deadline-aware Data center TCP}
\newacronym{d3}{D$^3$}{Deadline-Driven Delivery}
\newacronym{dac}{DAC}{Digital to Analog Converter}
\newacronym{dag}{DAG}{Directed Acyclic Graph}
\newacronym{das}{DAS}{Distributed Antenna System}
\newacronym{dash}{DASH}{Dynamic Adaptive Streaming over HTTP}
\newacronym{dc}{DC}{Dual Connectivity}
\newacronym{dccp}{DCCP}{Datagram Congestion Control Protocol}
\newacronym{dce}{DCE}{Direct Code Execution}
\newacronym{dci}{DCI}{Downlink Control Information}
\newacronym{dctcp}{DCTCP}{Data Center TCP}
\newacronym{dl}{DL}{Downlink}
\newacronym{dmr}{DMR}{Deadline Miss Ratio}
\newacronym{dmrs}{DMRS}{DeModulation Reference Signal}
\newacronym{drlcc}{DRL-CC}{Deep Reinforcement Learning Congestion Control}
\newacronym{drs}{DRS}{Discovery Reference Signal}
\newacronym{du}{DU}{Distributed Unit}
\newacronym{e2e}{E2E}{end-to-end}
\newacronym{earfcn}{EARFCN}{E-UTRA Absolute Radio Frequency Channel Number}
\newacronym{ecaas}{ECaaS}{Edge-Cloud-as-a-Service}
\newacronym{ecn}{ECN}{Explicit Congestion Notification}
\newacronym{edf}{EDF}{Earliest Deadline First}
\newacronym{embb}{eMBB}{Enhanced Mobile Broadband}
\newacronym{empower}{EMPOWER}{EMpowering transatlantic PlatfOrms for advanced WirEless Research}
\newacronym{enb}{eNB}{evolved Node Base}
\newacronym{endc}{EN-DC}{E-UTRAN-\gls{nr} \gls{dc}}
\newacronym{epc}{EPC}{Evolved Packet Core}
\newacronym{eps}{EPS}{Evolved Packet System}
\newacronym{es}{ES}{Edge Server}
\newacronym{etsi}{ETSI}{European Telecommunications Standards Institute}
\newacronym[firstplural=Estimated Times of Arrival (ETAs)]{eta}{ETA}{Estimated Time of Arrival}
\newacronym{eutran}{E-UTRAN}{Evolved Universal Terrestrial Access Network}
\newacronym{faas}{FaaS}{Function-as-a-Service}
\newacronym{fapi}{FAPI}{Functional Application Platform Interface}
\newacronym{fdd}{FDD}{Frequency Division Duplexing}
\newacronym{fdm}{FDM}{Frequency Division Multiplexing}
\newacronym{fdma}{FDMA}{Frequency Division Multiple Access}
\newacronym{fed4fire}{FED4FIRE+}{Federation 4 Future Internet Research and Experimentation Plus}
\newacronym{fir}{FIR}{Finite Impulse Response}
\newacronym{fit}{FIT}{Future \acrlong{iot}}
\newacronym{fpga}{FPGA}{Field Programmable Gate Array}
\newacronym{fr2}{FR2}{Frequency Range 2}
\newacronym{fs}{FS}{Fast Switching}
\newacronym{fscc}{FSCC}{Flow Sharing Congestion Control}
\newacronym{ftp}{FTP}{File Transfer Protocol}
\newacronym{fw}{FW}{Flow Window}
\newacronym{ge}{GE}{Gaussian Elimination}
\newacronym{gnb}{gNB}{Next Generation Node Base}
\newacronym{gop}{GOP}{Group of Pictures}
\newacronym{gpr}{GPR}{Gaussian Process Regressor}
\newacronym{gpu}{GPU}{Graphics Processing Unit}
\newacronym{gtp}{GTP}{GPRS Tunneling Protocol}
\newacronym{gtpc}{GTP-C}{GPRS Tunnelling Protocol Control Plane}
\newacronym{gtpu}{GTP-U}{GPRS Tunnelling Protocol User Plane}
\newacronym{gtpv2c}{GTPv2-C}{\gls{gtp} v2 - Control}
\newacronym{gw}{GW}{Gateway}
\newacronym{harq}{HARQ}{Hybrid Automatic Repeat reQuest}
\newacronym{hetnet}{HetNet}{Heterogeneous Network}
\newacronym{hh}{HH}{Hard Handover}
\newacronym{hol}{HOL}{Head-of-Line}
\newacronym{hqf}{HQF}{Highest-quality-first}
\newacronym{hss}{HSS}{Home Subscription Server}
\newacronym{http}{HTTP}{HyperText Transfer Protocol}
\newacronym{ia}{IA}{Initial Access}
\newacronym{iab}{IAB}{Integrated Access and Backhaul}
\newacronym{ic}{IC}{Incident Command}
\newacronym{ietf}{IETF}{Internet Engineering Task Force}
\newacronym{imsi}{IMSI}{International Mobile Subscriber Identity}
\newacronym{imt}{IMT}{International Mobile Telecommunication}
\newacronym{iot}{IoT}{Internet of Things}
\newacronym{ip}{IP}{Internet Protocol}
\newacronym{itu}{ITU}{International Telecommunication Union}
\newacronym{kpi}{KPI}{Key Performance Indicator}
\newacronym{kpm}{KPM}{Key Performance Measurement}
\newacronym{kvm}{KVM}{Kernel-based Virtual Machine}
\newacronym{los}{LoS}{Line of Sight}
\newacronym{lsm}{LSM}{Link-to-System Mapping}
\newacronym{lstm}{LSTM}{Long Short Term Memory}
\newacronym{lte}{LTE}{Long Term Evolution}
\newacronym{lxc}{LXC}{Linux Container}
\newacronym{m2m}{M2M}{Machine to Machine}
\newacronym{mac}{MAC}{Medium Access Control}
\newacronym{manet}{MANET}{Mobile Ad Hoc Network}
\newacronym{mano}{MANO}{Management and Orchestration}
\newacronym{mc}{MC}{Multi-Connectivity}
\newacronym{mcc}{MCC}{Mobile Cloud Computing}
\newacronym{mchem}{MCHEM}{Massive Channel Emulator}
\newacronym{mcs}{MCS}{Modulation and Coding Scheme}
\newacronym{mec2}{MEC}{Multi-access Edge Computing}
\newacronym{mec}{MEC}{Mobile Edge Computing}
\newacronym{mfc}{MFC}{Mobile Fog Computing}
\newacronym{mgen}{MGEN}{Multi-Generator}
\newacronym{mi}{MI}{Mutual Information}
\newacronym{mib}{MIB}{Master Information Block}
\newacronym{miesm}{MIESM}{Mutual Information Based Effective SINR}
\newacronym{mimo}{MIMO}{Multiple Input, Multiple Output}
\newacronym{ml}{ML}{Machine Learning}
\newacronym{mlr}{MLR}{Maximum-local-rate}
\newacronym[plural=\gls{mme}s,firstplural=Mobility Management Entities (MMEs)]{mme}{MME}{Mobility Management Entity}
\newacronym{mmtc}{mMTC}{Massive Machine-Type Communications}
\newacronym{mmwave}{mmWave}{millimeter wave}
\newacronym{mpdccp}{MP-DCCP}{Multipath Datagram Congestion Control Protocol}
\newacronym{mptcp}{MPTCP}{Multipath TCP}
\newacronym{mr}{MR}{Maximum Rate}
\newacronym{mrdc}{MR-DC}{Multi \gls{rat} \gls{dc}}
\newacronym{mse}{MSE}{Mean Square Error}
\newacronym{mss}{MSS}{Maximum Segment Size}
\newacronym{mt}{MT}{Mobile Termination}
\newacronym{mtd}{MTD}{Machine-Type Device}
\newacronym{mtu}{MTU}{Maximum Transmission Unit}
\newacronym{mumimo}{MU-MIMO}{Multi-user \gls{mimo}}
\newacronym{mvno}{MVNO}{Mobile Virtual Network Operator}
\newacronym{nalu}{NALU}{Network Abstraction Layer Unit}
\newacronym{nas}{NAS}{Network Attached Storage}
\newacronym{nat}{NAT}{Network Address Translation}
\newacronym{nbiot}{NB-IoT}{Narrow Band IoT}
\newacronym{nfv}{NFV}{Network Function Virtualization}
\newacronym{nfvi}{NFVI}{Network Function Virtualization Infrastructure}
\newacronym{ni}{NI}{Network Interfaces}
\newacronym{nic}{NIC}{Network Interface Card}
\newacronym{now}{NOW}{Non Overlapping Window}
\newacronym{nsm}{NSM}{Network Service Mesh}
\newacronym{nr}{NR}{New Radio}
\newacronym{nrf}{NRF}{Network Repository Function}
\newacronym{nsa}{NSA}{Non Stand Alone}
\newacronym{nse}{NSE}{Network Slicing Engine}
\newacronym{nssf}{NSSF}{Network Slice Selection Function}
\newacronym{o2i}{O2I}{Outdoor to Indoor}
\newacronym{oai}{OAI}{OpenAirInterface}
\newacronym{oaicn}{OAI-CN}{\gls{oai} \acrlong{cn}}
\newacronym{oairan}{OAI-RAN}{\acrlong{oai} \acrlong{ran}}
\newacronym{oam}{OAM}{Operations, Administration and Maintenance}
\newacronym{ofdm}{OFDM}{Orthogonal Frequency Division Multiplexing}
\newacronym{olia}{OLIA}{Opportunistic Linked Increase Algorithm}
\newacronym{omec}{OMEC}{Open Mobile Evolved Core}
\newacronym{onap}{ONAP}{Open Network Automation Platform}
\newacronym{onf}{ONF}{Open Networking Foundation}
\newacronym{onos}{ONOS}{Open Networking Operating System}
\newacronym{oom}{OOM}{\gls{onap} Operations Manager}
\newacronym{opnfv}{OPNFV}{Open Platform for \gls{nfv}}
\newacronym{oran}{O-RAN}{Open Radio Access Network}
\newacronym{orbit}{ORBIT}{Open-Access Research Testbed for Next-Generation Wireless Networks}
\newacronym{os}{OS}{Operating System}
\newacronym{oss}{OSS}{Operations Support System}
\newacronym{pa}{PA}{Position-aware}
\newacronym{pase}{PASE}{Prioritization, Arbitration, and Self-adjusting Endpoints}
\newacronym{pawr}{PAWR}{Platforms for Advanced Wireless Research}
\newacronym{pbch}{PBCH}{Physical Broadcast Channel}
\newacronym{pcef}{PCEF}{Policy and Charging Enforcement Function}
\newacronym{pcfich}{PCFICH}{Physical Control Format Indicator Channel}
\newacronym{pcrf}{PCRF}{Policy and Charging Rules Function}
\newacronym{pdcch}{PDCCH}{Physical Downlink Control Channel}
\newacronym{pdcp}{PDCP}{Packet Data Convergence Protocol}
\newacronym{pdsch}{PDSCH}{Physical Downlink Shared Channel}
\newacronym{pdu}{PDU}{Packet Data Unit}
\newacronym{pf}{PF}{Proportional Fair}
\newacronym{pgw}{PGW}{Packet Gateway}
\newacronym{phich}{PHICH}{Physical Hybrid ARQ Indicator Channel}
\newacronym{phy}{PHY}{Physical}
\newacronym{pmch}{PMCH}{Physical Multicast Channel}
\newacronym{pmi}{PMI}{Precoding Matrix Indicators}
\newacronym{powder}{POWDER}{Platform for Open Wireless Data-driven Experimental Research}
\newacronym{ppo}{PPO}{Proximal Policy Optimization}
\newacronym{ppp}{PPP}{Poisson Point Process}
\newacronym{prach}{PRACH}{Physical Random Access Channel}
\newacronym{prb}{PRB}{Physical Resource Block}
\newacronym{psnr}{PSNR}{Peak Signal to Noise Ratio}
\newacronym{pss}{PSS}{Primary Synchronization Signal}
\newacronym{pucch}{PUCCH}{Physical Uplink Control Channel}
\newacronym{pusch}{PUSCH}{Physical Uplink Shared Channel}
\newacronym{qam}{QAM}{Quadrature Amplitude Modulation}
\newacronym{qci}{QCI}{\gls{qos} Class Identifier}
\newacronym{qoe}{QoE}{Quality of Experience}
\newacronym{qos}{QoS}{Quality of Service}
\newacronym{quic}{QUIC}{Quick UDP Internet Connections}
\newacronym{rach}{RACH}{Random Access Channel}
\newacronym{ran}{RAN}{Radio Access Network}
\newacronym[firstplural=Radio Access Technologies (RATs)]{rat}{RAT}{Radio Access Technology}
\newacronym{rbg}{RBG}{Resource Block Group}
\newacronym{rcn}{RCN}{Research Coordination Network}
\newacronym{rc}{RC}{RAN Control}
\newacronym{rec}{REC}{Radio Edge Cloud}
\newacronym{red}{RED}{Random Early Detection}
\newacronym{renew}{RENEW}{Reconfigurable Eco-system for Next-generation End-to-end Wireless}
\newacronym{rf}{RF}{Radio Frequency}
\newacronym{rfc}{RFC}{Request for Comments}
\newacronym{rfr}{RFR}{Random Forest Regressor}
\newacronym{ric}{RIC}{RAN Intelligent Controller}
\newacronym{rlc}{RLC}{Radio Link Control}
\newacronym{rlf}{RLF}{Radio Link Failure}
\newacronym{rlnc}{RLNC}{Random Linear Network Coding}
\newacronym{rmr}{RMR}{RIC Message Router}
\newacronym{rmse}{RMSE}{Root Mean Squared Error}
\newacronym{rnis}{RNIS}{Radio Network Information Service}
\newacronym{rr}{RR}{Round Robin}
\newacronym{rrc}{RRC}{Radio Resource Control}
\newacronym{rrm}{RRM}{Radio Resource Management}
\newacronym{rru}{RRU}{Remote Radio Unit}
\newacronym{rs}{RS}{Remote Server}
\newacronym{rsrp}{RSRP}{Reference Signal Received Power}
\newacronym{rsrq}{RSRQ}{Reference Signal Received Quality}
\newacronym{rss}{RSS}{Received Signal Strength}
\newacronym{rssi}{RSSI}{Received Signal Strength Indicator}
\newacronym{rtt}{RTT}{Round Trip Time}
\newacronym{ru}{RU}{Radio Unit}
\newacronym{rw}{RW}{Receive Window}
\newacronym{rx}{RX}{Receiver}
\newacronym{s1ap}{S1AP}{S1 Application Protocol}
\newacronym{sa}{SA}{standalone}
\newacronym{sack}{SACK}{Selective Acknowledgment}
\newacronym{sap}{SAP}{Service Access Point}
\newacronym{sc2}{SC2}{Spectrum Collaboration Challenge}
\newacronym{scef}{SCEF}{Service Capability Exposure Function}
\newacronym{sch}{SCH}{Secondary Cell Handover}
\newacronym{scoot}{SCOOT}{Split Cycle Offset Optimization Technique}
\newacronym{sctp}{SCTP}{Stream Control Transmission Protocol}
\newacronym{sdap}{SDAP}{Service Data Adaptation Protocol}
\newacronym{sdk}{SDK}{Software Development Kit}
\newacronym{sdm}{SDM}{Space Division Multiplexing}
\newacronym{sdma}{SDMA}{Spatial Division Multiple Access}
\newacronym{sdn}{SDN}{Software-defined Networking}
\newacronym{sdr}{SDR}{Software-defined Radio}
\newacronym{seba}{SEBA}{SDN-Enabled Broadband Access}
\newacronym{sgsn}{SGSN}{Serving GPRS Support Node}
\newacronym{sgw}{SGW}{Service Gateway}
\newacronym{si}{SI}{Study Item}
\newacronym{sib}{SIB}{Secondary Information Block}
\newacronym{sinr}{SINR}{Signal to Interference plus Noise Ratio}
\newacronym{sip}{SIP}{Session Initiation Protocol}
\newacronym{siso}{SISO}{Single Input, Single Output}
\newacronym{sla}{SLA}{Service Level Agreement}
\newacronym{sm}{SM}{Service Model}
\newacronym{smo}{SMO}{Service Management and Orchestration}
\newacronym{smsgmsc}{SMS-GMSC}{\gls{sms}-Gateway}
\newacronym{snr}{SNR}{Signal-to-Noise-Ratio}
\newacronym{son}{SON}{Self-Organizing Network}
\newacronym{sptcp}{SPTCP}{Single Path TCP}
\newacronym{srb}{SRB}{Service Radio Bearer}
\newacronym{srn}{SRN}{Standard Radio Node}
\newacronym{srs}{SRS}{Sounding Reference Signal}
\newacronym{ss}{SS}{Synchronization Signal}
\newacronym{sss}{SSS}{Secondary Synchronization Signal}
\newacronym{st}{ST}{Spanning Tree}
\newacronym{svc}{SVC}{Scalable Video Coding}
\newacronym{tb}{TB}{Transport Block}
\newacronym{tcp}{TCP}{Transmission Control Protocol}
\newacronym{tdd}{TDD}{Time Division Duplexing}
\newacronym{tdm}{TDM}{Time Division Multiplexing}
\newacronym{tdma}{TDMA}{Time Division Multiple Access}
\newacronym{tfl}{TfL}{Transport for London}
\newacronym{tfrc}{TFRC}{TCP-Friendly Rate Control}
\newacronym{tft}{TFT}{Traffic Flow Template}
\newacronym{tgen}{TGEN}{Traffic Generator}
\newacronym{tip}{TIP}{Telecom Infra Project}
\newacronym{tm}{TM}{Transparent Mode}
\newacronym{to}{TO}{Telco Operator}
\newacronym{tr}{TR}{Technical Report}
\newacronym{trp}{TRP}{Transmitter Receiver Pair}
\newacronym{ts}{TS}{Technical Specification}
\newacronym{tti}{TTI}{Transmission Time Interval}
\newacronym{ttt}{TTT}{Time-to-Trigger}
\newacronym{tx}{TX}{Transmitter}
\newacronym{uas}{UAS}{Unmanned Aerial System}
\newacronym{uav}{UAV}{Unmanned Aerial Vehicle}
\newacronym{udm}{UDM}{Unified Data Management}
\newacronym{udp}{UDP}{User Datagram Protocol}
\newacronym{udr}{UDR}{Unified Data Repository}
\newacronym{ue}{UE}{User Equipment}
\newacronym{uhd}{UHD}{\gls{usrp} Hardware Driver}
\newacronym{ul}{UL}{Uplink}
\newacronym{um}{UM}{Unacknowledged Mode}
\newacronym{uml}{UML}{Unified Modeling Language}
\newacronym{upa}{UPA}{Uniform Planar Array}
\newacronym{upf}{UPF}{User Plane Function}
\newacronym{urllc}{URLLC}{Ultra Reliable and Low Latency Communications}
\newacronym{usa}{U.S.}{United States}
\newacronym{usim}{USIM}{Universal Subscriber Identity Module}
\newacronym{usrp}{USRP}{Universal Software Radio Peripheral}
\newacronym{utc}{UTC}{Urban Traffic Control}
\newacronym{vim}{VIM}{Virtualization Infrastructure Manager}
\newacronym{vm}{VM}{Virtual Machine}
\newacronym{vnf}{VNF}{Virtual Network Function}
\newacronym{volte}{VoLTE}{Voice over \gls{lte}}
\newacronym{voltha}{VOLTHA}{Virtual OLT HArdware Abstraction}
\newacronym{vr}{VR}{Virtual Reality}
\newacronym{vran}{vRAN}{Virtualized \gls{ran}}
\newacronym{vss}{VSS}{Video Streaming Server}
\newacronym{wbf}{WBF}{Wired Bias Function}
\newacronym{wf}{WF}{Waterfilling}
\newacronym{wg}{WG}{Working Group}
\newacronym{wlan}{WLAN}{Wireless Local Area Network}
\newacronym{osm}{OSM}{Open Source \gls{nfv} Management and Orchestration}
\newacronym{pnf}{PNF}{Physical Network Function}
\newacronym{drl}{DRL}{Deep Reinforcement Learning}
\newacronym{mtc}{MTC}{Machine-type Communications}
\newacronym{osc}{OSC}{O-RAN Software Community}
\newacronym{mns}{MnS}{Management Services}
\newacronym{ves}{VES}{\gls{vnf} Event Stream}
\newacronym{ei}{EI}{Enrichment Information}
\newacronym{fh}{FH}{Fronthaul}
\newacronym{fft}{FFT}{Fast Fourier Transform}
\newacronym{laa}{LAA}{Licensed-Assisted Access}
\newacronym{plfs}{PLFS}{Physical Layer Frequency Signals}
\newacronym{ptp}{PTP}{Precision Time Protocol}
\newacronym{lidar}{LiDAR}{Light Detection And Ranging}
\newacronym{dem}{DEM}{Digital Elevation Model}
\newacronym{dtm}{DEM}{Digital Terrain Model}
\newacronym{dsm}{DEM}{Digital Surface Models}
\newacronym{ota}{OTA}{Over-The-Air}
\newacronym{ns}{NS}{Network Slicing}
\newacronym{ne}{NE}{Nash Equilibrium}
\newacronym{hf}{HF}{High Frequency}
\newacronym{noma}{NOMA}{Non-Orthogonal Multiple Access}
\newacronym{sre}{SRE}{Smart Radio Environment}
\newacronym{ris}{RIS}{Reconfigurable Intelligent Surface}
\newacronym{inp}{InP}{Infrastructure Provider}
\newacronym{smf}{SMF}{Slicing Magangement Framework}
\newacronym{nsn}{NSN}{Network Slicing Negotiation}
\newacronym{sms}{SMS}{Slicing MAC Scheduler}
\newacronym{brd}{BRD}{Best Response Dynamics}
\newacronym{dssbr}{DSSBR}{Double Step Smoothed Best Response}
\newacronym{poa}{PoA}{Price of Anarchy}
\newacronym{pos}{PoS}{Price of Stability}
\newacronym{milp}{MILP}{Mixed Integer-Linear Program}
\newacronym{pod}{PoD}{Price of DSSBR}
\newacronym{roc}{ROC}{Radio Overload Control}
\newacronym{ciot}{cIoT}{critical Internet of Things}
\newacronym{embbpr}{eMBB Pr.}{enhanced Mobile BroadBand Premium}
\newacronym{embbbs}{eMBB Bs.}{enhanced Mobile BroadBand Basic}
\newacronym{en}{EN}{Edge Node}
\newacronym{ec}{EC}{Edge Computing}
\newacronym{sp}{SP}{Service Provider}
\newacronym{me}{ME}{Market Equilibrium}
\newacronym{so}{SO}{Social Optimum}
\newacronym{wso}{WSO}{Weighted Social Optimum}
\newacronym{wsn}{WSN}{Wireless Sensor Network}
\newacronym{ps}{PS}{Proportional Sharing}
\newacronym{eg}{EG}{Eisenberg-Gale program}
\newacronym{pe}{PE}{Pareto Efficiency}
\newacronym{nsw}{NSW}{Nash Social Welfare}
\newacronym{ef}{EF}{Envy-Freeness}
\newacronym{sub6}{sub6GHz}{Below 6GHz}
\newacronym{ncr}{NCR}{Network-Controlled Repeater}
\newacronym{nlos}{NLoS}{Non-LoS}
\newacronym{src}{SRC}{Smart Radio Connection}
\newacronym{srd}{SRD}{Smart Radio Device}
\newacronym{cs}{CS}{Candidate Site}
\newacronym{tp}{TP}{Test Point}
\newacronym{fov}{FoV}{Field of View}
\newacronym{nrric}{Near-RT RIC}{Near Real-time RAN Intelligent Controller}
\newacronym{e2ap}{E2AP}{E2 Application Protocol}
\newacronym{e2sm}{E2SM}{E2 Service Model}
\newacronym{nrtric}{Non-RT RIC}{Non-Real-Time Ran Intelligent Controller}
\newacronym{itti}{ITTI}{Inter-task Interface}
\newacronym{bap}{BAP}{Backhaul Adaptation Protocol}
\newacronym{iabest}{IABEST}{Integrated Access and Backhaul Experimental large-Scale Tetbed}
\newacronym{teid}{TEID}{Tunnel Endpoint Identifier}
\newacronym{dlsch}{DL-SCH}{Downlink Shared Channel }
\newacronym{ulsch}{UL-SCH}{Uplink Shared Channel }
\newacronym{opex}{OpEx}{Operational Expenditure}
\newacronym{capex}{CapEx}{Capital Expenditure}
\newacronym{mno}{MNO}{Mobile Network Operator}
\newacronym{fr}{FR}{Frequency Range}

\begin{document}
\bstctlcite{IEEEexample:BSTcontrol}
\title{Joint Routing, Resource Allocation, and Energy Optimization for Integrated Access and Backhaul with Open RAN
\thanks{This work has been submitted to the IEEE for possible publication. Copyright may be transferred without notice, after which this version may no longer be accessible.}
\thanks{R. Prasad, M. Elkael, G. Gemmi, M. Polese, and T. Melodia are with the Institute for Intelligent Networked Systems, Northeastern University, Boston, MA 02115. Email: \{re.prasad, m.elkael, g.gemmi, m.polese, melodia\}@northeastern.edu.}
\thanks{O. Bushnaq, D. Mishra, P. Raut, and J. Simonjan are with the Technology Innovation Institute Abu Dhabi, United Arab Emirates. Email: \{osama.bushnaq,
debashisha.mishra, prasanna.raut, jennifer.simonjan\}@tii.ae.}
\thanks{This article is based upon work partially supported by the Cooperative Networked UAVs project funded by Technology Innovation Institute and by OUSD(R\&E) through Army Research Laboratory Cooperative Agreement Number W911NF-24-2-0065. The views and conclusions contained in this document are those of the authors and should not be interpreted as representing the official policies, either expressed or implied, of the Army Research Laboratory or the U.S. Government. The U.S. Government is authorized to reproduce and distribute reprints for Government purposes notwithstanding any copyright notation herein.}}


\author{
%\IEEEauthorblockN{Maxime 
Reshma Prasad,
Maxime Elkael,
Gabriele Gemmi,
Osama M. Bushnaq, 
Debashisha Mishra,\\
Prasanna Raut, 
Jennifer Simonjan,
Michele Polese, 
Tommaso Melodia}
%\IEEEauthorblockA{
%\IEEEauthorblockA{\IEEEauthorrefmark{2}Twentieth Century Fox, Springfield, USA}
%\IEEEauthorblockA{\IEEEauthorrefmark{3}Starfleet Academy, San Francisco, CA 96678 USA}
%\IEEEauthorblockA{\IEEEauthorrefmark{4}Tyrell Inc., 123 Replicant Street, Los Angeles, CA 90210 USA}% <-this % stops an unwanted space
%\thanks{Manuscript received December 1, 2012; revised August 26, 2015. 
%Corresponding author: M. Shell (email: http://www.michaelshell.org/contact.html).}

    \maketitle

% \IEEEtitleabstractindextext{%
\begin{abstract}
%Energy consumption represents a major part of the operating expenses of mobile network operators. With the densification foreseen with 5G and beyond, energy optimization has become a problem of crucial importance. 
As networks evolve towards 6G, \glspl{mno} must accommodate diverse requirements and at the same time manage  rising energy consumption. \gls{iab} networks facilitate dense cellular deployments with reduced infrastructure complexity. However, the multi-hop wireless backhauling in \gls{iab} networks necessitates proper routing and resource allocation decisions to meet the performance requirements. At the same time, cell densification makes energy optimization crucial for sustainable network operation. 
This paper addresses the joint optimization of routing and resource allocation in \gls{iab} networks through two distinct objectives: energy minimization and throughput maximization. We develop a novel capacity model that links power levels to achievable data rates, offering more precise capacity estimates than theoretical upper bounds. We propose two practical large-scale approaches to solve the optimization problems and leverage the closed-loop control framework introduced by the \gls{oran} architecture to integrate the solutions.  The approaches are evaluated on diverse scenarios built upon open data of two months of traffic collected by network operators in the city of Milan, Italy. Optimality gap analysis on small instances confirms that our method achieves optimal or near-optimal performance compared to exact MILP solutions. Results show that the proposed approaches effectively reduce number of activated nodes to save energy and achieve $\sim100$ Mbps of minimum data rate per \gls{ue} during peak hours of the day using spectrum within the \gls{fr}3, or upper midband.  The results validate the practical applicability of our framework for next-generation \gls{iab} network deployment and optimization.
\end{abstract}
% }

% \IEEEdisplaynontitleabstractindextext

% \IEEEpeerreviewmaketitle

\begin{IEEEkeywords}
IAB, Open RAN, Optimization, Energy
\end{IEEEkeywords}

% \hl{let's use gls consistently}

\glsresetall
\glsunset{oran}

\section{Introduction}

%Future wireless networks must address the challenge of supporting a massive number of connected devices while meeting stringent data rate requirements\cite{chen2020vision,lee2023towards}. %Leveraging ultra-dense deployment  and \gls{mmwave} frequencies, is envisioned as a promising approach to address this challenge and to enhance network capacity.
%\gls{iab} was introduced in 3GPP Release 16 as a cost-effective deployment solution that enables flexible network densification. \gls{iab} enables flexible deployment of \glspl{gnb} called \gls{iab}-nodes that rely on in-band wireless communications to connect to a set of \glspl{gnb} called \gls{iab}-donors with wired backhaul links. Thus, \gls{iab} reduces infrastructure costs and deployment complexity compared to traditional fiber-based solutions\cite{polese2020integrated}. %Figure~\ref{fig:iab_system} presents a typical \gls{iab} network with one \gls{iab} donor with fiber connectivity towards the core network and three \gls{iab}-nodes serving a set of \glspl{ue}.

Future wireless networks must address the challenge of supporting a massive number of connected devices while meeting stringent data rate requirements~\cite{chen2020vision,lee2023towards}. Network densification through small cell deployment is a key enabler for meeting these demands, yet its widespread adoption is constrained by the high cost of fiber backhaul infrastructure. \gls{iab}, introduced in 3GPP Release 16, offers a compelling alternative by enabling flexible densification through wireless backhauling, eliminating the need for dedicated fiber connections to every base station. Despite its standardization, practical \gls{iab} deployments remain limited, largely due to unresolved challenges in jointly optimizing routing, resource allocation, and energy consumption at scale. The significance of these challenges is underscored by 3GPP's introduction of the Wireless Access Backhaul (WAB) architecture in Release 19, which directly evolved from \gls{iab} to address its implementation complexity and rigid resource management requirements~\cite{rubaltelli2025multi} confirming that joint optimization of wireless backhaul networks remains an active and critical open problem.  This represents a significant missed opportunity: \gls{iab} not only reduces infrastructure costs but also enables operators to rapidly adapt network capacity to dynamic demands, such as large-scale events or temporary densification scenarios, without the long lead times associated with fiber deployment. Addressing these practical challenges is therefore critical to realizing the full potential of wireless backhaul architectures for next-generation network densification.

As end-to-end performance of the network depends on number of hops between donor and the \gls{ue}, the route selection, and the resource allocation, it is necessary to jointly optimize  \gls{iab}-network topology, routing, and resource allocation to achieve optimal performance\cite{polese2020integrated,yin2022routing}. Beyond these optimization aspects, energy-aware network management is crucial for sustainable \gls{iab} network management~\cite{carvalho2024energy}. Fixed \gls{iab} nodes enable long-term network planning; however, they do not need to remain active continuously. Instead, they can be selectively activated based on network load, helping to minimize energy consumption. As shown in~\cite{feng2017base,gemmi2023joint,teng2021joint}, dynamically activating and deactivating \glspl{gnb} with respect to the current load of the network  reduces the energy footprint of the system upto 47\%. Power control is another aspect which is crucial for energy optimization, as nodes operating at optimal power levels decrease overall energy consumption. Additionally, power control is essential for interference management, which can affect network capacity. Consequently, the underlying capacity model becomes more dynamic and complex, necessitating a more nuanced optimization approach.

The implementation of these comprehensive optimization strategies requires flexible and programmable network architectures. Such capabilities are enabled by the \gls{ran} flexibility and programmability introduced by \gls{oran} ALLIANCE and the extension of \gls{oran} architecture to support \gls{iab} networks proposed in~\cite{moro2023toward}. The \gls{oran} architecture allows seamless integration of optimization logic directly into the RAN by exposing  interfaces such as E2 and O1, and enabling intelligent closed-loop control via the \gls{ric}~\cite{polese2023understanding}.

%\begin{figure}
 %   \centering
 %   \includegraphics[width=0.5\textwidth]{figures/iab.pdf}
 %   \caption{IAB network}
 %   \label{fig:iab_system}
%\end{figure}

\subsection{Related Works}
\label{sec:sota}

In recent years, the topic of \gls{iab} topology optimization and routing has been of significant interest to the research community. Recent works include genetic algorithms for throughput and coverage optimization in \gls{mmwave} \gls{iab}~\cite{madapatha2021topology}, dynamic programming for capacity optimization~\cite{simsek2021optimal} and reinforcement learning for latency optimization~\cite{yin2022routing}. 
In~\cite{ajayi2023machine}, the problem is modeled as a multi-commodity flow problem~\cite{ouorou2000survey} which is solved by a combination of machine learning and linear programming. Though the work focuses on throughput maximization and considers routing, it overlooks power control, which is a critical factor that impacts the achievable data rate. Similarly, \cite{orhan2021connection} explores connection management and throughput maximization in \gls{oran} architectures leveraging deep reinforcement learning. However, rather than \gls{iab} networks, the work focuses on traditional cellular networks.  

On the other hand, energy efficiency of cellular networks is a long-standing concern, with various works typically tackling the problem through techniques such as power allocation~\cite{zappone2015energy,he2013energy}, sleep mode optimization~\cite{salahdine2021survey}, and leveraging network heterogeneity~\cite{soh2013energy}.
Recently, a few works have explored \gls{iab} topology optimization with energy-related objectives. This mainly includes~\cite{shang2024energy}, where the problem is modeled considering beamforming and using the Shannon capacity model. The authors then formulate a mixed integer non-linear program, which they solve heuristically by reformulating the objective function, using continuous relaxations and a greedy algorithm. However, the work does not consider dynamic activation/deactivation of \gls{iab}-nodes, enabling sleep mode optimization.  Though the authors develop a detailed interference model, the use of Shannon capacity leads to theoretical upper bounds on achievable data rates.  Other important works on energy optimization include \cite{zhang2024freshness} where energy-efficiency is optimized with respect to the Age-of-Information, and our previous work \cite{gemmi2023joint} where the topology is optimized for energy in a simplified model which does not consider power allocation and interference.

A concept related to \gls{iab} is that of mesh and wireless sensor networks, where a set of sensors (typically \gls{iot} devices) are directly interconnected via wireless links, without a central entity such as the \gls{gnb} to coordinate the communication. In that context, numerous papers tackle energy efficiency~\cite{salhieh2001power, torkestani2013energy,rault2014energy}.
However, the optimization task typically is very different from \gls{iab}, as these networks are assumed to run on battery and have a mesh structure, compared to a tree typically used in \gls{iab} systems. This makes the problem mostly akin to keeping the network on as long as possible and routing more flexible. On the other hand, \gls{iab} networks typically have access to a stable power source, and the main concern with energy efficiency is to reduce the economic cost and carbon footprint of the network. These differences give rise to significantly diverging problems.

%One of the key limitations of existing \gls{iab} optimization works is the dependence on Shannon capacity models\cite{shang2024energy,carvalho2024energy}. This can lead to overestimation of achievable data rate. Thus there is a need to have realistic capacity model that can be leveraged for practical \gls{iab} deployments.

Finally beyond \gls{iab} networks, the problem of tackling \gls{sinr} constraints in optimization problems is long-standing issue which our article also tackles. In some simple cases, this type of formulations can be solved through convex reformulations \cite{boyd2004convex}, which has been leveraged in various related contexts such as beamforming \cite{shi2016sinr} and power allocation \cite{hao2017power}.
However, these approaches overlook inter dependencies between power allocation and link capacity, and treat \gls{sinr} constraints in isolation.

\begin{figure*}[!t]
    \centering
    \vspace{-0.1cm}
    \includegraphics[width=0.8\textwidth]{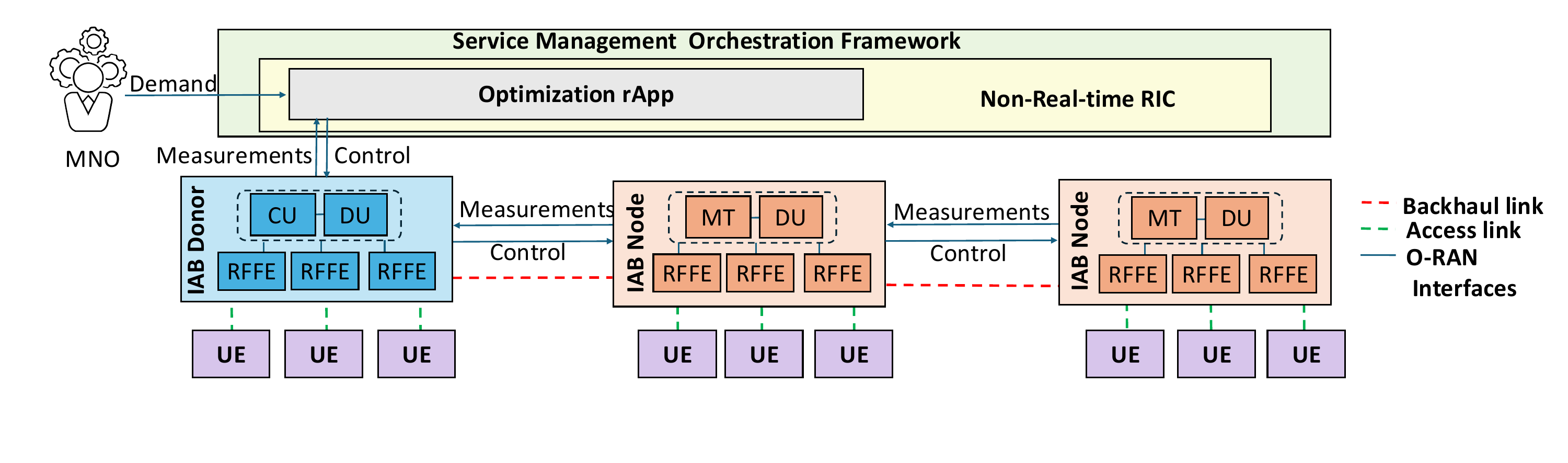}
     \vspace{-2cm}
    \caption{Example of an \gls{iab} Network optimized by O-RAN components. Each \gls{iab}-node includes a baseband unit (\gls{du} and \gls{mt}) and \gls{rf} frontends (RFFEs).}
    \label{fig:iab}
    \vspace{-0.5cm}
\end{figure*}

\subsection{Our Contributions}

In this paper, we address the challenge of designing dynamic \gls{iab} networks that aim to provide connectivity for a set of \glspl{ue} meeting the service requirements and minimizing energy consumption. We study the problem of identifying a tree from the network which has \gls{iab}-donor as root, \gls{iab}-nodes as intermediate nodes, and \glspl{ue} as leaves. To achieve this goal, we propose comprehensive optimization frameworks that include \gls{3gpp}-compliant realistic capacity modeling. 
To enable a practical deployment, the optimization solution that defines and controls the \gls{iab} network topology is designed to be deployed as an rApp that runs on the non-real-time RIC within the \gls{oran} architecture, as also shown in Fig.~\ref{fig:iab}. 

The main contributions of this paper are as follows:
\begin{itemize}
    \item \textbf{Joint energy minimization:} We formulate a joint routing, resource allocation, and energy minimization problem for IAB networks that simultaneously optimizes dynamic node activation, power control, and TDMA scheduling, while guaranteeing minimum per-UE data rate requirements.
    
    \item \textbf{Joint throughput maximization:} We formulate a joint routing, resource allocation, and power control problem that maximizes the minimum data rate across all UEs, ensuring fairness. Unlike fixed-power formulations, joint power allocation enables strictly better throughput performance.
    
    \item \textbf{3GPP-compliant capacity model:} We develop a novel capacity model based on 3GPP MCS tables that directly links transmission power to achievable data rates through a tractable linear reformulation of the non-convex SINR ratio constraint. This produces realistic capacity estimates and enables integration of power-dependent capacity constraints into a \gls{milp} framework, unlike Shannon capacity models used in prior work.
    
    \item \textbf{Two practical large-scale algorithms:} We develop a local search method based on fix-and-optimize decomposition, and a selective-reduction method with a feasibility guarantee theorem, both with computational complexity analysis.
    
    \item \textbf{Comprehensive real-world evaluation:} We evaluate both methods across two urban scenarios using two months of real network operator traffic data from Milan, Italy, at two frequency bands (FR1 and FR3), with comparison against exact MILP solutions and a literature baseline.
\end{itemize}
Our detailed analysis of the proposed solutions over realistic scenarios shows their distinct performance characteristics and trade-offs. This enables the network operators to make informed method selections according to the deployment and performance requirements and frequency band characteristics.

The rest of the paper is organized as follows. In Sec. \ref{sect:system}, we discuss the system model in detail, including the energy, channel, and capacity models. We formulate the optimization problems in Sec. \ref{sect:problem}. Sec. \ref{sect:heuristics} presents  practical solutions for the problems. In Sec. \ref{sect:setup}, we discuss the evaluation set up, followed by discussion of performance evaluation in Sec. \ref{sect:results}. Finally, in Sec. \ref{sect:conclusion}, we conclude the paper.

\section{System Model}
\label{sect:system}

\subsection{\gls{iab} Architecture}

We consider a downlink transmission scenario in an \gls{iab} network with \gls{iab}-nodes assisting the \gls{iab}-donor to provide service to a set of \glspl{ue}.  While we focus on the downlink direction, the proposed formulation is equally applicable to the uplink with appropriate adjustments to the interference model and power constraints, as the underlying routing, scheduling, and topology optimization structure is direction-agnostic.
%Each \gls{iab} node has three components: a \gls{mt} for wireless backhaul connection with other \gls{iab}-node or \gls{iab}-donor, a \gls{du} which is connected to the \gls{ru}, and multiple \gls{rf} front ends that provide wireless connectivity to the \glspl{ue}. Multiple \gls{rf} front ends enable \gls{ru} to support different sectors. 
Each \gls{iab}-node is divided into two main parts: the baseband processing unit, which includes both the \gls{du} and the \gls{mt}, and the radio frontend, which enables both the DU and the MT to communicate with other \gls{iab} nodes or \glspl{ue}.  In order to support spatial diversity, the \gls{rf} frontend is divided into up to three distinct sectors, which can be activated on demand. The \gls{iab}-donor is connected to the core network with a wired (fiber) backhaul and has a \gls{cu}, a \gls{du} connected to the \gls{cu}, and the \gls{rf} frontend, also divided into three distinct sectors. An example of \gls{iab} network is presented in Fig. \ref{fig:iab}.

Let us now introduce the problem formally. We start from a weighted directed graph $\pazocal G = (\pazocal V, \pazocal E)$, called measurements graph, constructed from \gls{ue} measurement reports.
The network topology consists of \glspl{ue}, \gls{iab}-nodes, and \gls{iab}-donor. For detailed and more accurate graph representation, we split an \gls{iab}-node to its \gls{mt}+\gls{du} unit and individual \gls{rf} front end units corresponding to each sector (we assume that each of the \gls{mt}+\gls{du} unit are interconnected to the \gls{rf} front end via a wired link, see Fig.~\ref{fig:iab}). As frontends serve different sectors with different transmission capabilities, this decomposition enables to optimize power and resource allocation individually.  Then, the vertices of the graph include:
%Thus, for detailed and more accurate graph representation, we consider the nodes as: \glspl{ue}; \gls{iab} front end radios, where each node corresponds to a \gls{iab}-node with a a specific front end.%with associated \gls{mt}, \gls{du}, \gls{ru} and . 

%

\begin{itemize}
    \item \glspl{ue}, denoted as $\pazocal U \subset \pazocal V$, which need to connect to the network in order to receive downlink traffic;
    \item \gls{iab} frontend radios, denoted as $\pazocal R \subset \pazocal V$, where each node corresponds to a \gls{rf} frontend in an \gls{iab}-node; 
    \item \gls{mt}+\gls{du} units, denoted as $\pazocal M \subset \pazocal V$. Each unit is connected to one or several \gls{rf} frontends.
    %We note that a single physical \gls{iab}-node with multiple \gls{rf} front ends will correspond to multiple nodes in our graph representation and in the rest of the paper, \gls{iab}-node refers to an individual front end node.
    %\item \glspl{ru}, which are nodes equipped with a radio. They are tasked to interconnect with other RUs to form the \gls{iab} topology and to provide wireless connectivity to the \glspl{ue}; 
    %
\end{itemize}

%Together, a DU and its connected RUs form an \gls{iab}-node. One of the \gls{iab}-nodes is the \gls{iab}-donor, which has access to a wired (fiber) backhaul behind which the core network is.

%We denote the set of \glspl{ue} as $\pazocal U \subset \pazocal V$, $\pazocal R \subset \pazocal V$ the set of front end nodes. %, and the DU of the \gls{iab}-donor as $t \in \pazocal V$.

 We denote by $\pazocal W \subset \pazocal E$ the set of wireless links, which are weighted by their pathloss, and by $\pazocal F \subset \pazocal E$ the internal connections which connect \gls{mt}+\gls{du} unit to their respective \gls{rf} frontends.  %Each edge $(u,v) \in \pazocal W$ of this graph represents a potentially usable wireless link between each node in the graph weighted by their pathloss.
Since the goal is to find a tree representing the routing from the donor  towards each \gls{ue}, the edges of the graph will be directed accordingly. Access links are always directed from \gls{iab} frontend radio to \glspl{ue}, with the \gls{iab} frontend radio as the source and the \gls{ue} as the destination. Backhaul links from other \gls{iab}-nodes or \gls{iab}-donor are always directed towards the \gls{mt}+\gls{du} unit. %always point towards its RUs. %The links between RUs instead can be used in one or the other direction to build the \gls{iab}-tree. Similarly, the links between non-donor DUs and RUs can also be used in both direction. 
%\cref{subfig:graph} and \cref{subfig:tree} report an example of a measurements graph and a possible \gls{iab}-tree.

%Local detailed information on the feasibility of wireless links between \glspl{ue} and \gls{iab}-nodes is available on each \gls{iab}-node. 
Each \gls{iab}-node maintains information about \glspl{ue} and the potential wireless links through measurements reports.  The \gls{oran} architecture allows extensions to standard interfaces so that we can assume that the local information can be collected by an rApp, running on the non-real-time \gls{ric}, which reconstructs the measurements graph we mentioned above.  The measurement reports are forwarded from \gls{iab}-node \glspl{du} to the non-real-time \gls{ric} via the E2 interface, where the rApp reconstructs the measurements graph $\pazocal{G}$.
Then, the optimization algorithm periodically runs and pushes the optimized topology to the \gls{ran} through the O1 interface (we discuss extensions of the O-RAN architecture to enable IAB in~\cite{moro2023toward}).  Note that we take into consideration periodic updates of the topology with a period in the order of minutes, so we assume that disabled nodes wake up to receive an updated topology with a similar schedule. This schedule is also perfectly compatible with the non-real-time \gls{ric} closed-loop time constraints.
Without loss of generality in the following model we will assume the optimization of a single tree, but the proposed optimization model can be trivially adapted to optimize multiple trees.

% \begin{figure}
%     % \hfill
%     \begin{subfigure}{.3\linewidth}
%       \centering
%       \input{figures/graph.tex}
%         \subcaption[]{}
%         \label{subfig:graph}
%     \end{subfigure}%
%     \hfill
%     \begin{subfigure}{.3\linewidth}
%       \centering
%       \input{figures/tree.tex}
%       \subcaption[]{}
%       \label{subfig:tree}
%     \end{subfigure}%
%     \hfill~
%     \vspace{5pt}
%     \caption{Example of a measurements graph $\pazocal G$ (a) and a possible \gls{iab} Tree $\pazocal T$ (b). \gls{iab}-donors are depicted in red, \gls{iab}-nodes in blue, \glspl{ue} in black, and deactivated \gls{iab}-nodes in light blue.}\label{fig:graph}
% \end{figure}

\subsection{Energy Model}
\subsubsection{Power consumption as a function of transmission power}
We adopt the EARTH  model\cite{auer2011much} to model energy consumed by \gls{rf} frontends formulated as:
\begin{equation}
    P_{in} = 
    \begin{cases}
    N_{TRX} \times P_0 + \Delta_p P_{tx}, \quad& 0 < P_{tx} \leq P_{max} \\
    N_{TRX} \times P_{sleep} & P_{tx} = 0
    \end{cases}    
\end{equation}
where $N_{TRX}$ is the number of RF transmission chains per frontend, $P_0$ is the power consumed when the \gls{rf} frontend is not transmitting, $\Delta_p$ is the rate of increase of the power consumption given the current transmission power $P_{tx}$, and $P_{max}$ is the maximum transmission power. $N_{TRX}, P_0, \Delta_p, P_{max}$ are constant parameters which depend on the \gls{rf} frontend configuration, hence this model is linear. %Furthermore, given that \gls{iab} nodes rely solely on wireless backhaul links, it is not feasible to completely turn them off. For this reason, 
Even if a node is not actively used in the topology, we assume it is in sleep mode, consuming a minimal amount of power $P_{sleep}$. Also, it wakes up periodically to check for potential inclusion in the updated \gls{iab} topology. Our model focuses on \gls{rf} frontend power consumption, as baseband processing power (i.e., \gls{du}/\gls{cu}) remains essentially constant  and therefore does not influence topology optimization decisions. This is directly validated by real-world experiments on disaggregated O-RAN small cell deployments~\cite{shirkhani2026tenoran}, which closely map to \gls{iab} node configurations.

\subsubsection{Scheduling Considerations}

Our model incorporates scheduling decisions to manage resources. Since the whole network operates using the same spectrum (\textit{i.e.,} in-band \gls{iab}), we use \gls{tdma} scheduler that allocates radio resources  to each radio link. This means we could obtain solutions where the radio is not transmitting for significant portions of the time. In those moments where the radio is not transmitting, our model deactivates the power amplifier of the associated \gls{rf} frontend to save energy \cite{razzac2023advanced}. 
% We assume that the radios we work with implement such a mechanism. 
Hence, we derive our energy model from the EARTH model as:
\begin{equation}
    P_{in} = 
    \begin{cases}
    N_{TRX} \times P_0 + \alpha \times \Delta_p P_{tx}, \quad& 0 < P_{tx} \leq P_{max} \\
    N_{TRX} \times P_{sleep} & P_{tx} = 0
    \end{cases}
    \label{eq:earth}
\end{equation}
where $\alpha$ is a number between 0 and 1 which indicates the proportion of the time when the radio is transmitting. %Also, note that this model is a lower bound: indeed, it assumes that it is possible to turn off the PA at any instant when the BS is not transmitting. This may not be the case, as the time to turn on and off the PA is non-zero (a full cycle lasts at least in the order of 0.15ms, and the consumption of the PA is non-zero during the transitions from the ON state to the OFF state and during the transition back to the ON state \cite{Tijani}.
\subsection{Channel Model}
Let us now introduce our channel model. %We assume that each \gls{gnb} is capable of measuring the path loss towards the other BS and the \glspl{ue}. 
%\todo{maxime: This requires more explanation, idk if the path loss is readily available or if we would have to reverse it from the measured SNR and noise estimate ?}
%This can be done using the measurements provided by the 5G \gls{gnb}.
%We assume that each \gls{iab}-node or \gls{iab}-donor obtains the path loss information towards the other \gls{iab}-nodes and the \glspl{ue}  through the following process. \glspl{ue} measure \gls{rsrp} and reports it to the serving node. %Each serving \gls{iab}-node or \gls{iab}-donor calculates pathloss from this reported \gls{rsrp} values. 
We assume that each \gls{iab}-node or \gls{iab}-donor obtains the receiving power at other \gls{iab}-nodes and \glspl{ue} through \gls{rsrp} measurement reports. 
% \glspl{ue} measure \gls{rsrp} and reports it to the serving node. 
In our simulations and model, the \gls{rsrp} follows the 3GPP UMi \cite{3gpp38901} channel, combined with ray tracing to evaluate whether \glspl{ue} and nodes are \gls{los} or not.  Indoor \glspl{ue} are considered always \gls{nlos} and we add an indoor-to-outdoor loss. Since the buildings are concrete, we use the O2I-high loss \cite{3gpp38901}.
% This information can be applied on the Friis formula to measure the power on the receiving end, 
% \begin{equation}
%     P^u_{rx}=P^u_{tx} \cdot G^u_{tx} \cdot G^u_{rx} \cdot (\dfrac{\omega}{4\pi d})^2,
%     %P^u_{rx}=\frac{P^u_{tx} \cdot G^u_{tx} \cdot G^u_{rx}}{PL_{v,u} ,
% \end{equation}
% where $G_{tx}$ and $G_{rx}$ are transmission and reception antenna gains, $d$ is the distance between them and $\omega$ is the effective aperture area of the receiving antenna.

\subsection{Capacity Model}
\label{sec:cap_model}
We design a capacity model with the goal to derive a function that defines the achievable link capacity, given a set of 5G \glspl{gnb} and \glspl{ue}, and the respective power levels.
Unlike Shannon capacity which provides theoretical upper bounds and creates non-convex logarithmic constraints in optimization, our model uses the discrete MCS structure from 3GPP specifications, yielding distinct capacity levels that match real 5G \gls{nr} performance. This discrete formulation enables reformulation into tractable linear constraints while maintaining realistic capacity estimates. The resulting model couples power allocation to achievable data rates in a computationally efficient manner suitable for large-scale IAB optimization. 

%We derive this function from the modulation encoding schemes used by \gls{oai}, which are defined by the 5G standard: from a given Bandwidth and Numerology, we derive the corresponding number of Physical Resource Blocks (which are implemented in \gls{oai} as a lookup table). Then, for each of the possible 28 \gls{mcs} of 5G, we use the formula defined by the \gls{3gpp} \cite{TS-38.306} to calculate the capacity of a link given the \gls{mcs}, the number of available \glspl{prb} and the number of \gls{mimo} layers. The formula is the following\bc{confirm}:
The capacity is calculated using the \gls{3gpp} standard formula defined in \cite{TS-38.306}  that considers \gls{mcs}, available physical resource blocks, and MIMO layers which is as follows:
\begin{align}
    C = 10^{-6}\sum_{j=1}^J \frac{Q_{j,m} \cdot f_j \cdot v_j \cdot C_{max} \cdot N_{B(j),\mu} \cdot 12}{T_{s,\mu}} \cdot (1 - OH_j).
    \label{eq:capacity}
\end{align}
Here, $Q_{j,m}$ is the modulation order, $J$ is the number of component carriers aggregated in a band, $f_j$ is the scaling factor  reported by \gls{ue} Capability information that is used to reflect maximum number of layers and modulation order with the band combination, $v_j$ is the number of MIMO layers, $N_{B(j),\mu}$ is the number of \glspl{prb} allocated in bandwidth $B(j)$ for numerology $\mu$, $T_{s,\mu}$ is the average symbol duration that depends on numerology, $OH_j$ is the overhead factor.
%\todo{add the formula}.

The capacity defined in \eqref{eq:capacity} depends on the modulation order $Q_{j,m}$, which is determined by the MCS index $m$. We can then leverage the table from \gls{3gpp} that associates \gls{sinr} thresholds with \gls{mcs} values to establish the relationship following a three-step process:
The first step is to calculate \gls{sinr} of the radio link between $u$ and $v$ using transmission power and interference defined as follows:
\begin{align}
    &SINR(u,v) = 10\cdot \log_{10} \frac{S_{u, v}}{I_{u, v}}
\end{align}
With
\begin{align}
    &S_{u, v} = P_{tx}^u \cdot 10^{G_{tx}^u/10} \cdot 10^{G_{rx}^v/10} \cdot 10^{PL_{u, v}/10} \label{eq:signal} \\
    &I_{u, v} = \sum\limits_{r \in \pazocal R \setminus \{ u, v \}} P^r_{tx} \cdot 10^{G_{tx}^v/10} \cdot 10^{G_{rx}^r/10} \cdot 10^{PL_{r, v}/10} \label{eq:interferences},
\end{align}
where $P_{tx}^i$ is the power of node $i$ in mW, $G_{tx}^i$ its transmission gain (in dBm), $G_{rx}^i$ its reception gain (in dBm) and $PL_{i,j}$ is the pathloss (in dBm) between $i$ and $j$.

The second step is to determine \gls{mcs} index m by mapping \gls{sinr} to \gls{3gpp} thresholds  by leveraging the table from \gls{3gpp} that associates \gls{sinr} thresholds with \gls{mcs} values. The third step is to apply corresponding modulation order $Q_{j,m}$ in \eqref{eq:capacity} to compute achievable capacity.

%Since this formula gives us the capacity of the link given the \gls{mcs} (the other parameters are assumed fixed for each \gls{gnb}), we leverage the table from \gls{3gpp} that associates \gls{sinr} thresholds with \gls{mcs} values to establish the relationship.

\begin{figure}[!t]
    \centering
        \setlength\fwidth{0.7\columnwidth} 
        \setlength\fheight{0.3\columnwidth} 
        
            % This file was created with tikzplotlib v0.10.1.post13.
\begin{tikzpicture}

\definecolor{darkgrey176}{RGB}{176,176,176}

\begin{axis}[
width=0.85\linewidth,
height=0.37\linewidth,
tick align=outside,
tick pos=left,
% title={5G NR Link Capacity X vs SINR
% 4x4 MIMO, 100 MHz Bandwidth},
x grid style={darkgrey176},
xlabel={SINR (dB)},
xmajorgrids,
xmin=-10, xmax=40,
xtick style={color=black},
y grid style={darkgrey176},
ylabel={Capacity [Mbps]},
ymajorgrids,
ymin=0, ymax=1349.6175693,
ytick style={color=black},
font=\scriptsize,
%y filter/.code {\pgfmathparse{#1/1000}\pgfmathresult}
]
\addplot [line width=1pt, blue]
table {%
-20 0
-19.6984924623116 0
-19.3969849246231 0
-19.0954773869347 0
-18.7939698492462 0
-18.4924623115578 0
-18.1909547738693 0
-17.8894472361809 0
-17.5879396984925 0
-17.286432160804 0
-16.9849246231156 0
-16.6834170854271 0
-16.3819095477387 0
-16.0804020100502 0
-15.7788944723618 0
-15.4773869346734 0
-15.1758793969849 0
-14.8743718592965 0
-14.572864321608 0
-14.2713567839196 0
-13.9698492462312 0
-13.6683417085427 0
-13.3668341708543 0
-13.0653266331658 0
-12.7638190954774 0
-12.4623115577889 0
-12.1608040201005 0
-11.8592964824121 0
-11.5577889447236 0
-11.2562814070352 0
-10.9547738693467 0
-10.6532663316583 0
-10.3517587939699 0
-10.0502512562814 0
-9.74874371859297 0
-9.44723618090452 0
-9.14572864321608 0
-8.84422110552764 0
-8.5427135678392 0
-8.24120603015075 0
-7.93969849246231 0
-7.63819095477387 0
-7.33668341708543 0
-7.03517587939699 0
-6.73366834170854 0
-6.4321608040201 0
-6.13065326633166 0
-5.82914572864322 0
-5.52763819095478 0
-5.22613065326633 0
-4.92462311557789 51.76899
-4.62311557788945 51.76899
-4.32160804020101 108.28347075
-4.02010050251256 108.28347075
-3.71859296482412 108.28347075
-3.41708542713568 108.28347075
-3.11557788944724 108.28347075
-2.81407035175879 108.28347075
-2.51256281407035 108.28347075
-2.21105527638191 108.28347075
-1.90954773869347 108.28347075
-1.60804020100502 108.28347075
-1.30653266331658 108.28347075
-1.00502512562814 108.28347075
-0.7035175879397 108.28347075
-0.402010050251256 108.28347075
-0.100502512562816 108.28347075
0.201005025125625 132.873741
0.502512562814069 132.873741
0.804020100502512 132.873741
1.10552763819095 132.873741
1.40703517587939 163.50372675
1.70854271356784 163.50372675
2.01005025125628 163.50372675
2.31155778894472 163.50372675
2.61306532663316 163.50372675
2.91457286432161 163.50372675
3.21608040201005 193.70230425
3.51758793969849 193.70230425
3.81909547738693 193.70230425
4.12060301507537 193.70230425
4.42211055276382 193.70230425
4.72361809045226 226.9207395
5.0251256281407 226.9207395
5.32663316582914 226.9207395
5.62814070351759 226.9207395
5.92964824120603 226.9207395
6.23115577889447 259.7077665
6.53266331658291 259.7077665
6.83417085427136 293.35761
7.1356783919598 293.35761
7.43718592964824 293.35761
7.73869346733668 293.35761
8.04020100502512 293.35761
8.34170854271357 326.144637
8.64321608040201 326.144637
8.94472361809045 326.144637
9.24623115577889 374.462361
9.54773869346733 374.462361
9.84924623115577 374.462361
10.1507537688442 374.462361
10.4522613065327 422.780085
10.7537688442211 422.780085
11.0552763819095 422.780085
11.356783919598 422.780085
11.6582914572864 422.780085
11.9597989949749 422.780085
12.2613065326633 477.1375245
12.5628140703518 477.1375245
12.8643216080402 477.1375245
13.1658291457286 477.1375245
13.4673366834171 531.494964
13.7688442211055 531.494964
14.070351758794 566.8704405
14.3718592964824 566.8704405
14.6733668341708 603.1087335
14.9748743718593 603.1087335
15.2763819095477 603.1087335
15.5778894472362 603.1087335
15.8793969849246 603.1087335
16.1809045226131 669.11419575
16.4824120603015 669.11419575
16.78391959799 669.11419575
17.0854271356784 733.82543325
17.3869346733668 733.82543325
17.6884422110553 733.82543325
17.9899497487437 797.242446
18.2914572864322 797.242446
18.5929648241206 797.242446
18.894472361809 797.242446
19.1959798994975 797.242446
19.4974874371859 797.242446
19.7989949748744 861.9536835
20.1005025125628 861.9536835
20.4020100502512 861.9536835
20.7035175879397 861.9536835
21.0050251256281 930.54759525
21.3065326633166 930.54759525
21.608040201005 930.54759525
21.9095477386935 930.54759525
22.2110552763819 930.54759525
22.5125628140703 999.141507
22.8140703517588 999.141507
23.1155778894472 999.141507
23.4170854271357 999.141507
23.7185929648241 1063.8527445
24.0201005025126 1063.8527445
24.321608040201 1063.8527445
24.6231155778894 1063.8527445
24.9246231155779 1063.8527445
25.2261306532663 1063.8527445
25.5276381909548 1063.8527445
25.8291457286432 1063.8527445
26.1306532663317 1063.8527445
26.4321608040201 1063.8527445
26.7336683417085 1129.85820675
27.035175879397 1129.85820675
27.3366834170854 1129.85820675
27.6381909547739 1129.85820675
27.9396984924623 1177.7445225
28.2412060301507 1177.7445225
28.5427135678392 1177.7445225
28.8442211055276 1177.7445225
29.1457286432161 1177.7445225
29.4472361809045 1177.7445225
29.748743718593 1177.7445225
30.0502512562814 1177.7445225
30.3517587939698 1177.7445225
30.6532663316583 1177.7445225
30.9547738693467 1177.7445225
31.2562814070352 1177.7445225
31.5577889447236 1177.7445225
31.8592964824121 1177.7445225
32.1608040201005 1177.7445225
32.4623115577889 1177.7445225
32.7638190954774 1177.7445225
33.0653266331658 1177.7445225
33.3668341708543 1226.925063
33.6683417085427 1226.925063
33.9698492462311 1226.925063
34.2713567839196 1226.925063
34.572864321608 1226.925063
34.8743718592965 1226.925063
35.1758793969849 1226.925063
35.4773869346734 1226.925063
35.7788944723618 1226.925063
36.0804020100502 1226.925063
36.3819095477387 1226.925063
36.6834170854271 1226.925063
36.9849246231156 1226.925063
37.286432160804 1226.925063
37.5879396984925 1226.925063
37.8894472361809 1226.925063
38.1909547738693 1226.925063
38.4924623115578 1226.925063
38.7939698492462 1226.925063
39.0954773869347 1226.925063
39.3969849246231 1226.925063
39.6984924623115 1226.925063
40 1226.925063
};
\end{axis}

\end{tikzpicture}
 
    \caption{Capacity as a function the SINR for 100 MHz link with 4 MIMO layers.}
    \label{fig:capacity_sinr}
\end{figure}
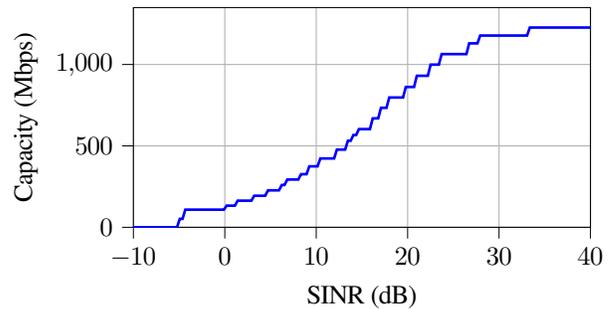

This  enables us to link power levels and capacity as it creates a direct relationship through sequential mapping of power level, \gls{sinr}, \gls{mcs} and capacity. We show an example of such a capacity function computed using the real \gls{oai} values in Fig. \ref{fig:capacity_sinr}. Figure~\ref{fig:capacity_sinr} suggests that the capacity function can be approximated by a simple model such as a piecewise-linear function of \gls{sinr}. However, the problem still cannot be solved by a linear solver since the \gls{sinr} is a ratio of power variables and thus introduces non-linearity. 
%\begin{figure}[!t]
%    \centering

 %   \begin{subfigure}{0.5\columnwidth}
 %   \centering
 %   \setlength\fwidth{\linewidth} 
 %   \setlength\fheight{0.9\linewidth}
 %   \input{figures/capacity}
  %  \caption{Capacity as function of  SINR.}
%\end{subfigure}
%\hfill
%\begin{subfigure}{0.46\columnwidth}
 %   \centering
  %  \setlength\fwidth{\linewidth} 
   % \setlength\fheight{0.9\linewidth}
    %\input{figures/ratio_mcs1}
    %\caption{Capacity function.}
%\end{subfigure}
    
   % \caption{Capacity analysis for a 100\,MHz link with 4 MIMO layers.}
   % \label{fig:capacity_combined}
%\end{figure}

\begin{figure}[!t]
    \centering
        \setlength\fwidth{0.8\columnwidth} 
        \setlength\fheight{0.4\columnwidth} 
        
            \input{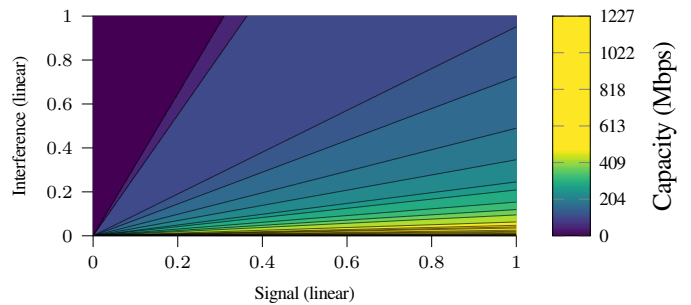}
 
    \caption{Capacity Function for 100 MHz link with 4 MIMO layers.}
    \label{fig:capacity_sinr_interf}
\end{figure}

Instead, we resort to analyzing the capacity as a bivariate function of the signal and of the interference. Let us now denote the capacity function as $C(S,I)$. Observe that $C(S,I)$ is a piecewise-constant function, and it is increasing with the ratio $S/I$.  Let us note the ordered possible values of $C(S,I)$ as $C_0, C_1, \cdots C_{27}$ where each $C_i$ corresponds to a specific \gls{mcs}. Since $C(S,I)$ is an increasing function of $S/I$, each capacity $C_i$ is associated with a threshold $th_i$, such that $S/I \geq th_i$ ensures the link has a capacity of at least $C_i$. This means we have:
\begin{align}
    C(S, I) \geq C_i \Leftrightarrow S \geq th_i \cdot I. \label{eq:cap_thr}
\end{align}
This is illustrated in Figure \ref{fig:capacity_sinr_interf}, where we plot $C(S, I)$. We can observe that each of the steps of this function is defined by two lines crossing at the point such that $S = I = 0$, these lines correspond to thresholds $th_{i-1}$ and $th_{i}$.

\section{Optimization Problems \label{sect:problem}}
In this section, we present two \gls{iab} topology design problem formulations: one aimed at maximizing throughput and the other at minimizing energy consumption. 
Our optimization models identify a tree, denoted as $\pazocal T$, which is a subgraph of $\pazocal G$, rooted in the \gls{iab}-donor, and whose leaves are the \glspl{ue}. The variables and notations used in this section is summarized in Tab. \ref{tab:variables}. 

\begin{table}[t!]
\centering
\caption{System Variables and Notation}
\label{tab:variables}
\begin{tabular}{@{}p{2.5cm}p{5.5cm}@{}}

\toprule
\multicolumn{2}{@{}l}{\textit{Graph Structure}} \\
\midrule
$\mathcal{G} = (\mathcal{V}, \mathcal{E})$ & Weighted directed measurements graph \\
$\mathcal{U} \subset \mathcal{V}$ & Set of user equipment (UEs) \\
$\mathcal{R} \subset \mathcal{V}$ & Set of \gls{iab} frontend radios \\
$t \in \mathcal{V}$ & DU of the IAB-donor node \\
$\mathcal{W} \subset \mathcal{E}$ & Set of wireless links (weighted by pathloss) \\
$\mathcal{F} \subset \mathcal{E}$ & Set of wired links (infinite capacity) \\
$\mathcal{N}_{out}(v)$  & Outgoing neighbors of node $v$ \\
$\mathcal{N}_{in}(v)$  & Incoming neighbors of node $v$ \\
$\mathcal{N}_{all}(v)$  & All neighbors: $\mathcal{N}_{in}(v) \cup \mathcal{N}_{out}(v)$ \\
$out(v) \in \mathbb{Z}^+$ & Outgoing degree: $|\mathcal{N}_{out}(v)|$ \\
$in(v) \in \mathbb{Z}^+$ & Incoming degree: $|\mathcal{N}_{in}(v)|$ \\
$deg(v) \in \mathbb{Z}^+$ & Total degree: $out(v) + in(v)$ \\
\midrule
\multicolumn{2}{@{}l}{\textit{Commodity and Routing}} \\
\midrule
$\mathcal{K}$ & Set of commodities to be routed \\
$k \in \mathcal{K}$ & Commodity defined as tuple $(s_k, t_k)$ \\
$s_k \in \mathcal{U}$ & Source node of commodity $k$ (UE) \\
$t_k = t$ & Destination node of commodity $k$ (IAB-donor) \\
$d_k \in \mathbb{R}^+$ & Traffic demand for commodity $k$\\
\midrule
\multicolumn{2}{@{}l}{\textit{Flow Variables}} \\
\midrule
$f_k(u,v) \in \mathbb{R}^+$ & Flow rate of commodity $k$ on edge $(u,v)$  \\
$f(u,v) \in \{0,1\}$ & Binary indicator: edge $(u,v)$ used by any commodity \\
\midrule
\multicolumn{2}{@{}l}{\textit{Node State Variables}} \\
\midrule
$a(u) \in \{0,1\}$ & Binary indicator: node $u$ is turned on (1) or sleeping (0) \\
\midrule
\multicolumn{2}{@{}l}{\textit{Power and Resource Variables}} \\
\midrule
$P^u_{tx} \in [0, P_{\max}^u]$ & Transmission power of node $u$  \\
%$P^u_{in}\geq 0$ & Total input power consumption of node $u$ \\
%$P_{rx}$ & Received power \\
$P_0$ & Power consumed when \gls{gnb} is not transmitting \\
$P_{sleep}$ & Power consumption in sleep mode \\
$\Delta_p$ & Power consumption increase rate \\
$\alpha(u,v) \in [0,1]$ & Time fraction allocated by node $u$ to node $v$ \\
\midrule
\multicolumn{2}{@{}l}{\textit{Capacity and Quality Variables}} \\
\midrule
$\phi_i(u,v) \in \{0,1\}$ & Binary: SINR condition in Eq.~\eqref{eq:cap_thr} satisfied \\
$c(u,v) \geq 0$ & Link capacity of edge $(u,v)$  \\
\bottomrule
\end{tabular}
\end{table}
%Since the whole network operates using the same spectrum (\textit{i.e.} in-band \gls{iab}), we assume that each node has a 
The \gls{tdma} scheduler operates using a weighted round-robin policy to schedule the inbound traffic and a dedicated radio device to relay the outbound traffic. This additional constraint---which follows guidance from \gls{3gpp} technical documents~\cite{3GPP_iab}---differentiates our model from a classical multicommodity flow problem, where adjacent edges do not have to share the same time resources as in a wireless network.

We formulate the problem as a mixed integer multicommodity flow problem. In such a problem, we have to route a set of $\pazocal K$ commodities on the graph, each using a single path. A commodity $k \in \pazocal K$ is defined as a tuple $s_k, t_k$ where $s_k$ is the source node (in our case, the \gls{iab}-donor $t$) and $t_k$ is the destination node (in our case, a \gls{ue} ). Additionally, a demand $d_k \in \mathbb{R}$ can be defined as the amount of traffic to route between $s_k$ to $t_k$.
These commodities are decided by the \gls{mno} beforehand, depending on the minimum capacity it wants to guarantee to its customers, and might be differentiated by different classes. The \gls{mno} can feed this information to the rApp running the optimization problem. 

The information about each \gls{iab} node is also available and can be fed to the rApp: we denote by $G_{tx}^u$ + $G_{rx}^u$ the transmission and reception gain of each node. Furthermore, we assume that it is possible to measure the pathloss $PL^{u,v}$ between nodes $u$ and $v$.

We denote by $\pazocal N_{out}(v)$ the outer neighbors of node $v$ and by $\pazocal N_{in}(v)$ its inner neighbors. We also have $\pazocal N_{all}(v) = \pazocal N_{in}(v) \bigcup \pazocal N_{out}(v)$. The cardinality of these sets (\textit{e.g.,} the outer and inner degrees) are denoted by $out(v)$ and $in(v)$, and their sum (the degree of the node) $deg(v) = out(v) + in(v)$.

We now introduce the flow allocation variables.
$f_k(u,v)$ denote rate of data flow for commodity $k$ that uses edge $(u,v) \in \pazocal E$, and binary variable $f(u,v)$  indicate whether $(u,v)$ is used by any commodity. 
We also introduce variable $P_{tx}(v)$ which controls the transmission power of node $v$, and variables $\phi_i(u, v)$ indicating whether the Signal and Interferences of edge $(u, v)$ are such that the condition in equation \eqref{eq:cap_thr} holds. Furthermore we introduce variable $\alpha(u,v) \in [0,1]$ which decides what portion of the total transmission time of node $u$ is allocated to node $v$. Further, we denote the capacity of each edge $(u, v) \in \pazocal W $ as $c(u,v)$. %As discussed in the previous section, we note that the capacity is also a variable that depends on }.

\subsection{Throughput Maximization Problem}

In this section, we present throughput maximization problem. Instead of maximizing the aggregate throughput of the entire \gls{iab} network,  we maximize the minimum throughput achieved by any \gls{ue} in the \gls{iab} network. The goal of formulating such an objective is to make sure to maintain consistent \gls{qos} via minimum guaranteed throughput among the \glspl{ue} served. This formulation does not seek to maximize the total achievable throughput of the network, but rather to ensure that every \gls{ue} receives a fair and sufficient share of resources — prioritizing service guarantee over aggregate efficiency.
The throughput maximization problem can be formally stated as follows:

%\begin{small}
\allowdisplaybreaks
    \begin{align}
        &\hspace{30mm} \max_{\boldsymbol{f}, \boldsymbol{f_k}, \boldsymbol{\alpha}, \boldsymbol{P_{tx}}} \quad  Z \label{obj_max}\\
        & \hspace{5mm}  \textrm{Subject to: } \nonumber\\ 
        &\sum\limits_{v \in \pazocal V}f_k(u,v) - \sum\limits_{v \in \pazocal V}f_k(v, u) = 0~~~~~~~~~~~~~ \forall k \in \pazocal K, \forall u \in \pazocal{V} \setminus \{s_k, t_k\} \label{eq:equilibrium}\\
        &\sum\limits_{v \in \pazocal V} f_k(s_k,v) - \sum\limits_{v \in \pazocal V} f_k(v, s_k) =-(\sum\limits_{v \in \pazocal V} f_k(v, t_k)  - \sum\limits_{v \in \pazocal V} f_k(t_k, v))\nonumber\\
         &\hspace{60mm}~~~~~~~~~~~\forall k \in \pazocal K\label{eq:source}\\
        &\sum\limits_{v\in \pazocal N_{all}(u)} \alpha(u, v) \leq 1 \hspace{5mm}~~~~~~~~~~~~~~~~~~~~~~~~~~~~~~~~~~~~~~~~~~~~~~ \forall u \in \pazocal V \label{eq:sched} \\
        &f(u,v) \geq \alpha(u,v) \hspace{10mm}~~~~~~~~~~~~~~~~~~~~~~~~~~~~~ \forall (u,v) \in \pazocal E, k \in \pazocal K \label{eq:aggregation}\\
        &\sum\limits_{v \in \pazocal N_{in}(u)} \mathbb{1}_{u \in \pazocal R} \cdot f(u,v) \leq 1 \hspace{10mm}~~~~~~~~~~~~~~~~~~~~~~ \forall u \in \pazocal R \cup \pazocal U \label{eq:tree} \\
        & \sum\limits_{u \in \pazocal N_{in}(v)} f_k(u,v) \geq Z \hspace{10mm}~~~~~~~~~~~~~~~~~~~~~~~~~~~~~~~~~~~~~ \forall v \in \pazocal U \label{eq:bound}\\
        & \sum_{k \in \pazocal K} f_k(u,v) \leq c(u,v) \hspace{10mm}~~~~~~~~~~~~~~~~~~~~~~~~~~~~~~ \forall (u, v) \in \pazocal W \label{eq:capacity_constraint}\\
        & \phi_i(u, v) = 1 \implies \{S_{u, v} \geq th_i(u,v) \times I_{u, v}\} \nonumber \\ &\hspace{10mm}~~~~~~~~~~~~~~~~~~~~~~~~~~~~~~~~~~~~~~~~~~~~~\forall i \in [0, 27], \forall(u,v) \in \pazocal W \label{eq:ind1}\\
        & \phi_i(u, v) = 0 \implies \{S_{u, v} \leq th_i(u,v) \times I_{u, v}\} \nonumber \\ &\hspace{10mm}~~~~~~~~~~~~~~~~~~~~~~~~~~~~~~~~~~~~~~~~~~~~~\forall i \in [0, 27], \forall(u,v) \in \pazocal W \label{eq:ind2}\\
        & c(u, v) \leq (\phi_i(u,v) * C_i(u, v) + (1 - \phi_i(u,v)) * C_{27}(u, v)) \cdot \alpha(u, v) \nonumber \\ &\hspace{10mm}~~~~~~~~~~~~~~~~~~~~~~~~~~~~~~~~~~~~~~~~~~~~~\forall i \in [0, 27], \forall(u,v) \in \pazocal W \label{eq:cap_set}\\
        & \eqref{eq:signal}\nonumber\\ & \eqref{eq:interferences}\nonumber\\
        & \phi_i(u,v) \in \{0,1\} ~~~~~~~~~~~~~~~~~~~~~~~~~~~~~~~~ \forall i \in [0, 27], \forall (u, v) \in \pazocal W \label{eq:phi}\\
        & P^u_{tx} \in [0, P_{\max}^u] ~~~~~~~~~~~~~~~~~~~~~~~~~~~~~~~~~~~~~~~~~~~~~~~~~~~~~~~~~~ \forall u \in \pazocal R \label{eq:Ptx}\\
        & f(u, v) \in \{0, 1\} ~~~~~~~~~~~~~~~~~~~~~~~~~~~~~~~~~~~~~~~~~~~~~~~~~~~~ \forall (u,v) \in \pazocal E \label{eq:binary}\\
        %& P_{tx}(v) \in [0, P_{max}^v] \quad\quad\quad\quad\quad\quad\quad\quad\quad\quad\quad~ \forall v \in \pazocal V \\
        & \alpha(u,v) \in [0, 1] ~~~~~~~~~~~~~~~~~~~~~~~~~~~~~~~~~~~~~~~~~~~~~~~~~~~~~ \forall (u,v) \in \pazocal E \label{eq:alpha}\\
        & f_k(u,v) \in \mathbb{R}^+~~~~~~~~~~~~~~~~~~~~~~~~~~~~~~~~~~~~~~~~~ \forall (u,v) \in \pazocal E, \forall~k \in \pazocal K \\
        & Z \in \mathbb R^+ \label{last_max}
      \end{align}
%\end{small}
The objective, $Z$ is bounded by the smallest throughput received by all \glspl{ue} due to constraint \eqref{eq:bound}. In this formulation, we do not impose $d_k$, the demand requirement of commodity $k$. 
Constraints \eqref{eq:equilibrium} and \eqref{eq:source} are flow conservation constraints which enforces that the total incoming flow must equal the total outgoing flow and all flow leaving the source must arrive at the destination. Constraint \eqref{eq:sched} ensures that we do not schedule more time resources than possible, constraint \eqref{eq:aggregation} ensures that link variables are equal to 1 if the link carries some flow, \eqref{eq:tree} prevents UEs and \gls{rf} frontends from having more than 1 inner neighbor, which forces the topology to be a tree. Constraint \eqref{eq:capacity_constraint} defines the maximum capacity of the wireless links. The expression of the capacity $c(u,v)$ is calculated according to the description of Section \ref{sec:cap_model}: we have one binary variable $\phi_i(u, v)$ per capacity value of each wireless link $(u,v)$. Then, the indicator constraints \eqref{eq:ind1} and \eqref{eq:ind2} ensure that this variable is equal to 1 if the link $(u, v)$ can achieve the $i^{th}$ capacity level $C_i$ and zero otherwise. Such implication constraints are standard in modern solvers which transform them into mixed binary formulations efficiently \cite{williams2013model}. Then, in constraint \eqref{eq:cap_set}, we ensure that the capacity of the link $(u, v)$ respects the capacity values given by the thresholds. Note that this value given by the thresholds is multiplied by the scheduling variable of the link $\alpha(u, v)$. This gives a non-linear constraint. However, since the non-linearity comes from the product of a binary and a continuous variable, modern solvers such as Gurobi, CPLEX or OR-Tools can linearize this equation.

\subsection{Energy Minimization Problem}

We now introduce the variant of the same problem which optimizes for energy consumption minimization. Unlike the throughput maximization problem in the previous section, we aim to minimize power consumption by dynamically activating only the nodes necessary to meet minimum throughput demands $d_k$ for all \glspl{ue}. To this end, we introduce the binary variables $a(v)~\forall v \in \pazocal R$ which indicate whether node $v$ is turned on or sleeping.
As discussed in the previous section, nodes consume a minimal amount of power $P_{sleep}$ when it is turned off. Then using Eq. \eqref{eq:earth}, the aggregate power consumption across the entire \gls{iab} network can be represented as:

\begin{align}
    P_{total} = &\sum_{v \in \pazocal{V}} (1 - a(v)) \times P_{sleep} + \sum_{v \in \pazocal{V}} a(v) \times P_0 +  \nonumber \\
    &\sum_{u \in \pazocal{V}}\sum_{v \in \pazocal{N}_{out}(u)} a(v) \times P^u_{tx} \times \alpha(u,v) \times \Delta_P. 
\end{align}
The first part of the  function is the static baseline power consumption of a \gls{rf} frontend. It is either $P_{sleep}$ if the \gls{rf} frontend is not activated or $P_0$ if it is activated.

Based on this, the energy minimization problem can be formally stated as follows:
%\begin{small}
\allowdisplaybreaks
    \begin{align}
    & \hspace{30mm} \min_{\boldsymbol{f},\boldsymbol{f_k},\boldsymbol{\alpha}, \boldsymbol{P_{tx}},\boldsymbol{a}}\quad  %\sum_{v \in \pazocal{V}} (1 - a(v)) \times P_{sleep} + \sum_{v \in \pazocal{V}} a(v) \times P_0 + \nonumber \\
      % & ~~~~~~~~\sum_{u \in \pazocal{V}}\sum_{v \in \pazocal{N}_{out}(u)} P^u_{tx} \times \alpha(u,v) \times \Delta_P 
      P_{total} \label{eq:objective2} \\
      & \hspace{5mm}  \textrm{Subject to: } \nonumber\\
       % \phantom{i + j + k}
       & \sum\limits_{v \in \pazocal V}f_k(u,v) - \sum\limits_{v \in \pazocal V}f_k(v, u) = 0~~~~~~~~~~~~~~~ \forall u \in \pazocal{V}, \forall k \in \pazocal K \label{eq:equilibrium2}\\
       &\sum\limits_{v \in \pazocal V} f_k(s_k,v) - \sum\limits_{v \in \pazocal V} f_k(v, s_k) = 1~~~~~~~~~~~~~~~~~~~~~~~\forall k \in \pazocal K\label{eq:source2}\\
       &\sum\limits_{v \in \pazocal V} f_k(v, t_k)  - \sum\limits_{v \in \pazocal V} f_k(t_k, v) = -1 ~~~~~~~~~~~~~~~~~~~~~ \forall k \in \pazocal K \label{eq:destination2}\\
       & \sum_{k \in \pazocal K} f_k(u,v) \cdot d_k \leq c(u,v) ~~~~~~~~~~~~~~~~~~~~~~~~~~ \forall (u, v) \in \pazocal W \label{eq:capacity2}\\
       &f(u,v) \geq f_k(u,v) ~~~~~~~~~~~~~~~~~~~~~~~~~~~~~ \forall (u,v) \in \pazocal E, k \in \pazocal K \label{eq:activ}\\
       & \eqref{eq:signal}, \eqref{eq:interferences},\eqref{eq:sched}, \eqref{eq:aggregation}, \eqref{eq:tree}, \eqref{eq:ind1}, \eqref{eq:ind2}, \eqref{eq:cap_set}, \eqref{eq:phi}, \eqref{eq:Ptx}, \eqref{eq:binary}, \eqref{eq:alpha} \nonumber\\
       & f_k(u,v) \in \{0, 1\} ~~~~~~~~~~~~~~~~~~~~~~~~~~~~\forall (u,v) \in \pazocal E, \forall~k \in \pazocal K\label{eq:binary2} \\
       & a(v) \in \{0, 1\}~~~~~~~~~~~~~~~~~~~~~~~~~~~~~~~~~~~~~~~~~~~ \forall v \in \pazocal V - \pazocal U \label{min:last}
      \end{align}
%\end{small}
Equation \eqref{eq:objective2} is the objective function that considers node activation.  In this formulation, we consider the demand $d_k$ for each commodity $k$. Thus the variables $f_k(u,v)$ indicate whether commodity $k$ uses edge $(u,v) \in \pazocal E$ and thus are binary (instead of being continuous in the previous problem). Consequentially, the flow conservation constraints change (equations \eqref{eq:equilibrium2}, \eqref{eq:source2} and \eqref{eq:destination2}). \eqref{eq:activ} ensures that an edge is activated if any commodity uses it. Capacity constraints \eqref{eq:capacity2} now take the demand of the commodity $d_k$ into account because the goal of the problem at hand is to route all the demands while minimizing the energy consumption. The rest of the constraints of the problem remain the same. Note that the objective function of this problem is non-convex due to the product of continuous variables $\alpha(u,v)$ and $P_{tx}^u$.

\section{Practical Large-Scale Approaches \label{sect:heuristics}}
The throughput maximization and energy minimization problems generate a large search space that scales exponentially with the network size. In addition to this, non-linear relationships between variables and their tight coupling through constraints such as capacity and scheduling constraints  have a significant impact on problem complexity, making it challenging to obtain exact solutions. Therefore, in this section, we discuss two algorithms designed to address the throughput maximization and energy minimization problems. The first approach employs an iterative search strategy that refines solutions such that the optimization objective is met. The crux of the second approach is problem size reduction with the goal of simplifying the optimization process and leverages pruning techniques to achieve this goal. 

\subsection{Local Search Method}
%In this section, we present local search heuristics used to perform flow allocation with maximum throughput and with minimum energy. 

\begin{figure}
    \centering
        \includegraphics[width=0.5\textwidth]{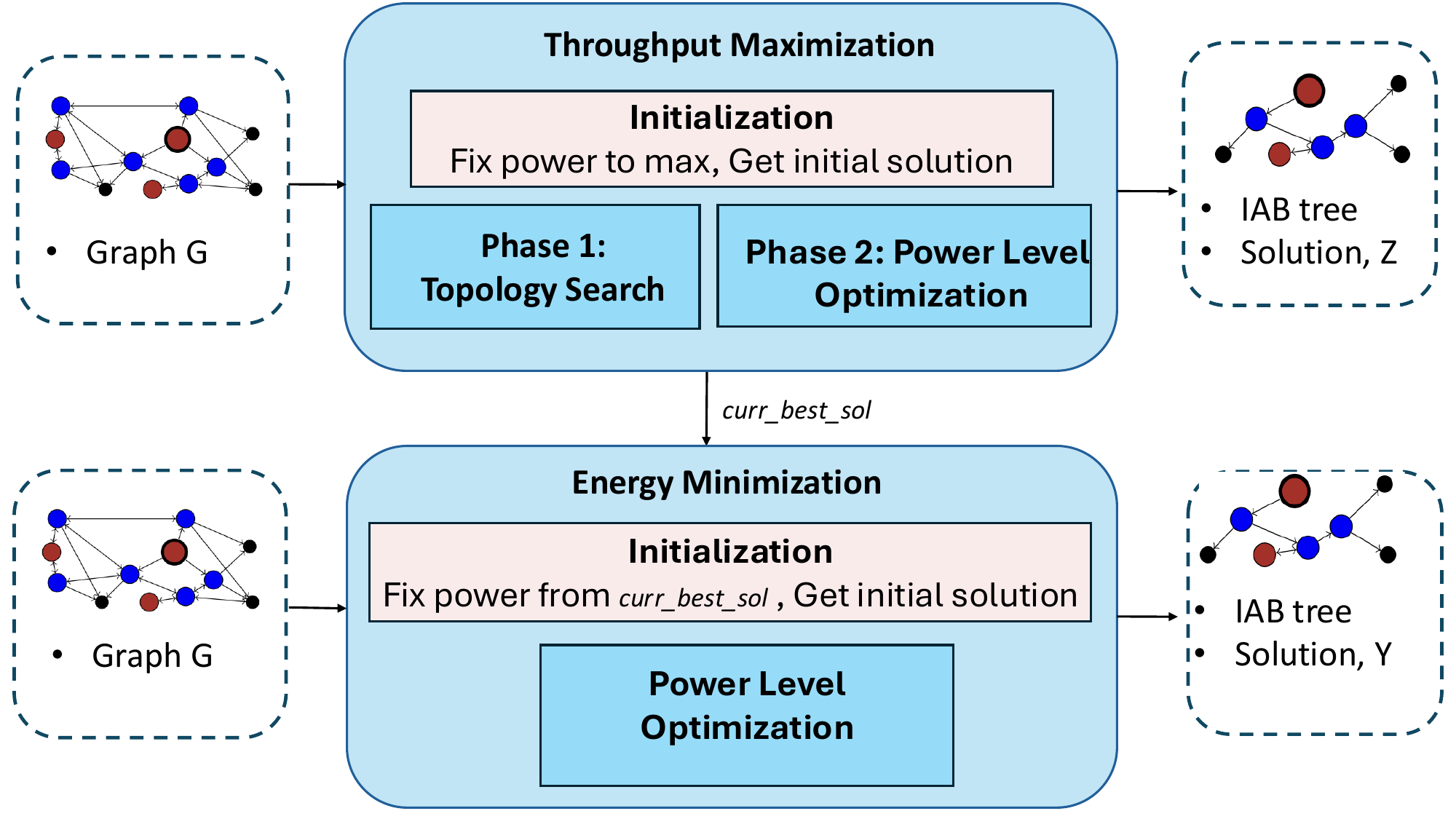}
        \caption{Local search method.}
        \label{fig:localsearch}
\end{figure}

The local search method is based on iterative fix-and-optimize strategy~\cite{danna2005exploring} toward the optimization objective while maintaining feasibility.  Instead of solving a complex global optimization problem directly, the algorithm explores different configurations locally by optimizing one frontend node $u \in \pazocal R$ at a time. A high-level overview of the local-search method algorithm for throughput maximization and energy minimization is provided in the Fig. \ref{fig:localsearch}.

The algorithm for solving the maximum throughput problem, outlined as Algorithm \ref{alg:max}, takes  measurement graph $\pazocal G = (\pazocal V, \pazocal E)$ as input. Initially, the transmission power for all frontend nodes $u \in \pazocal R$ are set to their maximum values $P^u_{max}$. The algorithm maintains a solution vector \emph{curr\_best\_sol}  that stores the best solution found so far for each node $u \in \pazocal R$. It also tracks the best objective value found, stored in \emph{curr\_best\_obj}. The method uses two phases of iterative search to converge toward a feasible optimal solution. The first phase finds an initial topology and the second phase varies power at each  node to further fine-tune the solution.

\begin{algorithm}
     \begin{algorithmic}[1]
     \renewcommand{\algorithmicrequire}{\textbf{Input:}}
     \renewcommand{\algorithmicensure}{\textbf{Output:}}
     \REQUIRE Measurement Graph 
     \ENSURE  Flow Allocation that Maximizes minimum Throughput of \glspl{ue}
     \STATE Fix all power variables $P^u_{tx}$ to $P^u_{max}$ $\forall u \in \pazocal R$
     \STATE Solve \eqref{obj_max} - \eqref{last_max} and obtain objective value $Z'$
     \STATE $curr\_best\_obj = Z'$
     \STATE $prev\_best\_obj = -1$
     \STATE $curr\_best\_sol = Dict()$
     \FOR{$u \in \pazocal R$}
     \STATE $curr\_best\_sol[u] = P^u_{max}$
     \ENDFOR
     \WHILE{$curr\_best\_obj \neq prev\_best\_obj$}
        \STATE $prev\_best\_obj$ = $curr\_best\_obj$
        \FOR{$u \in \pazocal R$}
           \IF{$curr\_best\_sol[u]$ == 0}
                \STATE $val$ = $P^u_{max}$
            \ELSE
                \STATE $val$ = 0
            \ENDIF
           \STATE Fix $P^u_{tx} = val$
           \STATE Fix $P^v_{tx} = curr\_best\_sol[v]$ $~~ \forall v \in \pazocal R - \{u\}$
           \STATE Solve optimization problem \eqref{obj_max} - \eqref{last_max}, obtain objective value $Z'$
           \IF{$Z' \geq curr\_best\_obj$}
                \STATE $curr\_best\_sol[u]$ = $val$
                \STATE $curr\_best\_obj = Z'$
           \ENDIF
        \ENDFOR
     \ENDWHILE
     \WHILE{$curr\_best\_obj \neq prev\_best\_obj$}
        \STATE $prev\_best\_obj = curr\_best\_obj$
        \FOR{$u \in \pazocal R$}
            \STATE Unfix $P^u_{tx}$ if it was fixed
            \STATE Fix $P^v_{tx} = curr\_best\_sol[v] ~~ \forall v \in \pazocal R - \{u\}$
            \STATE Solve \eqref{obj_max} - \eqref{last_max} and obtain objective $Z'$
            \IF{$Z' > curr\_best\_obj$}
                \STATE $curr\_best\_sol[u]$ = $P^u_{tx}$
                \STATE $curr\_best\_obj = Z'$
           \ENDIF
        \ENDFOR
     \ENDWHILE
     \RETURN Solution obtained by solving \eqref{obj_max} - \eqref{last_max} with $P_{tx}^u$ fixed to $curr\_best\_sol[u]$ $\forall u \in \pazocal R$. 
     \end{algorithmic} 
     \caption{Local search method for solving the maximum throughput problem}
     \label{alg:max}
\end{algorithm}

In each iteration of the first phase, the algorithm follows these steps for each node: it toggles the power between $0$ and $P^u_{max}$ (lines 11-15 of Algorithm \ref{alg:max}), fixes the power of all other nodes to the best solution found so far (line 16 of Algorithm \ref{alg:max}), solves the optimization problem (line 17 of Algorithm \ref{alg:max}), and checks if this leads to a higher throughput $Z'$. If a higher throughput is found, it updates the solution and objective variables, \emph{curr\_best\_sol} and \emph{curr\_best\_obj}, for the corresponding node (lines 18-20 of Algorithm \ref{alg:max}). Subsequently, the first phase results in an initial \gls{iab} network topology. 

We leverage this initial topology obtained from solving throughput maximization as a warm start for solving the energy problem. The second phase  further refines the solution space by following these steps: it fixes transmission power of all nodes except one at a time using the best solution found so far from \emph{curr\_best\_sol} (lines 24-25 of Algorithm \ref{alg:max}). Then it solves the optimization problem (line 26 of Algorithm \ref{alg:max}). Similar to the previous phase, the solution is checked for convergence followed by updating the best solution (lines 28-29 of Algorithm \ref{alg:max}). This process is repeated until no further improvements in throughput can be made and  the algorithm returns the final solution.

The algorithm for solving energy minimization problem is outlined as Algorithm \ref{alg:energy}. First, it obtains the maximum throughput solution using Algorithm \ref{alg:max}.
Then it checks if the demand constraints are met for all the commodities by $d_k \geq Z' \forall k \in \pazocal K$.
 The algorithm then fixes the power values from the solution of throughput maximization, obtained as \emph{curr\_best\_sol}, and solves the energy minimization problem  (lines 7-8 of Algorithm \ref{alg:energy}).
Similar to the second phase of Algorithm \ref{alg:max}, the solution is iteratively  refined by fixing transmission power of all nodes except one at a time using the best solution from the solution vector \emph{curr\_best\_sol} and solving the optimization problem (lines 12-19 of Algorithm \ref{alg:energy}). This is repeated until convergence and the algorithm returns the solution.

\begin{algorithm}[t]
         \begin{algorithmic}[1]
     \renewcommand{\algorithmicrequire}
     {\textbf{Input:}}
     \renewcommand{\algorithmicensure}{\textbf{Output:}}
     \REQUIRE Measurement Graph 
     \ENSURE  Flow Allocation with minimum energy
     \STATE Obtain a maximum throughput solution via Algorithm \ref{alg:max}
     
     \FOR{$u \in \pazocal R$}
        \STATE $curr\_best\_sol[u] = P^u_{tx}$
     \ENDFOR
     \FORALL{$k \in \pazocal K$}
        \IF{$d_k \geq Z'$}
            \STATE return infeasible
        \ENDIF
     \ENDFOR
     \STATE Fix $P^u_{tx}= curr\_best\_sol[u]~~\forall u \in \pazocal R$
     \STATE Solve \eqref{eq:objective2} - \eqref{min:last} and obtain energy consumption value $Y'$
     \STATE $curr\_best\_obj = Y'$
     \STATE $prev\_best\_obj = -1$
     \WHILE{$curr\_best\_obj \neq prev\_best\_obj$}
        \STATE $prev\_best\_obj = curr\_best\_obj$
        \FOR{$u \in \pazocal R$}
            \STATE Unfix $P^u_{tx}$ if it was fixed
            \STATE Fix $P^v_{tx} =curr\_best\_sol[v] ~~ \forall v \in \pazocal R - \{u\}$
            \STATE Solve \eqref{eq:objective2} - \eqref{min:last} and obtain objective $Y'$
            \IF{$Y' > curr\_best\_obj$}
                \STATE $curr\_best\_sol[u]$ = $P^u_{tx}$
                \STATE $curr\_best\_obj = Y'$
           \ENDIF
        \ENDFOR
     \ENDWHILE
     \RETURN Solution obtained by solving \eqref{eq:objective2} - \eqref{min:last} with $P_{tx}^u$ fixed to $curr\_best\_sol[u]$ $\forall u \in \pazocal R$. 
     \end{algorithmic}
     
     \caption{Local search method for solving the minimum energy problem}
     \label{alg:energy}
\end{algorithm}

\subsection{Selective-Reduction Method}

The selective-reduction method simplifies the original problem by preserving only a subset of edges from the \gls{iab} network and eliminating those that are least likely to appear in the optimal subset~\cite{kirov2018optimized}. This reduces the problem size and consequently makes it easier to solve.

\begin{figure}
        \centering
        \includegraphics[width=0.95\linewidth]{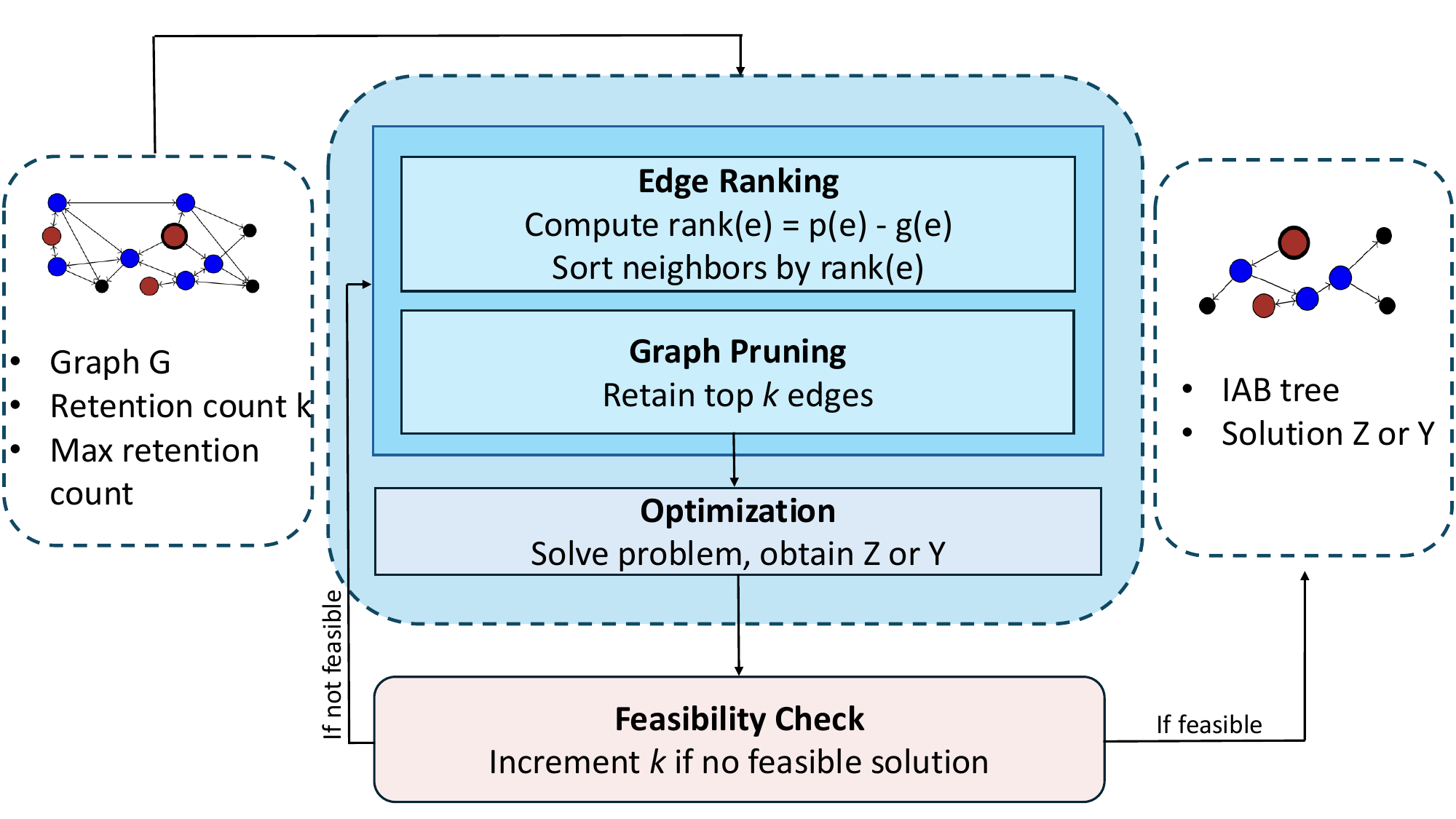}
        \caption{Selective-reduction method.}
        \label{fig:prune}
\end{figure}

%The algorithm iteratively prunes the graph and solves the optimization problem to achieve feasible solution. 
An overview of the approach is provided in Fig. \ref{fig:prune} and  algorithm is outlined as Algorithm \ref{alg:pruning}. We note that the method can be leveraged for solving either of the two optimization problems.

\begin{algorithm}
    \caption{Selective-reduction method }
    \label{alg:pruning}
    \begin{algorithmic}[1]
    \renewcommand{\algorithmicrequire}{\textbf{Input:}}
    \renewcommand{\algorithmicensure}{\textbf{Output:}}
    \REQUIRE Measurement Graph $G$, Initial Retention Count ($k$), Maximum Retention Count ($k_{max}$)
    \ENSURE  Flow Allocation that Maximizes minimum Throughput of \glspl{ue}
    %\STATE Initialize  $ret\_count$, and $max\_ret\_count$
    \STATE Set $feasible\_solution = \textbf{false}$
    \WHILE{$k \leq k_{max}$}
         
         \STATE Compute \textit{rank} for each edge $e \in \pazocal W$ as $g(e) - p(e)$ \COMMENT{Gain minus pathloss}
        \STATE Sort $N(u)$ in descending order of \textit{rank}
        \STATE Retain the top $k$ edges from $N(u)$ and add them to $E'$
        \STATE Solve optimization problem \eqref{obj_max} - \eqref{last_max}, obtain objective value $Z'$ or solve \eqref{eq:objective2} - \eqref{min:last}, obtain objective value $Y'$
        \IF{$Z'$ or $Y'$ is feasible}
            \STATE Store solution and set $feasible\_solution = \textbf{true}$
            \STATE \textbf{break}
        \ELSE
           
                \STATE Increase $k$ 
           
        \ENDIF
    \ENDWHILE
    
    \RETURN  solution with objective $Z$ or $Y$
    \end{algorithmic} 
    
\end{algorithm}
One of the key aspects in pruning-based approaches is to determine an effective pruning strategy. Another is to ensure that the pruning process is not aggressive and that the pruning process does not cause solution infeasibility. We address these as follows:
\begin{itemize}
    \item \textbf{Edge ranking}: We weight the edges using an appropriate metric and rank these weighted edges as a first step in the pruning process. 
    \item \textbf{Retention count parameter}: This variable controls the pruning intensity by specifying the number of neighbors to retain during the pruning process.
    \item \textbf{Adaptive Fine-Tuning}: The retention count is incremented  until a feasible solution is achieved and this ensures that the pruning process is not aggressive.
\end{itemize} 

The selective-reduction method given in Algorithm \ref{alg:pruning} takes  measurement graph $\pazocal G = (\pazocal V, \pazocal E)$ as input along with the initial retention count and maximum retention count. We rank each wireless link $e \in \pazocal W \subset \pazocal E$ using the metric $g(e)-p(e)$ where $g(e)$ is the gain of the link $e$ and $p(e)$ is the pathloss of link $e$. While pathloss gives an indication of distance and is often used as a pruning metric (e.g., \cite{kirov2018optimized}), we use gain-pathloss metric as it provides a representation of effective link quality when pruning edges in the network. The edges ranked by the metric are then sorted in descending order. For each node, all the neighboring edges other than top $k$ edges are removed from the graph. After this process, the graph with reduced size is used to solve the optimization problem. Algorithm returns the solution along with the corresponding objective value $Z'$ or $Y'$ if it is found. If no feasible solution is found, the algorithm increments retention count $k$ (with maximum retention count $k_{max}$) and the whole process is repeated to check for a feasible solution. 

\begin{theorem} \label{prop:feasibility}
If the original graph $G=(V,E)$ admits a feasible solution, then Algorithm 3 with $k_{max}= \max_{u \in V} |N(u)|$ is guaranteed to find a feasible solution.
\end{theorem}
\begin{IEEEproof}
Let $d_{max}= \max_{u \in V} |N(u)|$ denote the maximum degree of the graph $G$ and let the pruned graph be $G'$. As the algorithm is incremental, at iteration $k=d_{max}$,  the algorithm retains all neighbors for every node as $k \geq |N(u)|, \forall u \in \mathcal{U}$. Thus, the pruned graph $G'$ is the same as original graph $G$ and if $G$ has a feasible solution, $G'$ has feasible solution.
\end{IEEEproof}

\subsection{Computational Complexity Analysis}

The optimization problems (Section III) are NP-hard mixed-integer programs with large number of variables including $O(|\mathcal{K}| \cdot |\mathcal{E}|)$ flow variables, $O(|\mathcal{E}|)$ binary link activation and continuous scheduling variables, $O(|\mathcal{R}|)$ continuous power variables, and $O(|\mathcal{W}| \times 28)$ binary capacity indicator variables (one per MCS level per wireless link) that become computationally intractable for realistic network sizes. Our algorithms achieve tractability through problem decomposition: Local Search fixes $(|\mathcal{R}|-1)$ power variables per iteration while optimizing the remaining variables, requiring $O(|\mathcal{R}|)$ MILP solves in the typical case  and $O(|\mathcal{R}|^2)$ in the worst case. Selective-Reduction prunes the graph from $|\mathcal{E}|$ edges to retain only top-$k$ edges per node. This directly reduces flow variables from $O(|\mathcal{K}| \cdot |\mathcal{E}|)$ to $O(|\mathcal{K}| \cdot k|\mathcal{V}|)$, link variables from $O(|\mathcal{E}|)$ to $O(k|\mathcal{V}|)$, and capacity indicators from $O(|\mathcal{W}| \times 28)$ to $O(k|\mathcal{V}| \times 28)$. The algorithm requires $O(k_{max} - k_{init})$ MILP solves on progressively less-pruned graphs. Since MILP solve time scales linearly with problem size, this variable reduction translates directly to faster solver convergence in practice.
 %These decomposition strategies enable our methods to solve realistic network instances within practical timeframes suitable for non-RT RIC control loops, while exact methods become impractical as problem size grows. 

\section{Performance evaluation setup \label{sect:setup}}

%\subsection{Evaluation scenarios}
%We use datasets representing two distinct areas in the city of Milan, Italy. One represents a highly dynamic urban area surrounding a major train station, while the other corresponds to an area of 0.092 $km^2$ in the center of Milan.

We evaluate our approaches using publicly available traffic data from Milan, Italy, which include two months of network operator measurements \cite{barlacchi2015multi}. We select two representative areas for evaluation from this dataset: a highly dynamic urban area surrounding a major train station and an area  of 0.092 km$^2$ in the center of Milan. This dataset is used exclusively to derive the normalized temporal load shape capturing human activity rhythms such as commuting patterns and business hours, while all link capacities, SINR values, and energy consumption are computed using the 3GPP 38.901 channel model at 3.6 GHz (FR1) and 7 GHz (FR3) with current 5G NR parameters (Table~\ref{tab:parameters}).

We now discuss the techniques used to synthetically generate the set of measurement graphs $\pazocal G = (\pazocal V, \pazocal E)$, needed to evaluate the feasibility and effectiveness of our optimization model. We employ the same techniques that we use in our previous work for \gls{iab}-node placement and  data-driven time-varying UE density model \cite{gemmi2023joint}. However, for completeness, we describe both of them in this paper.
%To that end, we first present the state-of-the-art techniques used to place the \gls{iab}-nodes \cite{gemmi2022cost}, and the UEs \cite{3GPP_913}. Then, in the second subsection, we present our data-driven time-varying UE density model that enables us to generate different instances depending on the time of the day and the day of the week. 

\begin{table}[t]
    \centering
    \begin{tabular}{lll}
    \toprule
    & \textbf{Scenario 1}& \textbf{Scenario 2}\\ \midrule
         Area [km$^2]$ & $0.092$ &$0.132$ \\
         UE density [UE/km$^2$] & $0-900$ &$0-900$\\
         Path Loss Model & \multicolumn{2}{c}{3GPP 38.901} \\
         Frequency [GHz] & \multicolumn{2}{c}{$3.6$, $7$} \\
         gNB Lobes gain (Main/Side) [dBi] & \multicolumn{2}{c}{$24$/$-2$} \\
         UE Main lobe gain (Main/Side) [dBi] & \multicolumn{2}{c}{$0$/$-17.85$} \\
         gNB Type & \multicolumn{2}{c}{Micro-cell (see \cite{auer2011much})} \\
         Max tx power [W] & \multicolumn{2}{c}{$6.3$} \\
         
    \bottomrule
         
    \end{tabular}
    \caption{Parameters for the simulation}
    \label{tab:parameters}
\end{table}

The parameter values that we use for the simulations are presented in Table. \ref{tab:parameters}. 
Apart from the frequency bands FR1 and FR2 of 5G-NR, the upper mid-band with frequency range spanning from 7 GHz to 24 GHz, also known as FR3, has recently emerged as a frequency range of interest for 6G networks \cite{10634041,kang2024cellular}.
We analyze the results with an \gls{iab} network that operates at frequencies in the frequency range FR1 and FR3. One operates at 3.6 GHz in FR1 that is commonly used with bandwidth of 100 MHz. The other operates at 7 GHz, in FR3, with higher capacity (bandwidth of 400 MHz).

\subsection{Placement of \glspl{gnb} and \glspl{ue}}

\begin{figure}
    \centering
    \hfill
    \begin{subfigure}{0.46\linewidth}
        \includegraphics[width=\linewidth]{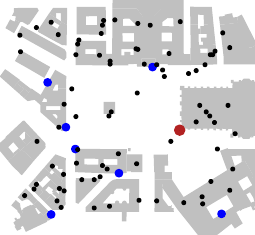}
        \caption{Scenario 1}
        \label{fig:sc1}
    \end{subfigure}
    \hfill
    \begin{subfigure}{0.43\linewidth}
        \includegraphics[width=\linewidth]{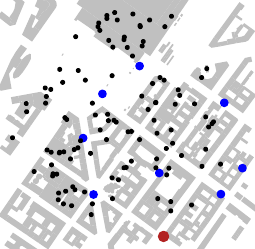}
        \caption{Scenario 2}
        \label{fig:sc2}
    \end{subfigure}
    \hfill
    \caption{Sample deployment of a network in center of Milan and train station.}
    \label{fig:scenarios}
    \vspace{-.5cm}
\end{figure}

The set of nodes of our graph $\pazocal V$ is comprised of both \gls{iab}-nodes, and \glspl{ue}, whose placement is done separately using two different techniques. \gls{iab}-nodes are placed on building facades with a given density $\lambda_{gNB}$. The exact position is computed by using a state-of-the-art placement heuristic \cite{gemmi2022cost} that leverages highly precise 3D models to place the gNB such that the number of potential \gls{ue} location in line of sight is maximized. \glspl{ue} are then randomly distributed both in public areas, such as streets, and inside buildings. Specifically, given a density of $\lambda_{UE}$, indoor \glspl{ue} are uniformly randomly distributed inside buildings with a density equal to $r_{i/o}$ $\lambda_{UE}$ and outdoor \glspl{ue} are uniformly distributed inside buildings with density $(1-r_{i/o}) \lambda_{UE}$. $r_{i/o}$ is a commonly used ratio of indoor to outdoor \gls{ue}, which we set to 0.8 based on \gls{3gpp} technical report~\cite{3GPP_913}. In short, we consider that in our simulations, 80\% of the \glspl{ue} are placed indoors. Figure~\ref{fig:scenarios} shows sample deployment of the network in the center of Milan and train station respectively.

\subsection{Time-varying UE Density Model}
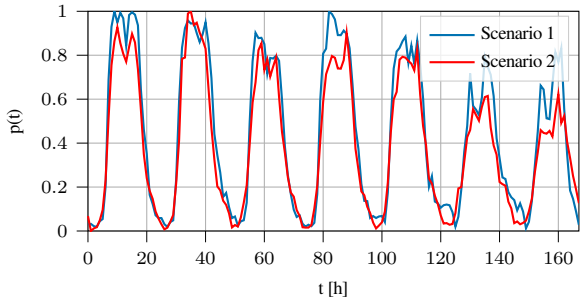
\begin{figure}[]
    \centering
     \setlength\fwidth{0.9\columnwidth} 
        \setlength\fheight{0.4\columnwidth} 
    % This file was created with tikzplotlib v0.10.1.
\begin{tikzpicture}

\definecolor{darkcyan1115178}{RGB}{1,115,178}
\definecolor{darkgray176}{RGB}{176,176,176}
\definecolor{lightgray204}{RGB}{204,204,204}
\pgfplotsset{every tick label/.append style={font=\scriptsize}}

\begin{axis}[
width=\fwidth,
height=\fheight,
legend cell align={left},
legend style={fill opacity=0.8, draw opacity=1, text opacity=1, draw=lightgray204},
tick align=outside,
tick pos=left,
x grid style={darkgray176},
xlabel={t [h]},
xmajorgrids,
xmin=0, xmax=167,
xtick style={color=black},
y grid style={darkgray176},
ylabel={p(t)},
ymajorgrids,
ymin=0, ymax=1,
ytick style={color=black},
font=\scriptsize,
xmajorgrids,
ymajorgrids,
xlabel style={font=\scriptsize},
ylabel style={font=\scriptsize}
]
\addplot [thick, darkcyan1115178]
table {%
0 0.029604121525941
1 0.0323529551452497
2 0.0219930056559067
3 0.0187875235249557
4 0.0440786788538376
5 0.0543742642149492
6 0.202921374014722
7 0.688145808444598
8 0.929651636160799
9 1
10 0.948922503938067
11 0.993877832785381
12 0.968196554673834
13 0.843374827244111
14 0.983110737068639
15 0.995763988828446
16 0.984686054959635
17 0.93616642857153
18 0.645581918014017
19 0.46169574604274
20 0.351478955685975
21 0.164204763692931
22 0.123289021644218
23 0.0714327601234381
24 0.0593026154100752
25 0.0587075014642561
26 0.0308588893722236
27 0.018906211491764
28 0.0404185789637163
29 0.0485174090489718
30 0.234291502454554
31 0.592572438325778
32 0.834962360032701
33 0.923580995091752
34 0.958002079962324
35 0.944276403272997
36 0.945366431348189
37 0.900588196597656
38 0.860298898410496
39 0.894893412186428
40 0.949817266916623
41 0.853384350427692
42 0.696630020799349
43 0.441563979388241
44 0.345632727571207
45 0.265747828675138
46 0.155723892833779
47 0.173429362806589
48 0.108309417704394
49 0.0586592052742137
50 0.0659007514511843
51 0.0244394834449676
52 0.0439749408150679
53 0.0488246404730871
54 0.128838333719815
55 0.492010092112408
56 0.772431813832361
57 0.902029800438084
58 0.878082334411951
59 0.879296448064037
60 0.850914788360196
61 0.715975987415759
62 0.787591785827268
63 0.78495752172421
64 0.797710582617449
65 0.774807588212435
66 0.598728819902927
67 0.383426019877734
68 0.253912666317232
69 0.168789642268641
70 0.0969071052582045
71 0.0637504689067984
72 0.0710465741074795
73 0.032860531991211
74 0.02
75 0.0284998656558022
76 0.0219245343729104
77 0.0374497767204225
78 0.169017597539563
79 0.593398840002466
80 0.807311683791842
81 0.829821458505043
82 0.998340316591948
83 0.987457135664106
84 0.963154577522626
85 0.891780191007409
86 0.891449800779422
87 0.912288744494961
88 0.896376800708387
89 0.831916690177811
90 0.665318132530848
91 0.475066236765353
92 0.373484558005454
93 0.213974445735711
94 0.137148678962553
95 0.135898241560377
96 0.0590526679344104
97 0.067894855654359
98 0.0590703492818237
99 0.0662293631096387
100 0.0683207702278372
101 0.0408903249566594
102 0.175053734493085
103 0.496201295229126
104 0.739517131510981
105 0.81708309375958
106 0.833312015154624
107 0.863812865373904
108 0.884438656891554
109 0.765150768282667
110 0.843723034219181
111 0.762706611278305
112 0.866749659685923
113 0.756999646544025
114 0.603441282396389
115 0.358588309847121
116 0.201565799638186
117 0.249292968853629
118 0.133978633558852
119 0.126279828145207
120 0.106066148168709
121 0.117430616784426
122 0.120414036842441
123 0.117186497238321
124 0.0853351491629645
125 0.02
126 0.0650675430846819
127 0.190027542460111
128 0.361674808524159
129 0.477959735107461
130 0.717000489810094
131 0.569568645684907
132 0.552080694126755
133 0.527434720719129
134 0.607748474734511
135 0.792451098712278
136 0.747431017294972
137 0.711135480914091
138 0.614292577354162
139 0.421467346665895
140 0.332969349451384
141 0.243726914648081
142 0.236828987701074
143 0.177189730762132
144 0.153595031165899
145 0.140803562534149
146 0.0940415162947444
147 0.125564417145263
148 0.0885521709314595
149 0.0122586681953313
150 0.0420035423981965
151 0.136837497232213
152 0.296255581792493
153 0.451475328103812
154 0.661565540958488
155 0.644034723745582
156 0.512855432229041
157 0.509173111016169
158 0.669542254120969
159 0.812851749234896
160 0.769383229050996
161 0.831946027908679
162 0.555523233773363
163 0.347658556617216
164 0.184307839533039
165 0.146450179141883
166 0.0902309077761664
167 0.027799951256201
};
\addlegendentry{Scenario 1}
\addplot [thick, red]
table {%
0 0.0678823523960923
1 0
2 0.0119206876672045
3 0.0173489637417998
4 0.0472166161023006
5 0.0958745123131113
6 0.232193688261056
7 0.512097358378056
8 0.754553776375287
9 0.864088862454121
10 0.926841635259843
11 0.855649291730102
12 0.806761820501458
13 0.770704172968663
14 0.838265740047749
15 0.897239952252276
16 0.858330734378181
17 0.736291273968785
18 0.553121054163315
19 0.350533326954702
20 0.240657551590513
21 0.182313820491738
22 0.146822305437137
23 0.0888157802204261
24 0.0566437216028947
25 0.0304864344650174
26 0.0075774084577184
27 0.015197953203032
28 0.0417202322723517
29 0.0729754355144866
30 0.216543209366722
31 0.413178324458858
32 0.784002966002704
33 0.789445104937212
34 0.989671268777033
35 1
36 0.937708925491769
37 0.94795083093682
38 0.903067961733911
39 0.870621914702277
40 0.829575295067768
41 0.649765164175006
42 0.469711136628211
43 0.324753132075832
44 0.202990514650392
45 0.185135312475746
46 0.137585840928943
47 0.118965376811424
48 0.0798760849481285
49 0.0162639963447423
50 0.0272032966168453
51 0.0206045283948659
52 0.0617617391264287
53 0.128809599949359
54 0.208034191976997
55 0.391890463626937
56 0.584511113701255
57 0.676776954956338
58 0.822995111791664
59 0.853967519645072
60 0.732762191944566
61 0.781217731986008
62 0.703394397857168
63 0.748768610345515
64 0.790171404315955
65 0.611236011224395
66 0.451330386007918
67 0.354644088251984
68 0.244541481931097
69 0.174861643749031
70 0.156204590799474
71 0.0953225561458465
72 0.0684196525028464
73 0.0170768095371376
74 0.0154761598538289
75 0.0156395771903841
76 0.0269345406762922
77 0.101787970770636
78 0.205054965402253
79 0.407145478598124
80 0.587469113262319
81 0.713515795227239
82 0.754593714783321
83 0.796471238197852
84 0.788418977601255
85 0.739795019756838
86 0.738812798456437
87 0.777655406872266
88 0.900939414147824
89 0.768320390420603
90 0.594295428977345
91 0.374346165648371
92 0.279492446047672
93 0.180138528573981
94 0.162601197756148
95 0.110501561747548
96 0.0810445949199369
97 0.0383451862622928
98 0.011705525951186
99 0.025087799896949
100 0.0395849968309755
101 0.0885828729207112
102 0.220169920757629
103 0.447224229572847
104 0.598830620395105
105 0.632990954249476
106 0.780072402798232
107 0.792937837225656
108 0.789105482964221
109 0.773898799697087
110 0.761805723332105
111 0.731500475837071
112 0.824586273633836
113 0.622042337990163
114 0.477141120392775
115 0.390774909108741
116 0.237841588697903
117 0.215774284686547
118 0.16036697407025
119 0.123469077195862
120 0.0710451316598983
121 0.0340245505958512
122 0.0359528246827411
123 0.0286009553814132
124 0.0317929006858094
125 0.0824880909852369
126 0.193709122222545
127 0.199773716500013
128 0.315761802636122
129 0.426918715793511
130 0.459405294594771
131 0.554809854761043
132 0.528085082420799
133 0.502004307900691
134 0.564895055307561
135 0.6095232505992
136 0.614702500140151
137 0.471070425611037
138 0.388778650663682
139 0.347389192336926
140 0.223191522262292
141 0.213158451929009
142 0.206255981306302
143 0.133493452706027
144 0.103103764755474
145 0.0357438223537628
146 0.0433598762573602
147 0.0307281554570788
148 0.0368158995552236
149 0.0472905898721924
150 0.0899645314350686
151 0.147418712973256
152 0.265992353352808
153 0.357224847844673
154 0.459085389210397
155 0.446329227495193
156 0.442976047732332
157 0.455563885952855
158 0.430261905737512
159 0.535026378271651
160 0.622841942365161
161 0.492097631945887
162 0.524135314242489
163 0.405233100154134
164 0.332109023057845
165 0.250826520655825
166 0.191269124661931
167 0.12652004577071
};
\addlegendentry{Scenario 2}
\end{axis}

\end{tikzpicture}
    \caption{Weekly normalized cell load profile of the two scenarios.}
    \label{fig:ues}
    \vspace{-.5cm}
\end{figure}
Most studies dealing with topology optimization focus their analysis on a single, or a handful, value of $\lambda_{UE}$. Since the energy optimization technique we devise tunes the IAB-node activation on the basis of the number of \glspl{ue} and their load, we need to evaluate our model on a varying values of UE density, ideally following a realistic trend. Therefore, we employ a technique used in similar research~\cite{baiocchi2017joint} to devise a time-varying UE density. First, we extract the cell load profile $p(t)$ related of each area from openly available datasets \cite{barlacchi2015multi}. This is then  normalized  in the range $(0,1]$. We model UE density as $\lambda_{UE}(t)=p(t)l\lambda_{gNB}$, a function of time, where $l=10$ is the number of \glspl{ue} per \glspl{gnb} taken from \gls{3gpp} technical report~\cite{3GPP_913}. 
Finally, based on this model, we generate a set of 168 graphs---one for each hour of the week---capturing the temporal variation at one-hour granularity. 
The hourly trend of $\lambda_{UE}(t)$ corresponding to the two areas of analysis are shown in Fig. \ref{fig:ues}.

\section{Results}
 \label{sect:results}

In this section, we discuss the results of our approaches in detail. We compare the results for both scenarios detailed in the previous section. The measurement graphs are generated for each hour of the week, totaling 168 hours and we run the algorithms on these.

We provide an optimality gap analysis comparing our methods against exact MILP solutions on small instances where optimal solutions are computable. We further compare the results of local search method and selective-reduction method against greedy approach from~\cite{shang2024energy}. We select this baseline over simpler sequential heuristics like shortest-path with fixed power allocation because such approaches assign routes independently per \gls{ue}, ignoring the shared-spectrum nature of in-band IAB where link capacity is determined by aggregate interference from all active nodes simultaneously (Eq. 4–5). Routing decisions that are individually optimal per \gls{ue} can raise the interference floor for all other links, making independent routing fundamentally unsuitable as a point of comparison. The greedy baseline from~\cite{shang2024energy} represents a stronger sequential approach with interference-aware metrics and IAB-specific constraints, providing a more meaningful comparison. 

The original algorithm uses frequency-division (subchannel allocation), while our IAB network operates with time-division (TDMA). We adapted the approach by: replacing subchannel allocation with TDMA time-slot allocation using water-filling to maximize minimum UE rate, maintaining the hierarchical processing of donor to IAB nodes to UEs and SNR-based prioritization within each level, and using gain-pathloss ranking for initial topology construction instead of SNR-based association to match our problem formulation. 
In the first part of the section, we compare the throughput obtained by different approaches listed above. Then, in the second part, we evaluate the energy consumption in terms of power, number of nodes activated and energy efficiency for both strategies.

\subsection{Results on Throughput Maximization}

%\begin{itemize}
    %\item Results of small-scale instance
    %\item To analyze the throughput achieved by the IAB networks, we show the hourly throughput measured over a week. 
    %\item Furthermore, we analyze the run-time required by each of the approaches. 
    %\item We also analyze the evolution of the solutions over time which provides an idea on how fast the approaches  arrive at the best solution.
%\end{itemize}
%\bc{We first present the results for small-scale instance such that the problem can be solved optimally?. }

\begin{figure}[!t]
    \centering
        \setlength\fwidth{0.9\columnwidth} 
        \setlength\fheight{0.5\columnwidth} 
        
            \input{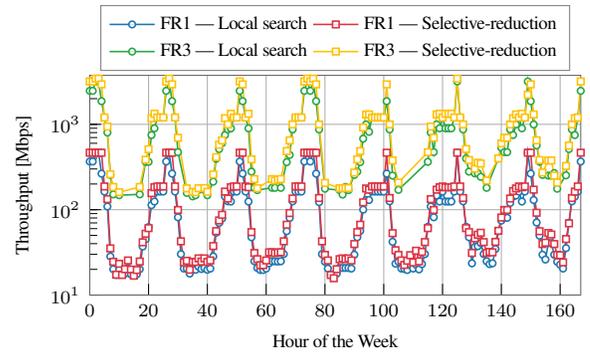}
 
    \caption{Throughput performance of two approaches in scenario 1.}
    \label{fig:throughput_week}
    \vspace{-.5cm}
\end{figure}

To analyze the throughput achieved by the \gls{iab} networks, we first show the hourly throughput measured over a week for scenario 1 using local search and selective-reduction approaches, which is presented in Fig.~\ref{fig:throughput_week}. We note that in the evaluations, initial retention count and maximum retention count for the selective-reduction method are set as 5 and 10 respectively.

The periodic fluctuation in the throughput achieved can be attributed to the varying user density corresponding to the peak and non-peak hours of a day. We observe that the selective-reduction method provides slightly higher throughput than the local search method and we note that the result is consistent across both FR1 and FR3. 
This improved performance is a consequence of the ranking and pruning of the links in the selective-reduction method based on effective link quality. 
Prioritizing link quality enables the optimization solvers to establish routes that maximize data transmission efficiency. 
We also observe that greedy baseline has the worst performance compared to the other two proposed methods. The greedy baseline exhibits failure cases at 7 GHz under high-load conditions, achieving 0 throughput in 27\% of hours (46 out of 168). This occurs because the greedy topology-building process, which iteratively connects nodes while enforcing tree constraints, fails to complete when network density is high. 
Our proposed algorithms maintain 100\% connectivity across all load conditions by solving joint optimization problems, rather than using greedy iterative construction. This demonstrates that joint optimization is necessary not only for efficiency but also for reliability at scale.

\begin{figure}[!t]
    \centering
    \setlength\fwidth{0.9\columnwidth} 
    \setlength\fheight{0.4\columnwidth}
        \input{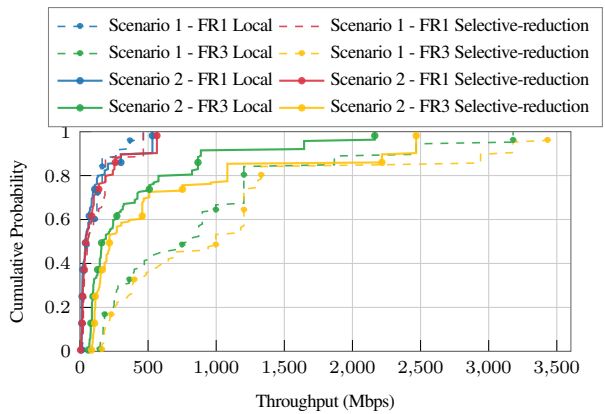}
     
    \caption{Throughput CDF for both scenarios.}
    \label{fig:throughput_cdf}
    \vspace{-.3cm}
\end{figure}

We also present the \gls{cdf} for throughput across all evaluated scenarios in Fig. \ref{fig:throughput_cdf}. \gls{cdf} converging  at lower throughput indicates a consistent performance ceiling, while convergence at higher throughput reflects greater peak capacity.
The plot highlights that the selective-reduction method outperforms the local-search method in terms of throughput and maintains consistent performance across both scenarios and for both FR1 and FR3.    
\begin{table}[t]
\centering
\caption{Optimality gap analysis on small instances.}
\label{tab:optimality}
\setlength{\tabcolsep}{2.5pt}
\begin{tabular}{cccccrrrrrrr}
\toprule
\textbf{Hour} & \textbf{UEs} & \textbf{Nodes} & \textbf{Freq} & \textbf{Exact} & \textbf{Local} & \textbf{Select.} & \textbf{Greedy} & \multicolumn{3}{c}{\textbf{Gap (\%)}} \\
\cmidrule{9-11}
 & & & & \textbf{(Mbps)} & \textbf{(Mbps)} & \textbf{(Mbps)} & \textbf{(Mbps)} & \textbf{L} & \textbf{S} & \textbf{G} \\
\midrule
\multirow{2}{*}{1} & \multirow{2}{*}{2} & \multirow{2}{*}{30} & FR1 & 465.3 & 366.9 & 465.3 & 128.9 & 21.1 & 0.0 & 72.3 \\
 & & & FR3 & 3179.1 & 2467.8 & 3179.1 & 870.1 & 22.4 & 0.0 & 72.6 \\
\midrule
\multirow{2}{*}{2} & \multirow{2}{*}{1} & \multirow{2}{*}{29} & FR1 & 465.3 & 465.3 & 465.3 & 128.9 & 0.0 & 0.0 & 72.3 \\
 & & & FR3 & 3432.1 & 2467.8 & 3432.1 & 870.1 & 28.1 & 0.0 & 74.6 \\
\midrule
\multirow{2}{*}{26} & \multirow{2}{*}{2} & \multirow{2}{*}{30} & FR1 & 465.3 & 366.9 & 465.3 & 128.9 & 21.1 & 0.0 & 72.3 \\
 & & & FR3 & 3179.1 & 2467.8 & 3179.1 & 870.1 & 22.4 & 0.0 & 72.6 \\
\midrule
\multirow{2}{*}{74} & \multirow{2}{*}{1} & \multirow{2}{*}{29} & FR1 & 465.3 & 465.3 & 465.3 & 128.9 & 0.0 & 0.0 & 72.3 \\
 & & & FR3 & 3432.1 & 2467.8 & 3432.1 & 870.1 & 28.1 & 0.0 & 74.6 \\
\midrule
\multirow{2}{*}{125} & \multirow{2}{*}{1} & \multirow{2}{*}{29} & FR1 & 465.3 & 465.3 & 465.3 & 128.9 & 0.0 & 0.0 & 72.3 \\
 & & & FR3 & 3432.1 & 2467.8 & 3432.1 & 870.1 & 28.1 & 0.0 & 74.6 \\
\bottomrule
\multicolumn{11}{l}{\scriptsize L=Local Search, S=Selective-Reduction, G=Greedy}
\end{tabular}
\vspace{-.5cm}
\end{table}

Table~\ref{tab:optimality} presents optimality gap analysis on small instances with 1-2 \glspl{ue}, 29-30 nodes. We observe high consistency: instances with identical \gls{ue} counts yield identical results across different hours validating result reproducibility. Selective-reduction achieves perfect optimality (0\% gap) on all 4 instance types. This is expected for low-load instances as the pruned graph retains all edges relevant to the optimal solution, making the reduced problem equivalent to the exact formulation. The solver therefore finds the exact optimum on the reduced problem. Local search exhibits 0-28\% gaps, achieving optimality on smallest instances but showing moderate gaps on larger or higher-frequency cases. The largest optimality gap of 28.1\% is observed for the local search method at FR3. This is attributed to the significantly larger capacity range at FR3 where the exact optimal reaches 3432.1 Mbps compared to 465.3 Mbps at FR1 meaning that suboptimal power configurations produce larger absolute gaps even when relative suboptimality is comparable across bands. The one variable at a time power fixing strategy of the local search method cannot capture joint power configurations across multiple nodes simultaneously, leading to convergence to a local optimum in high-capacity regimes. Nevertheless, local search still achieves 65\% higher throughput than the greedy baseline at FR3, confirming its practical utility despite the optimality gap relative to the exact solver. Greedy baseline consistently shows 72-75\% gaps, being approximately 3-4 times worse than optimal.

\begin{figure}[!t]
        \centering
        \setlength\fwidth{0.9\columnwidth} 
        \setlength\fheight{0.5\columnwidth} 
       
            \input{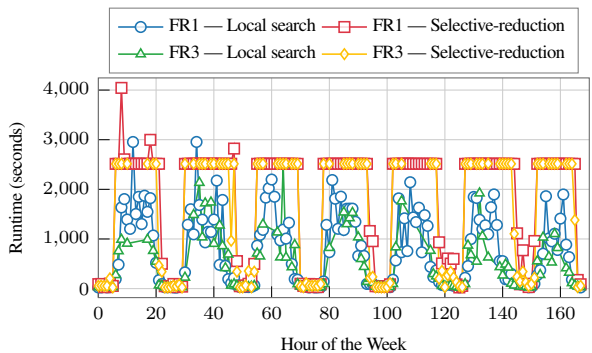}   
    \caption{Runtime performance of two approaches in scenario 1.}
    \label{fig:runtime_throughput}
    \vspace{-.5cm}
\end{figure}

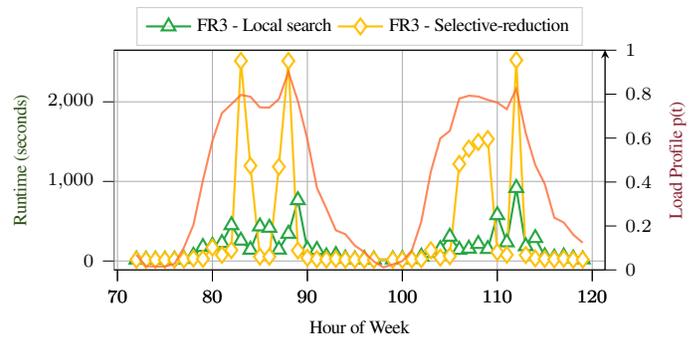
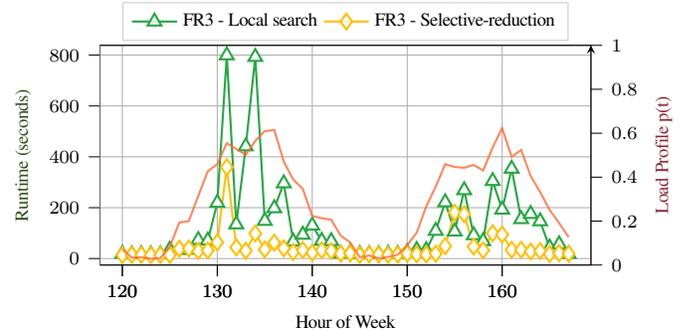
\begin{figure}[!t]
    \centering
    \begin{subfigure}{\linewidth}
        \centering
        \setlength\fwidth{0.9\columnwidth} 
        \setlength\fheight{0.4\columnwidth}
        % This file was created with tikzplotlib v0.10.1.
\begin{tikzpicture}

\definecolor{brown1393853}{RGB}{139,38,53}
\definecolor{darkgray176}{RGB}{176,176,176}
\definecolor{darkgreen458022}{RGB}{45,80,22}
\definecolor{dimgray102}{RGB}{102,102,102}
\definecolor{gold2551937}{RGB}{255,193,7}
\definecolor{lightgray204}{RGB}{204,204,204}
\definecolor{seagreen4016769}{RGB}{40,167,69}
\definecolor{tomato25510753}{RGB}{255,107,53}
\pgfplotsset{every tick label/.append style={font=\scriptsize}}

\begin{axis}[
width=\fwidth,
height=\fheight,
font=\scriptsize,
axis line style={dimgray102},
legend cell align={left},
legend style={
  fill opacity=0.8,
  draw opacity=1,
  text opacity=1,
  at={(0.5,1.02)},         % Centered above plot
  anchor=south,
  draw=lightgray204,
  legend columns=2,
},
tick align=outside,
tick pos=left,
x grid style={darkgray176},
xlabel={Hour of Week},
xmajorgrids,
xmin=69.65, xmax=121.35,
xtick style={color=black},
y grid style={darkgray176},
ylabel=\textcolor{darkgreen458022}{Runtime [s]},
ymajorgrids,
ymin=-112.127295696735, ymax=2648.67520567179,
ytick style={color=black},
xmajorgrids,
ymajorgrids,
xlabel style={font=\scriptsize},
ylabel style={font=\scriptsize}
]
\addplot [thick, seagreen4016769, mark=triangle*, mark size=3, mark options={solid,fill=white}]
table {%
72 18.2697191238403
73 16.5552408695221
74 15.9285190105438
75 15.8250818252563
76 16.5155560970306
77 24.7686719894409
78 65.6551599502563
79 176.374408006668
80 191.424275875092
81 226.228232860565
82 449.553988933563
83 255.784578084946
84 141.597519874573
85 434.272095918655
86 417.448131084442
87 142.811729192734
88 341.894912004471
89 764.019115924835
90 146.763411045074
91 138.625199079514
92 58.1433501243591
93 75.9059159755707
94 30.6543428897858
95 19.5041170120239
96 25.5531029701233
97 17.9235808849335
99 19.8381760120392
100 24.2650010585785
101 24.1933920383453
102 54.0198261737823
103 100.562776088715
104 149.143706083298
105 305.100582122803
106 140.164645195007
107 154.762257099152
108 213.88352394104
109 150.666967868805
110 576.047414779663
111 234.03110909462
112 917.291572093964
113 185.059120893478
114 288.398728132248
115 59.8068449497223
116 42.2413311004639
117 56.9322490692139
118 31.5789260864258
119 20.4655919075012
};
%\addlegendentry{7 GHz - local}
\addlegendentry{FR3 - Local search}
\addplot [thick, gold2551937, mark=diamond*, mark size=3, mark options={solid,fill=white}]
table {%
72 15.204715013504
73 13.4165370464325
74 14.2927877902985
75 13.3637270927429
76 14.3676650524139
77 18.0752079486847
78 31.6643888950348
79 26.1716160774231
80 168.383798122406
81 76.7552831172943
82 136.755750894547
83 2514.50417685509
84 1195.95370316505
85 54.6412999629974
86 54.9238181114197
87 1183.67834210396
88 2513.64954304695
89 132.328351974487
90 40.2177300453186
91 22.9290208816528
92 18.6039211750031
93 25.4859259128571
94 15.9527859687805
95 15.8794779777527
96 13.8149209022522
97 13.5620429515839
99 14.430764913559
100 13.6736681461334
101 18.5090620517731
102 23.163064956665
103 134.530338048935
104 42.2677340507507
105 60.5498278141022
106 1219.75591516495
107 1412.06402683258
108 1494.82217001915
109 1532.29182386398
110 111.468940973282
111 77.6303200721741
112 2523.18418288231
113 74.4072840213776
114 32.5826871395111
115 25.2998740673065
116 19.5316729545593
117 28.7912569046021
118 15.7783100605011
119 17.7828869819641
};
%\addlegendentry{7 GHz - Selective-reduction}
\addlegendentry{FR3 - Selective-reduction}

\end{axis}

\begin{axis}[
width=\fwidth,
height=\fheight,
font=\scriptsize,
axis y line=right,
tick align=outside,
x grid style={darkgray176},
xmin=69.65, xmax=121.35,
xtick pos=left,
xtick style={color=black},
y grid style={darkgray176},
ylabel=\textcolor{brown1393853}{Load Profile p(t)},
ymin=0, ymax=1,
ytick pos=right,
ytick style={color=black},
yticklabel style={anchor=west},
xlabel style={font=\scriptsize},
ylabel style={font=\scriptsize}
]
\addplot [thick, tomato25510753, opacity=0.8]
table {%
72 0.0684196525028464
73 0.0170768095371376
74 0.0154761598538289
75 0.0156395771903841
76 0.0269345406762922
77 0.101787970770636
78 0.205054965402253
79 0.407145478598124
80 0.587469113262319
81 0.713515795227239
82 0.754593714783321
83 0.796471238197852
84 0.788418977601255
85 0.739795019756838
86 0.738812798456437
87 0.777655406872266
88 0.900939414147824
89 0.768320390420603
90 0.594295428977345
91 0.374346165648371
92 0.279492446047672
93 0.180138528573981
94 0.162601197756148
95 0.110501561747548
96 0.0810445949199369
97 0.0383451862622928
98 0.011705525951186
99 0.025087799896949
100 0.0395849968309755
101 0.0885828729207112
102 0.220169920757629
103 0.447224229572847
104 0.598830620395105
105 0.632990954249476
106 0.780072402798232
107 0.792937837225656
108 0.789105482964221
109 0.773898799697087
110 0.761805723332105
111 0.731500475837071
112 0.824586273633836
113 0.622042337990163
114 0.477141120392775
115 0.390774909108741
116 0.237841588697903
117 0.215774284686547
118 0.16036697407025
119 0.123469077195862
};
\end{axis}

\end{tikzpicture}
        \caption{Runtime performance of two weekdays day Scenario 2}
        \label{fig:run_first}
    \end{subfigure}
    \hfill
    \begin{subfigure}{\linewidth}
        \centering
        \setlength\fwidth{0.9\columnwidth} 
        \setlength\fheight{0.4\columnwidth}
        % This file was created with tikzplotlib v0.10.1.
\begin{tikzpicture}

\definecolor{brown1393853}{RGB}{139,38,53}
\definecolor{darkgray176}{RGB}{176,176,176}
\definecolor{darkgreen458022}{RGB}{45,80,22}
\definecolor{dimgray102}{RGB}{102,102,102}
\definecolor{gold2551937}{RGB}{255,193,7}
\definecolor{lightgray204}{RGB}{204,204,204}
\definecolor{seagreen4016769}{RGB}{40,167,69}
\definecolor{tomato25510753}{RGB}{255,107,53}
\pgfplotsset{every tick label/.append style={font=\scriptsize}}

\begin{axis}[
width=\fwidth,
height=\fheight,
axis line style={dimgray102},
legend cell align={left},
legend style={
  fill opacity=0.8,
  draw opacity=1,
  text opacity=1,
  at={(0.5,1.02)},         % Centered above plot
  anchor=south, 
  draw=lightgray204,
  font=\scriptsize,
  legend columns=2,
},
tick align=outside,
tick pos=left,
x grid style={darkgray176},
xlabel={Hour of Week},
xmajorgrids,
xmin=117.65, xmax=169.35,
xtick style={color=black},
y grid style={darkgray176},
ylabel=\textcolor{darkgreen458022}{Runtime [s]},
ymajorgrids,
ymin=-25.4866423130035, ymax=838.372868490219,
ytick style={color=black},
xmajorgrids,
ymajorgrids,
xlabel style={font=\scriptsize},
ylabel style={font=\scriptsize}
]
\addplot [thick, seagreen4016769, mark=triangle*, mark size=3, mark options={solid,fill=white}]
table {%
120 18.5244359970093
121 16.6588969230652
122 17.8749260902405
123 16.6402587890625
124 16.7498388290405
125 35.1243650913239
126 34.2294781208038
127 34.5329349040985
128 72.9462599754333
129 70.459235906601
130 219.145797014236
131 799.106527090073
132 134.37948012352
133 441.240912914276
134 792.734397888184
135 148.011565208435
136 197.795040130615
137 295.625622987747
138 66.9719190597534
139 93.5893828868866
140 129.897373914719
141 69.6612849235535
142 68.5952930450439
143 19.323292016983
144 23.4284479618073
145 17.8234479427338
146 17.4637060165405
147 17.4293448925018
148 19.1243450641632
149 18.1843390464783
150 24.9939680099487
151 34.0040788650513
152 32.9460570812225
153 108.887366056442
154 221.086697101593
155 106.232445001602
156 268.674237966537
157 91.6768400669098
158 68.352814912796
159 304.862626075745
160 192.42782497406
161 353.210520982742
162 154.957018136978
163 175.244082212448
164 146.584302902222
165 43.1226360797882
166 55.989296913147
167 18.1295478343964
};
%\addlegendentry{7 GHz - Local search}
\addlegendentry{FR3 - Local search}
\addplot [thick, gold2551937, mark=diamond*, mark size=3, mark options={solid,fill=white}]
table {%
120 14.086198091507
121 14.6588201522827
122 14.642030954361
123 13.7796990871429
124 13.9951419830322
125 14.5116369724274
126 38.7723729610443
127 39.6185669898987
128 30.0129499435425
129 32.0633590221405
130 63.7776808738708
131 359.534832000732
132 44.3645570278168
133 30.7079210281372
134 98.6060440540314
135 35.7530279159546
136 62.8027491569519
137 40.1443300247192
138 25.4493989944458
139 33.4373111724854
140 22.5066518783569
141 32.4017400741577
142 29.3940839767456
143 19.291188955307
144 20.3380799293518
145 14.2829411029816
146 16.0133061408997
147 15.8395969867706
148 14.5449118614197
149 15.9243609905243
150 19.2795169353485
151 16.8993859291077
152 16.425940990448
153 19.164803981781
154 49.0690619945526
155 180.50487112999
156 174.471853971481
157 46.9733970165253
158 32.4162931442261
159 99.3765780925751
160 94.3825621604919
161 34.5146119594574
162 34.5680429935455
163 27.7222149372101
164 29.9673631191254
165 17.8556990623474
166 21.5327348709106
167 18.8414030075073
};
%\addlegendentry{7 GHz - Selective-reduction}
\addlegendentry{FR3 - Selective-reduction}

\end{axis}

\begin{axis}[
width=\fwidth,
height=\fheight,
font=\scriptsize,
axis y line=right,
tick align=outside,
x grid style={darkgray176},
xmin=117.65, xmax=169.35,
xtick pos=left,
xtick style={color=black},
y grid style={darkgray176},
ylabel=\textcolor{brown1393853}{Load Profile p(t)},
ymin=0, ymax=1,
ytick pos=right,
ytick style={color=black},
yticklabel style={anchor=west},
xlabel style={font=\scriptsize},
ylabel style={font=\scriptsize}
]
\addplot [thick, tomato25510753, opacity=0.8]
table {%
120 0.0710451316598983
121 0.0340245505958512
122 0.0359528246827411
123 0.0286009553814132
124 0.0317929006858094
125 0.0824880909852369
126 0.193709122222545
127 0.199773716500013
128 0.315761802636122
129 0.426918715793511
130 0.459405294594771
131 0.554809854761043
132 0.528085082420799
133 0.502004307900691
134 0.564895055307561
135 0.6095232505992
136 0.614702500140151
137 0.471070425611037
138 0.388778650663682
139 0.347389192336926
140 0.223191522262292
141 0.213158451929009
142 0.206255981306302
143 0.133493452706027
144 0.103103764755474
145 0.0357438223537628
146 0.0433598762573602
147 0.0307281554570788
148 0.0368158995552236
149 0.0472905898721924
150 0.0899645314350686
151 0.147418712973256
152 0.265992353352808
153 0.357224847844673
154 0.459085389210397
155 0.446329227495193
156 0.442976047732332
157 0.455563885952855
158 0.430261905737512
159 0.535026378271651
160 0.622841942365161
161 0.492097631945887
162 0.524135314242489
163 0.405233100154134
164 0.332109023057845
165 0.250826520655825
166 0.191269124661931
167 0.12652004577071
};
\end{axis}

\end{tikzpicture}
        \caption{Runtime performance of weekend scenario 2}
        \label{fig:run_last}
    \end{subfigure}
    \caption{Runtime performance- weekdays and weekend for scenario 2.}
    \label{fig:run_sc2}
    \vspace{-.5cm}
\end{figure}

Fig. \ref{fig:runtime_throughput} shows the runtime required for both algorithms. We observe that the local search algorithm requires less execution time than  the selective-reduction method. This advantage of the local search method stems from its problem decomposition strategy. While the selective-reduction method solves a single optimization problem with a reduced but still complex input space, the local search approach solves multiple smaller subproblems sequentially. 
To study the impact of network load intensity on runtime, we plot the sample runtime performance of weekday and weekend from scenario 2, where the cell load is lighter than scenario 1. The results are presented in Fig. \ref{fig:run_sc2}.
We observe that the runtime of the algorithms is heavily dependent on the network load. Selective-reduction achieves better runtime when the load is below 0.6, where the pruning effect has significant impact. This indicates the importance of fine-tuning pruning intensity by setting appropriate retention counts according to the load conditions, such as differentiating between weekend and weekdays. We note that the proposed approaches are ideal for periodic network reconfigurations on hourly timescale than real-time adaptations.

The runtime of the selective-reduction method can also be controlled explicitly by setting a limit on runtime.
%Thus  we can effectively control  runtime of selective-reduction method, and still achieve good performance. 
At the same time, runtime for local-search method is more variable and harder to control as it iteratively solves multiple  simplified problems according to the convergence criteria.

%\begin{figure}[!t]
%    \centering
%    \begin{subfigure}{\linewidth}
%        \centering
        %\setlength\fwidth{0.9\columnwidth} 
        %\setlength\fheight{0.5\columnwidth}
        %\input{figures/runtime_first_day_sc2.tex}
        %\caption{Run time performance of first day Scenario 2}
        %\label{fig:run_first}
    %\end{subfigure}
    %\hfill
    %\begin{subfigure}{\linewidth}
     %   \centering
        %\setlength\fwidth{0.9\columnwidth} 
        %\setlength\fheight{0.5\columnwidth}
        %\input{figures/runtime_last_day_sc2.tex}
        %\caption{Run time performance of last day scenario 2}
        %\label{fig:run_last}
    %\end{subfigure}
    %\caption{Energy efficiency CDF comparison across scenarios}
   % \label{fig:run_sc2}
%\end{figure}

\begin{figure}[h]
    \centering
    \begin{subfigure}{0.49\linewidth}
        \centering
         \setlength\fwidth{\columnwidth} 
         \setlength\fheight{0.8\columnwidth} % This file was created with tikzplotlib v0.10.1.
\begin{tikzpicture}

\definecolor{dimgray102}{RGB}{102,102,102}
\definecolor{gray}{RGB}{128,128,128}
\definecolor{lightgray204}{RGB}{204,204,204}
\definecolor{steelblue31119180}{RGB}{31,119,180}
\pgfplotsset{every tick label/.append style={font=\scriptsize}}

\begin{axis}[
width=\fwidth,
height=\fheight,
axis line style={dimgray102},
legend style={
  fill opacity=0.8,
  draw opacity=1,
  text opacity=1,
  at={(0.97,0.03)},
  anchor=south east,
  draw=gray,
  font=\scriptsize
},
tick align=inside,
tick pos=left,
x grid style={lightgray204},
xlabel={Execution Time (s)},
xmajorgrids,
xmin=6.59891724586487, xmax=15.914088010788,
xminorgrids,
xtick style={color=black},
y grid style={lightgray204},
ylabel={Throughput [Mbps]},
ymajorgrids,
ymin=247.450118987802, ymax=372.601411750581,
yminorgrids,
ytick style={color=black},
xmajorgrids,
ymajorgrids,
xlabel style={font=\scriptsize},
ylabel style={font=\scriptsize}
]
\addplot [semithick, steelblue31119180, forget plot]
table {%
7.02233409881592 253.138814113383
7.66460704803467 301.55436675
9.12061905860901 301.55436675
9.33160305023193 366.912716625
15.4906711578369 366.912716625
};
\addplot [draw=steelblue31119180, fill=white, forget plot, mark=*, only marks]
table{%
x  y
7.02233409881592 253.138814113383
7.66460704803467 301.55436675
9.12061905860901 301.55436675
9.33160305023193 366.912716625
15.4906711578369 366.912716625
};
\end{axis}

\end{tikzpicture}
        
        \subcaption[]{Non-peak hour: Hour 0}
        \label{subfig:hour0}
    \end{subfigure}
    \hfill
    \begin{subfigure}{0.49\linewidth}
        \centering
        \setlength\fwidth{\columnwidth} 
        \setlength\fheight{0.8\columnwidth} % This file was created with tikzplotlib v0.10.1.
\begin{tikzpicture}

\definecolor{dimgray102}{RGB}{102,102,102}
\definecolor{gray}{RGB}{128,128,128}
\definecolor{lightgray204}{RGB}{204,204,204}
\definecolor{steelblue31119180}{RGB}{31,119,180}
\pgfplotsset{every tick label/.append style={font=\scriptsize}}

\begin{axis}[
width=\fwidth,
height=\fheight,
axis line style={dimgray102},
legend style={
  fill opacity=0.8,
  draw opacity=1,
  text opacity=1,
  at={(0.97,0.03)},
  anchor=south east,
  draw=gray,
  font=\scriptsize
},
tick align=inside,
tick pos=left,
x grid style={lightgray204},
xlabel={Execution Time (s)},
xmajorgrids,
xmin=-23.2739530563355, xmax=1065.60102424622,
xminorgrids,
xtick style={color=black},
y grid style={lightgray204},
ylabel={Throughput [Mbps]},
ymajorgrids,
ymin=25.5102849838177, ymax=37.570673582105,
yminorgrids,
ytick style={color=black},
xmajorgrids,
ymajorgrids,
xlabel style={font=\scriptsize},
ylabel style={font=\scriptsize}
]
\addplot [semithick, steelblue31119180, forget plot]
table {%
26.2203640937805 26.058484465558
122.962609052658 27.3581120946399
264.587000131607 29.241991404843
279.153152942657 29.2419914078672
286.401230096817 31.2051970117375
304.40923500061 31.2275588016691
324.873220920563 32.2415352321993
340.166110038757 32.715125625
347.760055065155 33.477741709195
361.2906229496 33.6498435
374.070469141006 33.649843624715
385.125293016434 34.2369122774362
391.932203054428 34.3048115639226
399.301270961761 35.1889178651471
437.055093050003 35.4173767621723
444.394843101501 35.4173767658141
461.602518081665 35.4173768677587
477.924829006195 36.804516328125
639.285027980804 37.0224739717296
710.128831148148 37.0224741003646
1016.1067070961 37.0224739856556
};
\addplot [draw=steelblue31119180, fill=white, forget plot, mark=*, only marks]
table{%
x  y
26.2203640937805 26.058484465558
122.962609052658 27.3581120946399
264.587000131607 29.241991404843
279.153152942657 29.2419914078672
286.401230096817 31.2051970117375
304.40923500061 31.2275588016691
324.873220920563 32.2415352321993
340.166110038757 32.715125625
347.760055065155 33.477741709195
361.2906229496 33.6498435
374.070469141006 33.649843624715
385.125293016434 34.2369122774362
391.932203054428 34.3048115639226
399.301270961761 35.1889178651471
437.055093050003 35.4173767621723
444.394843101501 35.4173767658141
461.602518081665 35.4173768677587
477.924829006195 36.804516328125
639.285027980804 37.0224739717296
710.128831148148 37.0224741003646
1016.1067070961 37.0224739856556
};
\end{axis}

\end{tikzpicture}
        
        \subcaption[]{Peak hour: Hour 18}
        \label{subfig:hour18}
    \end{subfigure}

    \caption{Solution evolution using local search method at selected peak and non-peak hours, 3.6 GHz-100 MHz.}
    \label{fig:evolution}
    \vspace{-.3cm}
\end{figure}
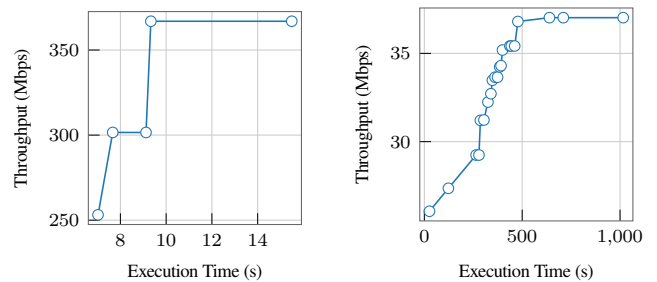

As such, we analyze the evolution of the solutions over time for local-search method which provides an idea on how fast the approaches  arrive at the best solution.  Fig. \ref{fig:evolution} presents the  convergence trajectory that shows the throughput achieved over execution time. The results are presented for one peak hour and one non-peak hour of a given day.
During the peak hour, improvements in throughput are marginal despite long execution times. This highlights the importance of imposing a runtime limit in  algorithms so that we can achieve trade-off on computational time and solution optimality. The solution evolution statistics for local-search method is given in Table~\ref{tab:hourly_performance}. 
The metrics captured include the initial and final throughput values, the relative improvement percentage, the time required to reach a near-final solution, and the total execution time.

\begin{table}[t]
  \centering
  \caption{Solution evolution statistics for local search method}
  \label{tab:hourly_performance}
  %\scriptsize
  \begin{tabular}{crrrrr}
    \toprule
    \textbf{Hour} & \textbf{Initial} & \textbf{Final} & \textbf{Improvement} & \textbf{Time to} & \textbf{Total} \\
    & \textbf{Solution} & \textbf{Solution} & \textbf{Percentage} & \textbf{Near-Final} & \textbf{Execution} \\
    &  &  &  &  & \textbf{Time} \\
    & \textbf{(Mbps)} & \textbf{(Mbps)} & \textbf{(\%)} & \textbf{(s)} & \textbf{(s)} \\
    \midrule
    0 & 253.14 & 366.91 & 44.95 & 9.33 & 15.49 \\
    1 & 253.14 & 366.91 & 44.95 & 8.09 & 14.64 \\
    2 & 374.46 & 465.27 & 24.25 & 8.01 & 16.80 \\
    3 & 374.46 & 465.27 & 24.25 & 7.82 & 15.76 \\
    4 & 187.23 & 264.59 & 41.32 & 8.12 & 19.92 \\
    5 & 163.07 & 163.07 & 0.00 & --- & 22.03 \\
    6 & 98.71 & 109.05 & 10.48 & 25.17 & 130.62 \\
    7 & 27.69 & 28.33 & 2.31 & 50.78 & 471.13 \\
    8 & 20.07 & 20.07 & 0.00 & 26.99 & 494.73 \\
    9 & 17.19 & 17.19 & 0.00 & --- & 27.46 \\
    10 & 19.77 & 19.80 & 0.14 & 28.82 & 760.64 \\
    11 & 17.43 & 17.43 & 0.00 & 27.89 & 1261.45 \\
    12 & 17.83 & 17.83 & 0.00 & 29.32 & 1312.08 \\
    13 & 20.45 & 20.45 & 0.00 & 26.38 & 440.68 \\
    14 & 17.51 & 17.51 & 0.00 & --- & 27.05 \\
    15 & 17.43 & 17.43 & 0.00 & 26.66 & 1193.39 \\
    16 & 17.51 & 17.51 & 0.00 & --- & 27.49 \\
    17 & 20.07 & 20.07 & 0.00 & 27.46 & 567.33 \\
    18 & 26.06 & 37.02 & 42.07 & 477.92 & 1016.11 \\
    19 & 41.40 & 44.93 & 8.53 & 193.23 & 1059.30 \\
    20 & 45.31 & 62.58 & 38.11 & 359.53 & 507.45 \\
    21 & 93.82 & 113.11 & 20.57 & 71.38 & 158.66 \\
    22 & 121.74 & 124.60 & 2.35 & 7.96 & 48.58 \\
    23 & 163.07 & 163.07 & 0.00 & --- & 35.20 \\
    \bottomrule
  \end{tabular}
  \vspace{-.2cm}
\end{table}
\begin{figure*}[htbp]
    \centering

    \begin{subfigure}{0.49\linewidth}
        \centering
        \setlength\fwidth{0.9\columnwidth} 
        \setlength\fheight{0.45\columnwidth}
        % This file was created with tikzplotlib v0.10.1.
\begin{tikzpicture}

\definecolor{darkorange25512714}{RGB}{255,127,14}
\definecolor{dimgray102}{RGB}{102,102,102}
\definecolor{gray}{RGB}{128,128,128}
\definecolor{lightgray204}{RGB}{204,204,204}
\definecolor{seagreen4016769}{RGB}{40,167,69}
\definecolor{steelblue31119180}{RGB}{31,119,180}

\begin{axis}[
width=\fwidth,
height=\fheight,
axis line style={dimgray102},
legend cell align={left},
legend style={fill opacity=0.8, draw opacity=1, text opacity=1, draw=gray},
font=\scriptsize,
tick align=outside,
tick pos=left,
x grid style={lightgray204},
xlabel={Hour of the Week},
xmajorgrids,
xmin=-8.35, xmax=175.35,
xminorgrids,
xtick style={color=black},
y grid style={lightgray204},
ylabel={Number of Activated nodes},
ymajorgrids,
ymin=1.175, ymax=19.325,
yminorgrids,
ytick style={color=black},
xmajorgrids,
ymajorgrids,
xlabel style={font=\scriptsize},
ylabel style={font=\scriptsize}
]
\addplot [semithick, darkorange25512714, forget plot]
table {%
0 17
1 17
2 17.6666666666667
3 18
4 17.75
5 17.75
6 17.25
7 16
8 16
9 15.25
10 14.25
11 14.25
12 14
13 14
14 13.75
15 13.75
16 13
17 13.5
18 14.25
21 14.25
22 16
23 16.25
24 16.75
25 17.5
26 17
27 17.5
28 17.25
29 17.25
30 16.75
32 15.75
33 15
34 14.5
35 14
36 13.5
37 14
38 13.5
39 14
40 14
41 13.75
42 13.75
43 13
45 13.5
47 13.25
48 14.75
49 16
50 16.5
51 17.5
52 16.75
53 16.75
54 17.25
55 16
56 15
57 15
58 13.5
59 14.25
60 14.5
61 14.5
62 14.25
63 13.5
64 13.75
65 13.5
67 13.25
68 13.5
69 13.25
70 13.75
71 15.25
72 16.25
73 17.25
74 17.5
75 17.5
76 18
77 17.75
78 16.25
80 15.5
81 14.25
82 14.25
83 14.5
84 14.75
85 15.5
86 15.75
87 15.75
88 15.5
90 14.75
91 14.25
92 13.5
93 14
95 14.25
96 14.75
97 16.25
98 16.5
99 17
100 17
102 16
103 16
104 15.75
105 14.75
106 15
107 15
108 14.75
110 15.25
111 15.25
112 15.25
113 15.75
114 15.75
115 15.75
116 15.75
117 15.25
118 15
119 16.25
120 16.25
121 17.25
122 18.5
123 18.5
124 18.5
125 18.5
126 18
127 16.75
128 16
130 15.5
131 15.5
132 15
133 14.75
135 13.5
136 13.25
137 14.5
138 15.25
139 15.75
141 15.5
142 15
143 14.5
146 15.25
147 16
148 16.5
149 17.75
150 17.5
151 16.5
152 16
154 13.75
155 13.5
156 13.75
157 14
158 14
159 13.5
160 13.25
161 13
162 14.75
164 15
166 15.75
167 16.25
};
\addplot [semithick, steelblue31119180, dashed, forget plot]
table {%
0 3
1 3
2 2.33333333333333
3 2
4 2
5 2
6 3.75
7 5.5
8 6.75
9 8
10 8
11 8
12 8
13 8
14 8
15 8
16 8
17 8
18 8
19 8
20 8
21 8
22 7.5
23 6.25
24 5
25 3.75
26 3
27 2.5
28 2.5
29 2.5
30 3.75
31 5.5
32 6.75
33 8
34 8
35 8
36 8
37 8
38 8
39 8
40 8
41 8
42 8
43 8
44 8
45 8
46 8
47 8
48 7.5
49 6.25
50 5
51 3.75
52 3
53 3
54 3.75
55 5
56 6.25
57 7.5
58 8
59 8
60 8
61 8
62 8
63 8
64 8
65 8
66 8
67 8
68 8
69 8
70 7.75
71 6.5
72 5.25
73 4
74 2.5
75 2.5
76 2
77 2
78 3.75
79 5
80 6.75
81 8
82 8
83 8
84 8
85 8
86 8
87 8
88 8
89 8
90 8
91 8
92 8
93 8
94 8
95 8
96 6.75
97 5.5
98 4.25
99 3
100 3
101 3
102 4.25
103 5.5
104 6.75
105 8
106 8
107 8
108 8
109 8
110 8
111 8
112 8
113 8
114 8
115 8
116 8
117 8
118 8
119 7.5
120 7.25
121 6.75
122 6.25
123 6.25
124 6.25
125 5
126 4.25
127 4.75
128 5
129 6.75
130 8
131 8
132 8
133 8
134 8
135 8
136 8
137 8
138 8
139 8
140 8
141 8
142 8
143 8
144 8
145 8
146 7.75
147 7.25
148 7
149 5.25
150 4.25
151 4.75
152 5
153 6.75
154 8
155 8
156 8
157 8
158 8
159 8
160 8
161 8
162 8
163 8
164 8
165 8
166 7.75
167 6.5
};
\addplot [semithick, seagreen4016769, dashed, forget plot]
table {%
0 14
1 14
2 12
3 12
4 11.25
5 10
6 10.25
7 8.75
8 8.75
9 11.25
10 10.5
11 13
12 12.75
13 10.25
14 12.5
15 10.25
16 10.75
17 11.5
18 9.75
19 9.75
20 8
21 7.75
22 9.75
23 11.5
24 12.5
25 13.25
26 12.75
27 11.25
28 11.5
29 10.25
30 8.75
31 8.25
32 8
33 9
34 10.75
35 11.25
36 11
37 10
38 8.5
39 9.25
40 9.25
41 9
42 8.25
44 7
45 6.5
46 7
47 8
48 10.5
49 12.5
51 13.5
52 13.25
53 12.75
54 13.25
55 11.75
56 11.5
57 9.75
58 8.25
59 8.75
60 8.75
61 8.75
62 8.75
63 8.25
64 8.25
65 9
66 8
67 7.75
68 8
69 8
70 9.25
71 9.75
72 9.75
73 10.75
74 10
75 11.25
76 12
77 10.75
78 11.25
79 10
80 8.5
81 8.25
82 10.5
83 12.25
84 13.75
85 13.5
86 11
87 8.75
88 8
89 8.75
90 8.75
91 9.5
92 9
93 8.5
94 10.25
95 11
96 12.75
97 14.25
98 13.75
99 14
100 14
101 13
102 12
103 10.5
104 8.5
105 8
106 7.5
107 8.5
108 9
109 8.75
110 9.25
111 8
112 8
113 8.25
114 8
115 7.5
116 8.25
117 8.25
118 10
119 12.25
120 12.25
121 13.75
122 13.75
123 14.25
124 14.5
125 14
126 12.25
127 10.75
128 9.75
129 8.75
131 8
132 7
133 7
134 7.25
135 8.25
136 8.5
137 8.25
140 7.75
141 7.25
142 7.25
143 8.5
144 9
145 10.75
146 13.25
147 13.75
148 14.5
149 13.75
151 13.5
152 11.75
153 10.75
154 9.5
155 7.5
156 7.75
157 7.25
158 7.25
160 7
161 7.25
162 8
163 7.75
164 9
165 9
166 9.75
167 11.75
};
\addplot [draw=darkorange25512714, fill=white, mark=*, mark size=1.25, only marks]
table{%
x  y
0 17
1 17
2 17.6666666666667
3 18
4 17.75
5 17.75
6 17.25
7 16
8 16
9 15.25
10 14.25
11 14.25
12 14
13 14
14 13.75
15 13.75
16 13
17 13.5
18 14.25
21 14.25
22 16
23 16.25
24 16.75
25 17.5
26 17
27 17.5
28 17.25
29 17.25
30 16.75
32 15.75
33 15
34 14.5
35 14
36 13.5
37 14
38 13.5
39 14
40 14
41 13.75
42 13.75
45 13.5
47 13.25
48 14.75
49 16
51 17.5
52 16.75
53 16.75
54 17.25
55 16
56 15
57 15
58 13.5
59 14.25
60 14.5
61 14.5
62 14.25
63 13.5
64 13.75
65 13.5
67 13.25
68 13.5
69 13.25
70 13.75
71 15.25
72 16.25
73 17.25
74 17.5
75 17.5
76 18
77 17.75
78 16.25
80 15.5
81 14.25
82 14.25
83 14.5
84 14.75
85 15.5
86 15.75
87 15.75
88 15.5
90 14.75
91 14.25
92 13.5
93 14
95 14.25
96 14.75
97 16.25
98 16.5
99 17
100 17
102 16
103 16
104 15.75
105 14.75
106 15
107 15
108 14.75
110 15.25
111 15.25
112 15.25
113 15.75
114 15.75
115 15.75
116 15.75
117 15.25
118 15
119 16.25
120 16.25
121 17.25
122 18.5
123 18.5
124 18.5
125 18.5
126 18
127 16.75
128 16
131 15.5
132 15
133 14.75
135 13.5
136 13.25
137 14.5
141 15.5
142 15
143 14.5
146 15.25
147 16
148 16.5
149 17.75
151 16.5
152 16
154 13.75
155 13.5
156 13.75
157 14
158 14
160 13.25
161 13
162 14.75
164 15
166 15.75
167 16.25
};
\addlegendentry{Local Search}
\addplot [draw=steelblue31119180, fill=white, mark=square*, mark size=1.25, only marks]
table{%
x  y
0 3
1 3
2 2.33333333333333
3 2
4 2
5 2
6 3.75
7 5.5
8 6.75
9 8
10 8
11 8
12 8
13 8
14 8
15 8
16 8
17 8
18 8
19 8
20 8
21 8
22 7.5
23 6.25
24 5
25 3.75
26 3
27 2.5
28 2.5
29 2.5
30 3.75
31 5.5
32 6.75
33 8
34 8
35 8
36 8
37 8
38 8
39 8
40 8
41 8
42 8
43 8
44 8
45 8
46 8
47 8
48 7.5
49 6.25
50 5
51 3.75
52 3
53 3
54 3.75
55 5
56 6.25
57 7.5
58 8
59 8
60 8
61 8
62 8
63 8
64 8
65 8
66 8
67 8
68 8
69 8
70 7.75
71 6.5
72 5.25
73 4
74 2.5
75 2.5
76 2
77 2
78 3.75
79 5
80 6.75
81 8
82 8
83 8
84 8
85 8
86 8
87 8
88 8
89 8
90 8
91 8
92 8
93 8
94 8
95 8
96 6.75
97 5.5
98 4.25
99 3
100 3
101 3
102 4.25
103 5.5
104 6.75
105 8
106 8
107 8
108 8
109 8
110 8
111 8
112 8
113 8
114 8
115 8
116 8
117 8
118 8
119 7.5
120 7.25
121 6.75
122 6.25
123 6.25
124 6.25
125 5
126 4.25
127 4.75
128 5
129 6.75
130 8
131 8
132 8
133 8
134 8
135 8
136 8
137 8
138 8
139 8
140 8
141 8
142 8
143 8
144 8
145 8
146 7.75
147 7.25
148 7
149 5.25
150 4.25
151 4.75
152 5
153 6.75
154 8
155 8
156 8
157 8
158 8
159 8
160 8
161 8
162 8
163 8
164 8
165 8
166 7.75
167 6.5
};
\addlegendentry{Greedy}
\addplot [draw=seagreen4016769, fill=white, mark=triangle*,mark size=1.25, only marks]
table{%
x  y
0 14
1 14
2 12
3 12
4 11.25
5 10
6 10.25
7 8.75
8 8.75
9 11.25
10 10.5
11 13
12 12.75
13 10.25
14 12.5
15 10.25
16 10.75
17 11.5
18 9.75
21 7.75
22 9.75
23 11.5
24 12.5
25 13.25
26 12.75
27 11.25
28 11.5
29 10.25
30 8.75
32 8
33 9
34 10.75
35 11.25
36 11
37 10
38 8.5
39 9.25
40 9.25
41 9
42 8.25
45 6.5
47 8
48 10.5
49 12.5
51 13.5
52 13.25
53 12.75
54 13.25
55 11.75
56 11.5
57 9.75
58 8.25
59 8.75
60 8.75
61 8.75
62 8.75
63 8.25
64 8.25
65 9
67 7.75
68 8
69 8
70 9.25
71 9.75
72 9.75
73 10.75
74 10
75 11.25
76 12
77 10.75
78 11.25
80 8.5
81 8.25
82 10.5
83 12.25
84 13.75
85 13.5
86 11
87 8.75
88 8
90 8.75
91 9.5
92 9
93 8.5
95 11
96 12.75
97 14.25
98 13.75
99 14
100 14
102 12
103 10.5
104 8.5
105 8
106 7.5
107 8.5
108 9
110 9.25
111 8
112 8
113 8.25
114 8
115 7.5
116 8.25
117 8.25
118 10
119 12.25
120 12.25
121 13.75
122 13.75
123 14.25
124 14.5
125 14
126 12.25
127 10.75
128 9.75
131 8
132 7
133 7
135 8.25
136 8.5
137 8.25
141 7.25
142 7.25
143 8.5
146 13.25
147 13.75
148 14.5
149 13.75
151 13.5
152 11.75
154 9.5
155 7.5
156 7.75
157 7.25
158 7.25
160 7
161 7.25
162 8
164 9
166 9.75
167 11.75
};
\addlegendentry{Selective-reduction}
\addplot [draw=red, fill=white, forget plot, mark=*,mark size=1.25, only marks]
table{%
x  y
43 13
50 16.5
130 15.5
138 15.25
139 15.75
150 17.5
159 13.5
};
\addplot [draw=red, fill=white, forget plot, mark=triangle*,mark size=1.25, only marks]
table{%
x  y
19 9.75
20 8
31 8.25
44 7
46 7
66 8
79 10
89 8.75
94 10.25
101 13
109 8.75
129 8.75
134 7.25
140 7.75
144 9
145 10.75
153 10.75
163 7.75
165 9
};
\end{axis}

\end{tikzpicture}
        \caption{RF activation – Scenario 1}
        \label{fig:ru_5}
    \end{subfigure}
    \hfill
    \begin{subfigure}{0.49\linewidth}
        \centering
        \setlength\fwidth{0.9\columnwidth} 
        \setlength\fheight{0.45\columnwidth}
        % This file was created with tikzplotlib v0.10.1.
\begin{tikzpicture}

\definecolor{darkorange25512714}{RGB}{255,127,14}
\definecolor{dimgray102}{RGB}{102,102,102}
\definecolor{gray}{RGB}{128,128,128}
\definecolor{lightgray204}{RGB}{204,204,204}
\definecolor{seagreen4016769}{RGB}{40,167,69}
\definecolor{steelblue31119180}{RGB}{31,119,180}

\begin{axis}[
width=\fwidth,
height=\fheight,
axis line style={dimgray102},
legend cell align={left},
legend style={fill opacity=0.8, draw opacity=1, text opacity=1, draw=gray},
font=\scriptsize,
tick align=inside,
tick pos=left,
x grid style={lightgray204},
xlabel={Hour of the Week},
xmajorgrids,
xmin=0, xmax=175.35,
xminorgrids,
xtick style={color=black},
y grid style={lightgray204},
ylabel={Number of Activated nodes},
ymajorgrids,
ymin=2.15, ymax=20.85,
yminorgrids,
ytick style={color=black},
xmajorgrids,
ymajorgrids,
xlabel style={font=\scriptsize},
ylabel style={font=\scriptsize}
]
\addplot [semithick, darkorange25512714, forget plot]
table {%
0 16
3 18.5
4 18
5 18.5
6 19
7 18.25
8 18
9 16.75
10 15.5
11 14.75
12 14.25
13 15
14 15.5
15 15
16 15.5
17 14.75
18 15
19 15.25
20 16
21 16.75
22 16.5
23 18
24 17.25
25 17
27 18.5
28 17.75
29 17.5
30 17.5
31 16.5
32 16.75
33 17.5
34 16.25
35 15.5
36 15.5
37 15.25
38 15.5
39 16.25
40 14.75
41 14.75
42 15.25
43 15
44 15.5
45 16
46 17
47 17.5
48 17.75
49 18.25
50 17.75
51 18.5
52 18.75
53 18.5
54 17.75
56 16.75
57 16.5
58 15.5
59 15.75
60 16
61 16.75
62 15.75
63 15
64 14.5
65 13.75
66 14.5
67 16.25
68 17.75
69 18.75
70 18.75
71 19
72 17.5
73 18
74 19.5
75 19.75
76 20
77 19.75
78 18
80 16.25
81 15.25
82 13
83 14
84 15
85 15.5
86 16.25
87 15.75
88 14.5
89 15
90 15.25
91 14.75
92 15.5
93 15.75
94 16.5
95 18
96 17.75
97 17.5
99 17.25
100 16.5
101 17.75
102 17.75
103 17.5
104 17
105 15
106 15.25
107 15.25
108 16
109 17
110 16.5
111 16.25
112 15.25
113 15.75
114 16.25
116 17.25
117 17.5
118 18.25
119 18.75
120 18
121 18.5
122 17.5
123 16.75
124 17
125 16.5
126 15.75
127 15
128 14
129 13.75
130 13.75
131 14.5
132 15.25
133 16
134 16.5
135 16.75
136 17.25
137 16.25
138 16.5
139 15.75
140 15.5
141 16
142 15.5
143 16.5
144 17.25
145 17.75
146 18.25
147 17.75
148 17
149 17
150 17.75
151 17.25
152 18
153 17.25
154 15.75
155 16.25
156 14.5
157 15
158 15.75
159 14.5
160 15.75
161 16
162 15.75
163 16.75
164 16.5
165 16.75
166 17.5
167 18.5
};
\addplot [semithick, steelblue31119180, dashed, forget plot]
table {%
0 6
3 4.5
4 4.66666666666667
5 5.25
6 5.25
7 6
8 6.5
9 6.5
10 6.75
11 7
12 7
13 7
14 7
15 7
16 7
17 7
18 7
19 6.75
20 6.5
21 6.5
22 6.25
23 6.5
24 6.5
25 5.5
27 4.75
28 4.25
29 4.5
30 5.5
31 6.5
32 7
33 7
34 7
35 7
36 7
37 7
38 7
39 7
40 7
41 7
42 6.75
43 6.75
44 6.75
45 6.75
46 6.75
47 6.5
48 6.5
49 5.5
50 4.75
51 4
52 3.75
53 4.5
54 5.5
55 6.25
56 6.5
57 6.75
58 6.75
59 7
60 7
61 7
62 7
63 7
64 7
65 7
66 7
67 6.75
68 6.5
69 6.5
70 6.25
71 6.5
72 6.5
73 5.5
74 4.75
75 3.75
76 3
77 4
78 5
79 5.75
80 6.75
81 6.75
82 6.75
83 7
84 7
85 7
86 7
87 7
88 7
89 7
90 7
91 6.75
92 6.5
93 6.5
94 6.25
95 6.25
96 6.5
97 6
99 5.25
100 5
101 5
102 5.5
103 6.5
104 7
105 7
106 7
107 7
108 7
109 7
110 7
111 7
112 7
113 7
114 6.75
115 6.5
116 6.25
117 6.25
118 6.25
119 6.25
120 6.25
121 5.25
122 4.5
123 3.75
124 3
125 4
126 5
127 6
128 7
129 7
130 7
131 7
132 7
133 6.75
134 6.75
135 6.75
136 6.75
137 6.75
138 6.5
139 6.5
140 6.5
141 6.75
142 7
143 6.75
144 6.75
145 5.75
146 5.25
147 4.5
148 4
149 4.5
150 5
151 5.75
152 6
153 6.25
154 6.25
155 6.5
156 6.75
157 7
158 7
159 7
160 7
161 6.75
162 6.75
163 6.5
164 6.5
165 6.5
166 6.5
167 6.5
};
\addplot [semithick, seagreen4016769, dashed, forget plot]
table {%
0 14
1 14
2 13.3333333333333
3 13
4 12.25
5 12.25
6 12
7 10.5
8 10.5
9 11.75
10 11
11 12
12 11.75
13 9.75
14 11.5
15 10.75
16 12
17 11
18 9.25
19 9.25
20 7.5
21 8.5
22 9.5
23 10
24 11.5
25 11
26 11.5
27 11.25
28 10
29 11.25
30 10.5
31 10.75
32 11
33 10.75
34 11
35 10.25
36 9.75
37 8.75
38 8.25
39 8.5
40 8
41 8
42 7
43 7
44 7
45 6.5
46 7
47 7.25
48 8.75
49 10.5
50 12
51 13.5
52 13
53 11.75
54 11.75
55 10.75
56 10.25
57 9.75
58 7.75
59 7.75
60 8.25
61 8.5
62 9
63 8.5
64 8
65 8
66 8.5
67 8.25
68 8
69 9
70 9.5
71 11.25
72 11.75
73 12.25
74 11.25
75 11.25
76 12
77 11
78 12
79 10.75
80 10.5
81 10.5
82 10
83 9.5
84 9.5
85 8.75
86 8
87 8
88 7
89 7.75
90 7.75
91 8.5
92 8.75
93 8.75
94 11
95 12.5
96 12.75
97 13.75
98 13.25
99 12.75
100 12.75
101 11.5
102 10
103 8.5
104 7.75
105 7.5
106 8
107 9
108 10
109 9.75
110 9.75
111 8.5
112 7.75
113 8
114 7.75
115 7.5
116 7.5
117 8.25
118 9.25
119 10.75
120 12
121 13.25
122 13
123 14
124 15.25
125 14.25
126 14.75
127 13.75
128 11.5
129 11
130 9.75
131 8.25
132 7.75
133 7.25
134 7.25
135 7.75
136 8.25
137 8.5
138 8
139 7.75
140 7.5
141 7
142 7
143 8.25
144 8.75
145 10.75
146 13.25
147 13.75
148 14.5
149 12.75
150 11
151 11.5
152 10.75
153 10.75
154 10
155 8
156 7.25
157 6.75
158 6.75
159 6.25
160 6.5
161 7.5
162 7.75
163 7.75
164 9
165 8.75
166 11
167 13
};
\addplot [draw=darkorange25512714, fill=white, mark=*, mark size=1.25, only marks]
table{%
x  y
0 16
3 18.5
4 18
5 18.5
6 19
7 18.25
8 18
9 16.75
10 15.5
11 14.75
12 14.25
13 15
14 15.5
15 15
16 15.5
17 14.75
18 15
19 15.25
20 16
21 16.75
22 16.5
23 18
24 17.25
25 17
27 18.5
28 17.75
29 17.5
30 17.5
31 16.5
32 16.75
33 17.5
34 16.25
35 15.5
36 15.5
37 15.25
38 15.5
39 16.25
40 14.75
41 14.75
42 15.25
43 15
44 15.5
45 16
46 17
47 17.5
48 17.75
49 18.25
50 17.75
51 18.5
52 18.75
53 18.5
54 17.75
56 16.75
57 16.5
58 15.5
59 15.75
60 16
61 16.75
62 15.75
63 15
64 14.5
65 13.75
66 14.5
67 16.25
68 17.75
69 18.75
70 18.75
71 19
72 17.5
73 18
74 19.5
75 19.75
76 20
77 19.75
78 18
80 16.25
81 15.25
82 13
83 14
84 15
85 15.5
86 16.25
87 15.75
88 14.5
89 15
90 15.25
91 14.75
92 15.5
93 15.75
94 16.5
95 18
96 17.75
97 17.5
99 17.25
100 16.5
101 17.75
102 17.75
103 17.5
104 17
105 15
106 15.25
107 15.25
108 16
109 17
110 16.5
111 16.25
112 15.25
113 15.75
114 16.25
116 17.25
117 17.5
118 18.25
119 18.75
120 18
121 18.5
122 17.5
123 16.75
124 17
125 16.5
126 15.75
127 15
128 14
129 13.75
130 13.75
131 14.5
132 15.25
133 16
134 16.5
135 16.75
136 17.25
137 16.25
138 16.5
139 15.75
140 15.5
141 16
142 15.5
143 16.5
144 17.25
145 17.75
146 18.25
147 17.75
148 17
149 17
150 17.75
151 17.25
152 18
153 17.25
154 15.75
155 16.25
156 14.5
157 15
158 15.75
159 14.5
160 15.75
161 16
162 15.75
163 16.75
164 16.5
165 16.75
166 17.5
167 18.5
};
\addlegendentry{Local Search}
\addplot [draw=steelblue31119180, fill=white, mark=square*, mark size=1.25, only marks]
table{%
x  y
0 6
3 4.5
4 4.66666666666667
5 5.25
6 5.25
7 6
8 6.5
9 6.5
10 6.75
11 7
12 7
13 7
14 7
15 7
16 7
17 7
18 7
19 6.75
20 6.5
21 6.5
22 6.25
23 6.5
24 6.5
25 5.5
27 4.75
28 4.25
29 4.5
30 5.5
31 6.5
32 7
33 7
34 7
35 7
36 7
37 7
38 7
39 7
40 7
41 7
42 6.75
43 6.75
44 6.75
45 6.75
46 6.75
47 6.5
48 6.5
49 5.5
50 4.75
51 4
52 3.75
53 4.5
54 5.5
55 6.25
56 6.5
57 6.75
58 6.75
59 7
60 7
61 7
62 7
63 7
64 7
65 7
66 7
67 6.75
68 6.5
69 6.5
70 6.25
71 6.5
72 6.5
73 5.5
74 4.75
75 3.75
76 3
77 4
78 5
79 5.75
80 6.75
81 6.75
82 6.75
83 7
84 7
85 7
86 7
87 7
88 7
89 7
90 7
91 6.75
92 6.5
93 6.5
94 6.25
95 6.25
96 6.5
97 6
99 5.25
100 5
101 5
102 5.5
103 6.5
104 7
105 7
106 7
107 7
108 7
109 7
110 7
111 7
112 7
113 7
114 6.75
115 6.5
116 6.25
117 6.25
118 6.25
119 6.25
120 6.25
121 5.25
122 4.5
123 3.75
124 3
125 4
126 5
127 6
128 7
129 7
130 7
131 7
132 7
133 6.75
134 6.75
135 6.75
136 6.75
137 6.75
138 6.5
139 6.5
140 6.5
141 6.75
142 7
143 6.75
144 6.75
145 5.75
146 5.25
147 4.5
148 4
149 4.5
150 5
151 5.75
152 6
153 6.25
154 6.25
155 6.5
156 6.75
157 7
158 7
159 7
160 7
161 6.75
162 6.75
163 6.5
164 6.5
165 6.5
166 6.5
167 6.5
};
\addlegendentry{Greedy}
\addplot [draw=seagreen4016769, fill=white, mark=triangle*, mark size=1.25, only marks]
table{%
x  y
0 14
3 13
4 12.25
5 12.25
6 12
7 10.5
8 10.5
9 11.75
10 11
11 12
12 11.75
13 9.75
14 11.5
15 10.75
16 12
17 11
18 9.25
19 9.25
20 7.5
21 8.5
22 9.5
23 10
24 11.5
25 11
27 11.25
28 10
29 11.25
30 10.5
31 10.75
32 11
33 10.75
34 11
35 10.25
36 9.75
37 8.75
38 8.25
39 8.5
40 8
41 8
42 7
43 7
44 7
45 6.5
46 7
47 7.25
48 8.75
49 10.5
50 12
51 13.5
52 13
53 11.75
54 11.75
56 10.25
57 9.75
58 7.75
59 7.75
60 8.25
61 8.5
62 9
63 8.5
64 8
65 8
66 8.5
67 8.25
68 8
69 9
70 9.5
71 11.25
72 11.75
73 12.25
74 11.25
75 11.25
76 12
77 11
78 12
80 10.5
81 10.5
82 10
83 9.5
84 9.5
85 8.75
86 8
87 8
88 7
89 7.75
90 7.75
91 8.5
92 8.75
93 8.75
94 11
95 12.5
96 12.75
97 13.75
99 12.75
100 12.75
101 11.5
102 10
103 8.5
104 7.75
105 7.5
106 8
107 9
108 10
109 9.75
110 9.75
111 8.5
112 7.75
113 8
114 7.75
116 7.5
117 8.25
118 9.25
119 10.75
120 12
121 13.25
122 13
123 14
124 15.25
125 14.25
126 14.75
127 13.75
128 11.5
129 11
130 9.75
131 8.25
132 7.75
133 7.25
134 7.25
135 7.75
136 8.25
137 8.5
138 8
139 7.75
140 7.5
141 7
142 7
143 8.25
144 8.75
145 10.75
146 13.25
147 13.75
148 14.5
149 12.75
150 11
151 11.5
152 10.75
153 10.75
154 10
155 8
156 7.25
157 6.75
158 6.75
159 6.25
160 6.5
161 7.5
162 7.75
163 7.75
164 9
165 8.75
166 11
167 13
};
\addlegendentry{Selective-reduction}
\addplot [draw=red, fill=white, forget plot, mark=triangle*, mark size=1.25, only marks]
table{%
x  y
1 14
2 13.3333333333333
26 11.5
55 10.75
79 10.75
98 13.25
115 7.5
};
\end{axis}

\end{tikzpicture}
        \caption{RF activation – Scenario 2}
        \label{fig:ru_10}
    \end{subfigure}
    
   % \vspace{0.4cm}
   
    \begin{subfigure}{0.49\linewidth}
        \centering
        \setlength\fwidth{0.9\columnwidth} 
        \setlength\fheight{0.45\columnwidth}
        \input{figures/power_consumption_3.6GHz_5Mbps_new.tex}
        \caption{Power consumption – Scenario 1}
        \label{fig:power5}
    \end{subfigure}
    \hfill
    \begin{subfigure}{0.49\linewidth}
        \centering
        \setlength\fwidth{0.9\columnwidth} 
        \setlength\fheight{0.45\columnwidth}
        \input{figures/station_power_consumption_3.6GHz_5Mbps_new.tex}
        \caption{Power consumption – Scenario 2}
        \label{fig:power10}
    \end{subfigure}

    %\vspace{0.4cm}

    \begin{subfigure}{0.49\linewidth}
        \centering
        \setlength\fwidth{0.9\columnwidth} 
        \setlength\fheight{0.45\columnwidth}
        % This file was created with tikzplotlib v0.10.1.
\begin{tikzpicture}

\definecolor{dimgray102}{RGB}{102,102,102}
\definecolor{gray}{RGB}{128,128,128}
\definecolor{lightgray204}{RGB}{204,204,204}
\definecolor{steelblue31119180}{RGB}{31,119,180}

\pgfplotsset{every tick label/.append style={font=\scriptsize}}

\begin{axis}[
width=\fwidth,
height=\fheight,
axis line style={dimgray102},
legend cell align={left},
legend style={fill opacity=0.8, draw opacity=1, text opacity=1, draw=gray},
font=\scriptsize,
tick align=outside,
tick pos=left,
x grid style={lightgray204},
xlabel={Hour of the Week},
xmajorgrids,
xmin=-8.35, xmax=175.35,
xminorgrids,
xtick style={color=black},
y grid style={lightgray204},
ylabel={Number of UEs failed},
ymajorgrids,
ymin=-1.075, ymax=22.575,
yminorgrids,
ytick style={color=black},
xmajorgrids,
ymajorgrids,
xlabel style={font=\scriptsize},
ylabel style={font=\scriptsize}
]
\addplot [semithick, steelblue31119180, dashed, forget plot]
table {%
0 0
1 0
2 0
3 0
4 0
5 0
6 0
7 2
8 6
9 11.75
10 15.75
11 19.5
12 21
13 19.75
14 21.25
15 21.25
16 21.25
17 20.75
18 16.75
19 11
20 5.5
21 1.5
22 0
23 0
24 0
25 0
26 0
27 0
28 0
29 0
30 0
31 1.5
32 6
33 10
34 15.5
35 18
36 17.5
37 17.5
38 16.5
39 16.5
40 16.5
41 17
42 14.5
43 10.5
44 6.5
45 2
46 0
47 0
48 0
49 0
50 0
51 0
52 0
53 0
54 0
55 1
56 4
57 8
58 12.75
59 16.5
60 18
61 16.5
62 14.75
63 13
64 11.5
65 12
66 10.5
67 7.5
68 4.5
69 1.5
70 0
71 0
72 0
73 0
74 0
75 0
76 0
77 0
78 0
79 1.5
80 4.5
81 9
82 14.75
83 19
84 21.5
85 21
86 19.25
87 17.5
88 16
89 16.5
90 14.5
91 11.5
92 7.5
93 3
94 1
95 0
96 0
97 0
98 0
99 0
100 0
101 0
102 0
103 1
104 3.75
105 6.75
106 11.25
107 14.75
108 16.75
109 16.75
110 16.75
111 15.25
112 15.25
113 15.25
114 12.25
115 9.25
116 4.5
117 1.5
118 0
119 0
120 0
121 0
122 0
123 0
124 0
125 0
126 0
127 0
128 0
129 1
130 3.5
131 4.75
132 5.75
133 5.75
134 4.75
135 6.5
136 8.5
137 10
138 10
139 7
140 4
141 1.5
142 0
143 0
144 0
145 0
146 0
147 0
148 0
149 0
150 0
151 0
152 0
153 0
154 2
155 3.5
156 4.5
157 5.5
158 5.5
159 7
160 9
161 12.5
162 11.5
163 8.5
164 5.5
165 1
166 0
167 0
};
\addplot [draw=steelblue31119180, fill=white, mark=square*,mark size=1.25, only marks]
table{%
x  y
0 0
1 0
2 0
3 0
4 0
5 0
6 0
7 2
8 6
9 11.75
10 15.75
11 19.5
12 21
13 19.75
14 21.25
15 21.25
16 21.25
17 20.75
18 16.75
19 11
20 5.5
21 1.5
22 0
23 0
24 0
25 0
26 0
27 0
28 0
29 0
30 0
31 1.5
32 6
33 10
34 15.5
35 18
36 17.5
37 17.5
38 16.5
39 16.5
40 16.5
41 17
42 14.5
43 10.5
44 6.5
45 2
46 0
47 0
48 0
49 0
50 0
51 0
52 0
53 0
54 0
55 1
56 4
57 8
58 12.75
59 16.5
60 18
61 16.5
62 14.75
63 13
64 11.5
65 12
66 10.5
67 7.5
68 4.5
69 1.5
70 0
71 0
72 0
73 0
74 0
75 0
76 0
77 0
78 0
79 1.5
80 4.5
81 9
82 14.75
83 19
84 21.5
85 21
86 19.25
87 17.5
88 16
89 16.5
90 14.5
91 11.5
92 7.5
93 3
94 1
95 0
96 0
97 0
98 0
99 0
100 0
101 0
102 0
103 1
104 3.75
105 6.75
106 11.25
107 14.75
108 16.75
109 16.75
110 16.75
111 15.25
112 15.25
113 15.25
114 12.25
115 9.25
116 4.5
117 1.5
118 0
119 0
120 0
121 0
122 0
123 0
124 0
125 0
126 0
127 0
128 0
129 1
130 3.5
131 4.75
132 5.75
133 5.75
134 4.75
135 6.5
136 8.5
137 10
138 10
139 7
140 4
141 1.5
142 0
143 0
144 0
145 0
146 0
147 0
148 0
149 0
150 0
151 0
152 0
153 0
154 2
155 3.5
156 4.5
157 5.5
158 5.5
159 7
160 9
161 12.5
162 11.5
163 8.5
164 5.5
165 1
166 0
167 0
};
\addlegendentry{Greedy}
\end{axis}

\end{tikzpicture}
        \caption{UEs not served - Scenario 1}
        \label{fig:plot5}
    \end{subfigure}
    \hfill
    \begin{subfigure}{0.49\linewidth}
        \centering
        \setlength\fwidth{0.9\columnwidth} 
        \setlength\fheight{0.45\columnwidth}
        % This file was created with tikzplotlib v0.10.1.
\begin{tikzpicture}

\definecolor{dimgray102}{RGB}{102,102,102}
\definecolor{gray}{RGB}{128,128,128}
\definecolor{lightgray204}{RGB}{204,204,204}
\definecolor{steelblue31119180}{RGB}{31,119,180}
\pgfplotsset{every tick label/.append style={font=\scriptsize}}

\begin{axis}[
width=\fwidth,
height=\fheight,
axis line style={dimgray102},
legend cell align={left},
legend style={fill opacity=0.8, draw opacity=1, text opacity=1, draw=gray},
font=\scriptsize,
tick align=inside,
tick pos=left,
x grid style={lightgray204},
xlabel={Hour of the Week},
xmajorgrids,
xmin=0, xmax=175.35,
xminorgrids,
xtick style={color=black},
y grid style={lightgray204},
ylabel={Number of UEs failed},
ymajorgrids,
ymin=-2.525, ymax=53.025,
yminorgrids,
ytick style={color=black},
xmajorgrids,
ymajorgrids,
xlabel style={font=\scriptsize},
ylabel style={font=\scriptsize}
]
\addplot [semithick, steelblue31119180, dashed, forget plot]
table {%
0 0
3 0
4 0.333333333333333
5 0.75
6 1.75
7 4.25
8 11.75
9 21.75
10 32.75
11 40.75
12 42
13 40
14 38.25
15 39.25
16 40.75
17 39.75
18 33.5
19 23.25
20 13.75
21 6.75
22 3.25
23 2.5
24 1.5
25 1
27 0.5
28 0.25
29 0.5
30 1.25
31 3.75
32 12
33 20.25
34 32.5
35 43.25
36 46.75
37 50.5
38 49
39 46.5
40 43.75
41 37.5
42 28.25
43 19.25
44 10.75
45 5.25
46 3.5
47 2.25
48 1.75
49 1.25
50 0.75
51 0.25
52 0
53 0.5
54 1.25
55 2.75
56 7
57 12.75
58 21.25
59 30.25
60 33.25
61 35.5
62 33
63 30.25
64 31.5
65 27.75
66 23.75
67 17.25
68 9.75
69 5.5
70 3.25
71 2.5
72 1.5
73 1
74 0.5
75 0
76 0
77 0.5
78 1.25
79 2.75
80 7
81 13.75
82 20.75
83 28
84 32.25
85 32.5
86 32.25
87 32
88 35
89 35.5
90 32.75
91 25.5
92 15
93 7.5
94 3.25
95 2.5
96 1.75
97 1.5
99 1
100 0.75
101 1
102 1.5
103 4.25
104 8.75
105 13.5
106 21.25
107 27
108 30.75
109 34
110 33.5
111 32.25
112 33
113 29.5
114 23.75
115 18
116 9.75
117 5.5
118 3.75
119 2.75
120 1.75
121 1
122 0.5
123 0
124 0
125 0.25
126 1
127 1.75
128 3.25
129 5.5
130 7.75
131 11
132 12.75
133 12.75
134 13.75
135 14.5
136 16.25
137 16
138 13.5
139 10.5
140 6.25
141 4.75
142 4
143 2.75
144 2.5
145 1.75
146 1.25
147 0.75
148 0.5
149 0.75
150 1
151 1.5
152 2.25
153 3.25
154 5.75
155 8
156 9.75
157 11.25
158 10.75
159 11.75
160 14
161 13.5
162 14.25
163 12
164 8.75
165 7.5
166 4.75
167 3.75
};
\addplot [draw=steelblue31119180, fill=white, mark=square*,mark size=1.25, only marks]
table{%
x  y
0 0
3 0
4 0.333333333333333
5 0.75
6 1.75
7 4.25
8 11.75
9 21.75
10 32.75
11 40.75
12 42
13 40
14 38.25
15 39.25
16 40.75
17 39.75
18 33.5
19 23.25
20 13.75
21 6.75
22 3.25
23 2.5
24 1.5
25 1
27 0.5
28 0.25
29 0.5
30 1.25
31 3.75
32 12
33 20.25
34 32.5
35 43.25
36 46.75
37 50.5
38 49
39 46.5
40 43.75
41 37.5
42 28.25
43 19.25
44 10.75
45 5.25
46 3.5
47 2.25
48 1.75
49 1.25
50 0.75
51 0.25
52 0
53 0.5
54 1.25
55 2.75
56 7
57 12.75
58 21.25
59 30.25
60 33.25
61 35.5
62 33
63 30.25
64 31.5
65 27.75
66 23.75
67 17.25
68 9.75
69 5.5
70 3.25
71 2.5
72 1.5
73 1
74 0.5
75 0
76 0
77 0.5
78 1.25
79 2.75
80 7
81 13.75
82 20.75
83 28
84 32.25
85 32.5
86 32.25
87 32
88 35
89 35.5
90 32.75
91 25.5
92 15
93 7.5
94 3.25
95 2.5
96 1.75
97 1.5
99 1
100 0.75
101 1
102 1.5
103 4.25
104 8.75
105 13.5
106 21.25
107 27
108 30.75
109 34
110 33.5
111 32.25
112 33
113 29.5
114 23.75
115 18
116 9.75
117 5.5
118 3.75
119 2.75
120 1.75
121 1
122 0.5
123 0
124 0
125 0.25
126 1
127 1.75
128 3.25
129 5.5
130 7.75
131 11
132 12.75
133 12.75
134 13.75
135 14.5
136 16.25
137 16
138 13.5
139 10.5
140 6.25
141 4.75
142 4
143 2.75
144 2.5
145 1.75
146 1.25
147 0.75
148 0.5
149 0.75
150 1
151 1.5
152 2.25
153 3.25
154 5.75
155 8
156 9.75
157 11.25
158 10.75
159 11.75
160 14
161 13.5
162 14.25
163 12
164 8.75
165 7.5
166 4.75
167 3.75
};
\addlegendentry{Greedy}
\end{axis}

\end{tikzpicture}
        \caption{UEs not served - Scenario 2}
        \label{fig:plot6}
    \end{subfigure}

    \caption{RF activation patterns and power consumption over a week at 3.6\,GHz with 100\,MHz bandwidth across different scenarios.}
    \label{fig:combined_results_fr1}
    \vspace{-.4cm}
\end{figure*}

\begin{figure}[!t]
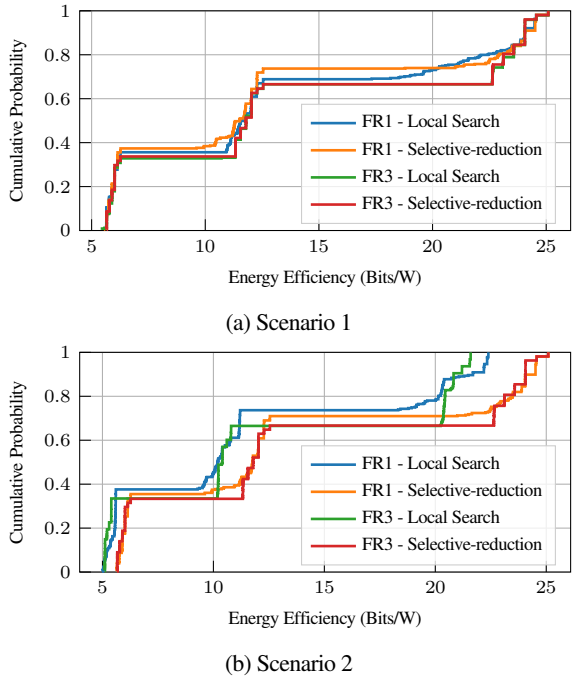

    \centering
    \begin{subfigure}{\linewidth}
        \centering
        \setlength\fwidth{0.9\columnwidth} 
        \setlength\fheight{0.35\columnwidth}
        \input{figures/empirical_cdf_sc1_new.tex}
        \caption{Scenario 1}
        \label{fig:cdf_sc1}
    \end{subfigure}
    \hfill
    \begin{subfigure}{\linewidth}
        \centering
        \setlength\fwidth{0.9\columnwidth} 
        \setlength\fheight{0.35\columnwidth}
        \input{figures/empirical_cdf_sc2_new.tex}
        \caption{Scenario 2}
        \label{fig:cdf_sc2}
    \end{subfigure}
    \caption{Energy efficiency CDF comparison across scenarios.}
    \label{fig:cdf_comparison}
    \vspace{-0.5cm}
\end{figure}

The results reveal a distinct contrast between peak and non-peak hours. During non-peak hours, for instance, hours 0-4 or 18-21, the practical algorithm consistently achieves substantial improvements, with high throughput gains. Also, the algorithm converges relatively quickly.
In contrast, there is little or no improvement in the solution despite extended execution times during peak hours. This implies that as the solution space becomes constrained under heavy load, time to reach an initial solution is large and the potential for the proposed method to identify significant improvements gets limited.

\subsection{Results on Energy Minimization}
%\begin{itemize}
    %\item Results of small-scale instance
    %\item To evaluate the performance in terms of energy, we analyze the power and number of activated nodes for different UE throughput requirements.
%    \item Run-time
    %\item We also measure the energy efficiency as the energy spent on transmission, defined as the ratio of total power consumption to the system's achieved throughput.

%\bc{Similar to throughput optimization analysis, we first present the results for small-scale instance such that the problem can be solved optimally?. }

To evaluate the performance in terms of energy, we analyze number of activated nodes and power consumption for different UE throughput requirements.  The minimum throughput requirement is set as 5 Mbps. 
The results for FR1 is presented in  Fig. \ref{fig:combined_results_fr1}. Results for FR3 are omitted due to space constraints but follow similar trends.
Red markers indicate hours where only one large-scale method yields a solution.
From the figures, we observe that the number of activated \gls{rf} frontends are fewer for the selective-reduction method compared to local-search despite higher power consumption. The result is consistent for both scenarios. This outcome is expected  as selective-reduction method operates on a reduced set  of connections in the \gls{iab} network. The greedy baseline activates the fewest nodes as it builds a minimal tree using only gain-pathloss ranking.

%This inverse relationship follows from the input space of both the approaches.  
%selective-reduction method operates on a reduced set  of nodes requiring higher power allocation. Meanwhile, local-search method can distribute power over a larger number of activated nodes leading to lower per user power levels. 

Interestingly, we observe that the power consumption under selective-reduction method is almost similar to that of local-search method, with slight differences, especially at the peak time of the day in scenario 1.  But at the same time, power consumption under selective-reduction method is much lower than that of local-search method in scenario 2. The observed variation in performance between two scenarios can be attributed to the cell-load profiles. The heavy load conditions of scenario 1, together with reduction in the edges when using  selective-reduction method leads to the slightly higher power allocation. Meanwhile, under lighter load conditions, reduction in network size becomes advantageous. local search method, on the other hand, distributes power over a larger number of activated nodes, leading to higher overall power consumption. Despite activating the fewest nodes, greedy consumes the highest power  because all active nodes operate at fixed power. This fixed-power approach is not a design choice but a necessity: power optimization on the greedy topology is infeasible because greedy link selection during  tree construction creates connections that only work at maximum power as reducing power would violate capacity constraints. Beyond power consumption, Fig.~\ref{fig:plot5},~\ref{fig:plot6} show the number of \glspl{ue} not served by the greedy baseline across the week. By construction, our proposed methods either serve all \glspl{ue} or return no solution, they cannot produce a partial solution that leaves \glspl{ue} unserved. The greedy baseline, however, always produces a topology regardless of load but cannot guarantee minimum throughput for all \glspl{ue}, leaving an increasing number of \glspl{ue} unserved as network load grows. This underscores that joint optimization is necessary not only for energy efficiency but also for guaranteed service reliability. %This demonstrates that topology and power must be optimized jointly: topology selection determines which power allocations are feasible. 

%smooth and predictable load variations, is more amenable to local search approaches, which can iteratively refine solutions without being hindered by high variability. selective-reduction method—which aggressively reduce the search space—demonstrate superior performance in terms of energy efficiency and throughput in th

To get more insights on the network performance-power trade-offs, we present the energy efficiency as the energy spent on transmission, defined as the ratio of achieved throughput to the total power consumption \cite{li2024energy,quek2011energy}.  This can be represented as
\begin{equation}
    \eta(t) = \frac{B_k}{P_{total}(t)},
\end{equation} where $P_{total}(t)$ is the total power required at the time period $t$ and $B_k$ is the throughput requirement that was guaranteed for commodity $k$.

Figure \ref{fig:cdf_comparison} presents the energy efficiency CDF of the algorithms under both scenarios. The CDF is computed by measuring energy efficiency across different throughput requirements to assess the consistency of algorithm performance. 
The energy efficiency CDF indicates that the selective-reduction method outperforms local-search method in terms of higher energy efficiency values and thus yields better throughput per unit power.  The difference is quite evident in scenario 2 than in scenario 1. This result shows how well the energy is utilized by the selective-reduction method.%But in contrast to the scenario 1, the trend reverses in scenario 2 where selective-reduction method have higher energy efficiency than local-search method. %This divergence suggests that the relative performance of these approaches is influenced by the underlying variance in traffic. Local search offer less power consumption under smoother, more predictable conditions, while pruning-based strategies are better suited for scenarios where load conditions have higher variance, allowing for more effective optimization in terms of power and RU node activation.

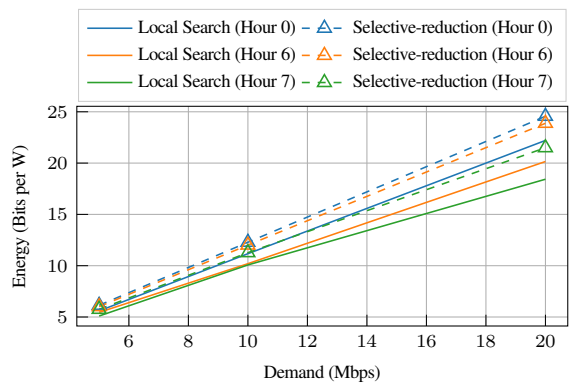
\begin{figure}[!h]
    \centering
    %\begin{subfigure}[c]{\linewidth}
    %    \centering
    %    \setlength\fwidth{0.5\columnwidth} 
    %    \setlength\fheight{0.5\columnwidth}
    %    \input{figures/energy_demand.tex}
        
    %    \caption{Scenario 1}
    %    \label{fig:energy_demand}
    %\end{subfigure}
    %\hfill
    %\begin{subfigure}[c]{\linewidth}
    %    \centering
        \setlength\fwidth{0.9\columnwidth} 
        \setlength\fheight{0.35\columnwidth}
        % This file was created with tikzplotlib v0.10.1.
\begin{tikzpicture}
\definecolor{darkgray176}{RGB}{176,176,176}
\definecolor{darkorange25512714}{RGB}{255,127,14}
\definecolor{forestgreen4416044}{RGB}{44,160,44}
\definecolor{lightgray204}{RGB}{204,204,204}
\definecolor{steelblue31119180}{RGB}{31,119,180}
\pgfplotsset{every tick label/.append style={font=\scriptsize}}
\begin{axis}[
width=\fwidth,
height=\fheight,
legend cell align={left},
legend style={
  fill opacity=0.8,
  draw opacity=1,
  text opacity=1,
  at={(0.5,1.02)},
  anchor=south,
  draw=lightgray204,
  font=\scriptsize,
  legend columns=2
},
tick align=inside,
tick pos=left,
x grid style={darkgray176},
xlabel={Demand [Mbps]},
xmajorgrids,
xmin=4.25, xmax=20.75,
xtick style={color=black},
y grid style={darkgray176},
ylabel={Energy [Bits/W])},
ymajorgrids,
ymin=4.12297742415866, ymax=25.5351384408252,
ytick style={color=black},
xmajorgrids,
ymajorgrids,
xlabel style={font=\scriptsize},
ylabel style={font=\scriptsize}
]
\addplot [semithick, steelblue31119180, mark=, mark size=3, mark options={solid}]
table {%
5 5.5949157842071
10 11.1689788438516
20 22.2066056472473
};
\addlegendentry{Local Search (Hour 0)}
\addplot [semithick, steelblue31119180, dashed, mark=triangle, mark size=3, mark options={solid}]
table {%
5 6.14177844469217
10 12.282776689708
20 24.561858394613
};
\addlegendentry{Selective-reduction (Hour 0)}
\addplot [semithick, darkorange25512714, mark=, mark size=3, mark options={solid}]
table {%
5 5.44964277603352
10 10.1960965035162
20 20.1526284358277
};
\addlegendentry{Local Search (Hour 6)}
\addplot [semithick, darkorange25512714, dashed, mark=triangle, mark size=3, mark options={solid}]
table {%
5 6.01017652086476
10 12.007783378747
20 23.8771576630387
};
\addlegendentry{Selective-reduction (Hour 6)}
\addplot [semithick, forestgreen4416044, mark=, mark size=3, mark options={solid}]
table {%
5 5.09625747037077
10 10.073913532401
20 18.4269495990445
};
\addlegendentry{Local Search (Hour 7)}
\addplot [semithick, forestgreen4416044, dashed, mark=triangle, mark size=3, mark options={solid}]
table {%
5 5.74847620881888
10 11.2707071491242
20 21.4971052254306
};
\addlegendentry{Selective-reduction (Hour 7)}
\end{axis}
\end{tikzpicture}
    %    \caption{Scenario 2}
    %    \label{fig:energy_demand_station}
    %\end{subfigure}
    \caption{Impact of data rate demand on energy efficiency (scenario 2, FR3).}
    \label{fig:energy}
    \vspace{-0.3cm}
\end{figure}

We now examine the sensitivity of energy efficiency to the minimum throughput demand in peak and non-peak hours.  Figure~ \ref{fig:energy} presents the variation of  energy efficiency with minimum throughput requirement at 3.6 GHz.
The plot shows that both algorithms become more energy efficient as minimum throughput requirements increase from 5 to 20 Mbps. This indicates that the fixed power overhead becomes less significant with higher bandwidth requirements and the power cost per additional bandwidth decreases as scale increases. %We observe that both approaches result in similar performance. 
At the same time, the performance gap between light and heavy load conditions widens significantly as throughput requirements increase in scenario 2. At 5 Mbps, both algorithms show minimal sensitivity to load conditions.
At 20 Mbps, load sensitivity increases. 
%\begin{figure}[h]
    
%        \centering
%        \includegraphics[width=0.5\textwidth]{figures/combined_energy_vs_demand_3.6GHz.png}
 %       \caption{Energy efficiency vs demand}
 %       \label{fig:energy}
%\end{figure}

%\begin{figure}[h]
    
 %       \centering
  %      \includegraphics[width=0.5\textwidth]{figures/efficiency_gap_3.6.png}
  %      \caption{Energy efficiency gap}
   %     \label{fig:energy_gap}
%\end{figure}

\section{Conclusion \label{sect:conclusion}}

In this paper, we formulated two optimization problems: a joint routing, resource allocation, and energy minimization problem, and a joint routing, resource allocation, and throughput maximization problem. To address these, we then presented two practical approaches and evaluated performance of the approaches under two realistic  scenarios. Our results show that selective-reduction method achieves better energy efficiency and throughput under dynamic conditions by activating fewer nodes in the \gls{iab} network. %Runtime of selective-reduction method can be effectively controlled, while still achieving good performance.  
Although the runtime of the local search method cannot be controlled like selective-reduction, the results show that it achieves better runtime under FR1 with the trade-off of throughput and number of activated nodes. This makes it a good candidate for lower frequency band deployments. Optimality gap analysis on small instances shows that our method achieves optimal or near-optimal solutions compared to exact MILP formulations. Comparison against a greedy baseline further confirms that sequential approaches, which independently determine topology, routing, and power allocation, cannot jointly guarantee energy efficiency and service reliability, underscoring the need for the proposed joint optimization framework in practical \gls{iab} deployment.
%\bc{Extending the proposed framework to \gls{dag}-based \gls{iab} topologies to support redundant routing paths and improve resilience against link failures represents a promising direction for future work.}
%In future, we plan to design heuristic algorithms with further reduction in runtime that enable real-time decision making.

\ifCLASSOPTIONcaptionsoff
  \newpage
\fi



%\bibliographystyle{IEEEtran}
\bibliographystyle{IEEEtran}
\bibliography{reference}
\bstctlcite{IEEEexample:BSTcontrol}


%\end{IEEEbiography}


%\newpage



%\vfill


\end{document}